\documentstyle[12pt,preprint,aps,epsfig]{revtex}
\tighten
\newcommand{\beq}{\begin{eqnarray}}
\newcommand{\eeq}{\end{eqnarray}}

\newcommand{\bsga}{ b \to s \gamma}
\newcommand{\bsg}{ b \to s g}

\newcommand{\brbsg}{ {\cal  B}( b \to s g) }
\newcommand{\bsl}{  {\cal B}_{SL}}

\newcommand{\mw}{ M_W }
\newcommand{\mhp}{ M_{H^{+}}}
\def\etap{\eta^{\prime}}
\def\etapp{\eta^{(')}}

\def\lsim{ {\ \lower-1.2pt\vbox{\hbox{\rlap{$<$}\lower5pt\vbox{\hbox{$\sim$}}}}\ } }
\def\gsim{ {\ \lower-1.2pt\vbox{\hbox{\rlap{$>$}\lower5pt\vbox{\hbox{$\sim$}}}}\ } }

\newcommand{\tab}[1]{Table \ref{#1}}
\newcommand{\fig}[1]{Fig.\ref{#1}}

\newcommand{\non}{\nonumber\\ }

\title{\bf{ Charmless hadronic decays $B \to PP, PV, VV$ and new
physics effects in the general two-Higgs doublet models} }
\author{ Zhenjun Xiao$^{(1,2)}$ \thanks{E-mail: zxiao@ibm320h.phy.pku.edu.cn},
Chong Sheng Li$^{(1)}$ \\
{\small 1. Department of Physics, Peking University, Beijing,
 100871, People's Republic of China }\\
{\small 2. Department of Physics, Henan Normal University, Xinxiang, Henan,
453002, People's Republic of China}  \\
 Kuang-Ta Chao \\
{\small  CCAST(World Laboratory), P.O.Box 8730, Beijing 100080,
People's Republic of China }\\
{\small and Department of Physics, Peking University, Beijing, 100871,
People's Republic of China}    }
\date{\today}
\begin{document}
\maketitle
\begin{abstract}
Based on the low-energy effective Hamiltonian with the generalized factorization,
we calculate the new physics contributions to the branching ratios of the
two-body charmless hadronic decays of $B_u$ and $B_d$ mesons induced by
the new gluonic and electroweak charged-Higgs penguin diagrams in the
general two-Higgs doublet models ( models I, II and  III). Within the
considered parameter space, we find that:
(a) the new physics effects from new gluonic penguin diagrams strongly dominate
over those from the new $\gamma$- and $Z^0$- penguin diagrams;
(b) in models I and II, new physics contributions to most studied B meson
decay channels are rather small in size: from $-15\%$ to $20\%$;
(c) in model III, however, the new physics enhancements to the penguin-dominated
decay modes can be significant, $\sim (30 -200)\%$, and therefore are measurable
in forthcoming high precision B experiments;
(d) the new physics enhancements to ratios ${\cal B}(B \to K \etap)$
are significant in model III, $\sim (35 -70)\%$, and hence provide a
simple and plausible new physics interpretation for the observed
unexpectedly large $B \to K \etap $ decay rates;
(e) the theoretical predictions for ${\cal B}(B \to K^+ \pi)$ and
${\cal B}(B \to K^0 \pi^+)$ in  model III are still consistent with
the data within $2\sigma$ errors;
(f) the significant new physics enhancements to the branching ratios
of $B \to K^0 \pi^0, K^* \eta,  K^{*+} \pi^-, K^+ \phi, K^{*0} \omega, K^{*+}
\phi$ and $K^{*0} \phi$ decays are helpful to improve the agreement
between the data and the theoretical predictions;
(g) the theoretical predictions of ${\cal B}(B \to PP, PV, VV )$ in the
2HDM's are generally consistent with experimental measurements and upper
limits ($90\% C.L.$)
\end{abstract}

\vspace{.5cm}

\newpage
\section{ Introduction } \label{sec:1}

The main objective of B experiments is to explore in detail the physics of
CP violation, to determine many of the flavor parameters of the
standard model (SM) at high precision, and to probe for possible
effects of new physics beyond the SM \cite{slac504,belle94,fj99}.
Precision measurements of B meson system can provide an insight into very
high energy scales via the indirect loop effects of new physics.
The B system therefore offers a complementary probe to the searches  for new
physics at Tevatron, LHC and NLC colliders \cite{slac504}.

In B experiments, new physics beyond the SM may manifest itself,
for example,  in the following two ways\cite{slac504,fj99}: (a) decays which are
expected to be rare in the SM are found to have large branching ratios;
(b) CP-violating asymmetries which are expected to vanish or be very small in the
SM are found to be significantly large or with a very different
pattern with what predicted in the SM. These potential deviations may be induced
by the virtual effects of new physics through loop diagrams.

It is well known that the two-body charmless hadronic decays $B \to h_1 h_2$
( where $h_1$ and $h_2$ are the light pseudo-scalar (P) and/or vector(V)
mesons ) play a very important role in studying CP violation and the
heavy flavor physics\cite{buras2008,cheng00b}. Several groups
\cite{du97,ali9804,ali9805,chen99} recently presented their
systematic calculations for these B decay channels in the SM by using the
low energy effective Hamiltonian \cite{bh1h2,buchalla96a,buras98}
with the generalized factorization approach
\cite{ali9804,bsw87,cheng98,ali98}.

Theoretically, the effective Hamiltonian  is our  basic tool to
calculate the branching ratios and CP-violating asymmetry $A_{CP}$
of B meson decays. The short and long distance quantum chromo-dynamical
(QCD) effects in the hadronic decays are
separated by means of the operator product expansion\cite{wilson69}.
The short-distance QCD corrected Lagrangian at next-to-leading order
(NLO) is available now, but we still do not know how to calculate
hadronic matrix element from the first
principles. One conventionally resort to the factorization ansatz
\cite{bsw87}.
However, we also know that the non-factorizable contribution
really exists and can not be neglected numerically for most hadronic B
decay channels. To remedy factorization hypothesis, some authors
\cite{ali9804,cheng98,ali98} introduced a
phenomenological parameter $N^{eff}$ (i.e. the effective number of color)
to model the non-factorizable contribution to hadronic matrix element,
which is commonly called the generalized factorization.
On the other hand, as pointed by Buras and Silvestrini \cite{bs98},
such generalization suffered from the problems of gauge  and infrared
dependence since the constant matrix $\hat{r}_V$ appeared
in the expressions of effective Wilson coefficients $C_i^{eff}$
depends on both the gauge chosen
and the external momenta. Very recently, Cheng {\it et al.} \cite{cheng99a}
studied and resolved above controversies on the gauge dependence and
infrared singularity of $C_i^{eff}$ by using the perturbative QCD
factorization theorem. In addition to the generalized factorization approach, a
new approach, called as the QCD factorization \cite{bbns2007},
appeared recently \cite{bbns2007,kls2004}, in which the decay amplitude is
described by a
kernel containing the `hard` interaction given by a perturbatively
evaluated effective Hamiltonian folded with form factors, decay constants
and light-cone distributions of mesons into which the long distance effects
are lumped. And some two-body hadronic
B meson decays, such as $B \to \pi \pi$ and $K\pi$ modes,
have been calculated in this approach\cite{bbns2007,kls2004,du2005}.

On the experimental side, CLEO collaboration reported the observations of
thirteen $B\to PP, PV $ decay channels and set new upper limits for many
other decay modes \cite{cleo99,cleo9912,cleo2001,cleosum}. The BaBar
and Belle collaboration at SLAC and KEK also presented their first observation
for some $B\to PP, PV$ decays at the ICHEP 2000 conference
\cite{babar2000,belle2000}.
Except for the decay channels $B \to K \etap$, the measured branching ratios
for $B \to h_1 h_2$ decays are generally in good agreement with the SM theoretical
predictions based on the effective Hamiltonian with factorization.
Unexpectedly large $B \to K \etap$ rate was
firstly reported by CLEO in 1997\cite{cleo98}, and confirmed very recently by
CLEO and BaBar Collaborations \cite{cleo9912,cleo-t0020,babar2000}.
Although many possible mechanisms such as gluon and/or charm content in $\etap$ and
the hairpin diagram have been considered in order to increase the theoretical
predictions of ${\cal B}(B \to \etap)$, it is now still difficult to explain the
observed large rate for $B \to K \etap $ decays
\cite{cleo9912,babar2000,cleo-t0020}.
This fact strongly suggests the requirement for additional contributions unique to
the $\etap $ meson in the framework of the SM, or large enhancements from new
physics beyond the SM.

According to the studies in Refs.\cite{hou94,kagan95,hou98,lu99}, we know
that (a) an enhanced $b \to sg$ can lead to a large ${\cal B}(B \to \eta' X_s)$,
and (b) the possible contributions to the ratio $\bsg$ in both type I and II
two-Higgs-doublet models (2HDM) are not large enough to meet the requirement
\cite{hou94,kagan95}. Very recently, we calculated \cite{xiao201,xiaonc} the new
physics enhancements to the branching ratios $\brbsg$ and ${\cal B}(b \to q' q\bar{q} )$
with $q'\in \{d,s\}$ and $q\in \{u,d,s\}$ induced by charged-Higgs gluonic
penguin diagrams in  model III (the third type of 2HDM ) with inclusion of
NLO QCD corrections \cite{lenz97}, and found that the rate of $b \to s g$ in model
III can be enhanced significantly. The predicted charm multiplicity
$n_c$ consequently become consistent with the measured $n_c$,
while the agreement between the theoretical predictions and the data of
$\bsl$ is  also improved by inclusion of the new physics effects.

In this paper  we calculate the new physics contributions to the
branching ratios of exclusive two-body charmless hadronic decays
$B\to PP, PV, VV~ $\footnote{In the following, $B$ always means $B_u$ or
$B_d$ mesons. We here do not consider the decays
of $B_s$ meson.} from new gluonic and electroweak charged-Higgs penguin
diagrams in  the general two-Higgs-doublet models (models I, II and III).
We try to check
the size and pattern of new physics effects on the exclusive two-body
charmless B meson decays and to see if the new physics contributions
in the model III can be large enough to provide the required
enhancements for $B \to K \etap$ decay modes.
We will present our systematic calculation of
branching ratios for seventy six $B \to h_1 h_2$ decay modes
by employing the effective Hamiltonian with the generalized factorization
\cite{ali9804,chen99}. We evaluate
analytically all new strong and electroweak penguin diagrams induced by
exchanges of charged Higgs bosons in the quark level processes
$b \to q V^*$ with $q \in \{d,s \}$ and $V\in \{ gluon,\; \gamma, Z\} $,
and then combine the new physics
contributions with their SM counterparts and finally calculate the branching
ratios for all seventy six exclusive $B\to h_1 h_2$ decay modes.

This paper is organized as follows. In Sec.\ref{sec:2hdm}, we describe the basic
structures  of the 2HDM's and examine the allowed parameter space of the general
2HDM's from  currently available data. In Sec. \ref{sec:heff}, we evaluate
analytically the new penguin diagrams, combine the new physics contributions with
their SM counterparts and find the effective Wilson coefficients
$C_i^{eff}$. In Sec. \ref{sec:bsw},
we present the formulae needed to calculate the branching ratios
${\cal B} (B \to h_1 h_2)$. In the following three sections, we calculate and
show numerical results of branching  ratios for $B \to PP, PV,$ and $VV$ decay modes,
respectively. We concentrate on those decay modes with well-measured branching
ratios and sizable yields. The conclusions and discussions are included in the final
section.

\section{ The general 2HDM and experimental constraints } \label{sec:2hdm}

The simplest extension of the SM is the so-called two-Higgs-doublet
models\cite{2hdm}. In such models, the tree level flavor changing
neutral currents (FCNC's) are absent if one introduces an {\it ad
hoc} discrete symmetry to constrain the 2HDM scalar potential and
Yukawa Lagrangian. Lets consider a Yukawa Lagrangian
of the form\cite{atwood97}
\beq
{\cal L}_Y &=&
\eta^U_{ij}\bar{Q}_{i,L} \tilde{\phi_1}U_{j,R} +
\eta^D_{ij}\bar{Q}_{i,L} \phi_1 D_{j,R}
+\xi^U_{ij}\bar{Q}_{i,L} \tilde{\phi_2}U_{j,R}
+\xi^D_{ij}\bar{Q}_{i,L} \phi_2 D_{j,R}+ h.c., \label{leff}
\eeq
where $\phi_{i}$ ($i=1,2$) are the two Higgs doublets of a
two-Higgs-doublet model, $\tilde{\phi}_{1,2}=
i\tau_2 \phi^*_{1,2}$, $Q_{i,L}$ ($U_{j,R}$) with $i=(1,2,3)$ are the
left-handed isodoublet quarks (right-handed   up-type quarks),
$D_{j,R}$  are the right-handed  isosinglet  down-type quarks,
while $\eta^{U,D}_{i,j}$  and $\xi^{U,D}_{i,j}$ ( $i,j=1,2,3$ are
family index ) are generally the non-diagonal matrices of the  Yukawa
coupling. By imposing the discrete symmetry
\beq
\phi_1 \to - \phi_1, \phi_2 \to \phi_2,
D_i \to - D_i, U_i \to  \mp U_i
\eeq
one obtains the so called models I and II. In model I the third
and fourth term in Eq.(\ref{leff}) will be dropped by the discrete
symmetry, therefore, both the up- and down-type quarks get mass from
Yukawa couplings to the same Higgs doublet $\phi_1$, while the
$\phi_2$ has no Yukawa couplings to the quarks. For model II, on
the other hand, the first and fourth term in
Eq.(\ref{leff}) will be dropped by imposing the discrete symmetry.
Model II has, consequently the up- and down-type quarks getting mass
from Yukawa couplings to two different scalar doublets $\phi_1$ and
$\phi_2$.

During past years, models I and II have been studied extensively in
literature and tested experimentally, and model II has been very
popular since it is the building block of the minimal supersymmetric
standard model. In this paper, we focus on  the third
type of the two-Higgs-doublet model\cite{hou92}, usually known as model III
\cite{atwood97,hou92}. In model III, no discrete symmetry is
imposed and both up- and down-type quarks then may have diagonal
and/or flavor changing couplings with $\phi_1$ and $\phi_2$.
As described in Ref.\cite{atwood97}, one can choose a suitable
basis $(H^0, H^1, H^2, H^\pm)$ to express two Higgs
doublets \cite{atwood97}
\beq
\phi_1 = \frac{1}{\sqrt{2}}\left (
\begin{array}{c} \sqrt{2} \chi^+ \\
 v + H^0 + i \chi^0\\ \end{array} \right ),\ \
\phi_2=\frac{1}{\sqrt{2}}\left ( \begin{array}{c}
\sqrt{2} H^+ \\ H^1 + i H^2 \\  \end{array} \right ),
\label{phi12}
\eeq
and take their vacuum expectation values as the form
\beq
\langle\phi_1 \rangle &=& \left ( \begin{array}{c}
0 \\ v/\sqrt{2}\\ \end{array} \right ), \ \  \langle \phi_2 \rangle =0,
\label{vev}
\eeq
where $v=(\sqrt{2}G_F)^{-1/2}=246\; GeV$. The transformation relation between $(H^0,H^1,H^2)$
and the mass eigenstates $(\overline{H}^0, h^0, A^0)$ can be found
in Ref.\cite{atwood97}. The $H^\pm$ are the physical charged Higgs
boson, $H^0$ and $h^0$ are the physical CP-even neutral Higgs boson
and the $A^0$ is the physical CP-odd neutral Higgs boson. After the
rotation of quark fields, the Yukawa Lagrangian of quarks are of the
form \cite{atwood97}
\beq
{\cal L}_Y^{III} =
\eta^U_{ij}\bar{Q}_{i,L} \tilde{\phi_1}U_{j,R} +
\eta^D_{ij}\bar{Q}_{i,L} \phi_1 D_{j,R}
+\hat{\xi}^U_{ij}\bar{Q}_{i,L} \tilde{\phi_2}U_{j,R}
+\hat{\xi}^D_{ij}\bar{Q}_{i,L} \phi_2 D_{j,R} + H.c.,
\label{lag3}
\eeq
where $\eta^{U,D}_{ij}$ correspond to the diagonal mass matrices of
up- and down-type quarks, while the neutral and charged flavor changing
couplings will be \cite{atwood97}
\footnote{We make the same ansatz on the $\xi^{U,D}_{ij}$ couplings as
the Ref.\cite{atwood97}. For more details about the definition of
$\hat{\xi}^{U,D}$ one can see Ref.\cite{atwood97}. }
\beq
\xi^{U,D}_{ij}=\frac{\sqrt{m_im_j}}{v} \lambda_{ij}, \ \
\hat{\xi}^{U,D}_{neutral}= \xi^{U,D}, \ \
\hat{\xi}^{U}_{charged}= \xi^{U}V_{CKM}, \ \
\hat{\xi}^{D}_{charged}= V_{CKM} \xi^{D}, \label{cxiud}
\eeq
where $V_{CKM}$ is the Cabibbo-Kobayashi-Maskawa mixing matrix
\cite{ckm}, $i,j=(1,2,3)$ are the generation index. The coupling
constants $\lambda_{ij}$ are free parameters to be determined by
experiments, and they may also be complex.

In model II and setting $\tan{\beta}=v_2/v_1 \geq 1$ ( $v_1$ and $v_2$ are the
vacuum expectation values of the Higgs doublet $\phi_1$ and $\phi_2$),
the constraint on
the mass $\mhp$ due to CLEO data of $\bsga$ \cite{cleobsga} is
$\mhp \gsim 200$ GeV  for the charged Higgs boson in the 2HDM at the
NLO level \cite{nlo2hdm}. For model I, however, the limit can
be much weaker due to the possible destructive interference with the SM
amplitude.

For model III, the situation is not as clear as model II because there
are more free parameters here. As pointed in Ref.\cite{atwood97},
the data of $K^0-\bar{K}^0$ and $B_d^0-\bar{B}_d^0$ mixing
processes put severe constraints on the FC couplings involving the first
generation of quarks. Imposing the limit $\lambda_{1j}=0$ for $j=(1,2,3)$ and
assuming all other $\lambda_{ij}$ parameters are of order 1, Atwood {\it et al.}
\cite{atwood96} found a very strong constraint of  $\mhp > 600 $ GeV by using the
CLEO data of $\bsga$ decay available in 1995. But this constraint can be lowered to
$\mhp \geq 400$ GeV by using the new CLEO data of $\bsga$ decay\cite{xiaonc}.
In Ref.\cite{aliev99}, Aliev {\it et al.} studied the $\bsga$ decay in
model III by extending the NLO results of model II \cite{nlo2hdm} to the case of
model III, and found the constraint on the FC couplings.

In a recent paper \cite{chao99}, Chao {\it et al.} studied the decay $\bsga$ by
assuming that only the couplings $\lambda_{tt}=|\lambda_{tt}| e^{i\theta_t}$
and $\lambda_{bb}=|\lambda_{bb}|e^{i\theta_b}$ are
nonzero. They found that the constraint on $\mhp$ imposed by the CLEO
data of $\bsga$ can be greatly relaxed by considering the phase
effects of $\lambda_{tt}$ and $\lambda_{bb}$. From the studies of
Refs.\cite{xiaonc,chao99}, we know that for model III the parameter space
\beq
&& \lambda_{ij}=0, \ \ for \ \ ij\neq tt,\ \ or \ \  bb, \nonumber\\
&& |\lambda_{tt}|= 0.3,\ \ |\lambda_{bb}|=35,\ \
\theta=(0^0 - 30^0),\ \ \mhp=(200 \pm 100 ){\rm GeV}, \label{eq:lm3}
\eeq
are allowed by the available data, where $\theta=\theta_{b}-\theta_{t}$.

From the LEP and Tevatron searches for charged Higgs bosons \cite{cdf00,gross99},
the new combined constraint in the $(\mhp, \tan{\beta})$ plane has been given,
for example, in Ref.\cite{pdg2000}: the direct lower limit is $\mhp > 77 GeV $,
while $0.5 \leq \tan{\beta} \leq 60 $ for a relatively light charged Higgs boson
with $\mhp \sim 100$ GeV.
Combining the direct and indirect limits together, we here conservatively
consider the range of $100 {\rm GeV } \leq \mhp \leq 300$ GeV, while take
$\mhp=200$ GeV as the typical value for models  I, II and III. For models I
and II we consider the range of $1 \leq \tan{\beta} \leq 50$, while take
$\tan{\beta}=2$ as the typical value. In the following sections, we
calculate the new physics
contributions to the exclusive two-body charmless decays of B meson in the
Chao-Cheung-Keung (CCK) scenario of model III \cite{chao99}. Model III
in the CCK scenario has the following advantages:

\begin{enumerate}
\item
Since we keep only the couplings $\lambda_{tt}$ and $\lambda_{bb}$ nonzero,
the neutral Higgs bosons do not contribute at tree level or one-loop
level. The new contributions therefore come only from the charged Higgs
penguin diagrams with the heavy internal top quark.

\item
The new operators $O_{9,10}$ and all flipped chirality partners of
operators $O_{1, \cdots,10}$ as defined in Ref.\cite{aliev99} do not
contribute to the decay $\bsga$ and the exclusive two-body charmless hadronic B
decays under study in this paper.

\item
The free parameters are greatly reduced to $\lambda_{tt}$,
$\lambda_{bb}$ and $\mhp$ in model III, and $\tan{\beta}$ and $\mhp$ in
models I and II.

\end{enumerate}

\section{ Effective Wilson coefficients in the SM and 2HDM's} \label{sec:heff}

In this section we evaluate the new gluonic and electroweak penguin diagrams
and present the well-known effective Hamiltonian for the two-body charmless
decays $B \to h_1 h_2$ with the inclusion of new physics contributions.
For more details about the effective Hamiltonian with
generalized factorization for B decays one can see, for example,
Refs.\cite{ali9804,chen99}.

\subsection{Operators and Wilson coefficients}

The standard theoretical frame to calculate the inclusive three-body decays
$b \to s \bar{q} q $~\footnote{For $b \to d \bar{q} q$ decays, one simply makes the
replacement $s \to d$.} is based on the effective Hamiltonian
\cite{buras98,ali9804},
\beq
{\cal H}_{eff}(\Delta B=1) = \frac{G_F}{\sqrt{2}} \left \{
\sum_{j=1}^2 C_j \left ( V_{ub}V_{us}^* Q_j^u  + V_{cb}V_{cs}^* Q_j^c \right )
- V_{tb}V_{ts}^* \left [ \sum_{j=3}^{10}  C_j Q_j  + C_{g} Q_{g} \right ] \right \}
\label{heff2}
\eeq
where $C_j$ and $C_g$ are Wilson coefficients, and the operator basis reads:
\beq
Q_1&=& (\bar{s}q)_{V-A} (\bar{q}b)_{V-A},\ \
Q_2=   (\bar{s}_\alpha q_{\beta})_{V-A} (\bar{q}_{\beta}b_{\alpha})_{V-A},
\label{q1-q2}
\eeq
with $q=u$ and  $q=c$, and
\beq
Q_3&=& (\bar{s}b)_{V-A} \sum_{q'} (\bar{q'}q')_{V-A}, \ \
Q_4 = (\bar{s}_\alpha b_{\beta})_{V-A} \sum_{q'} (\bar{q'}_{\beta}q'_{\alpha})_{V-A}, \label{q3-q4} \\
Q_5&=& (\bar{s}b)_{V-A} \sum_{q'}
(\bar{q'}q')_{V+A}, \ \
Q_6= (\bar{s}_\alpha b_{\beta})_{V-A} \sum_{q'}
(\bar{q'}_{\beta}q'_{\alpha})_{V+A}, \label{q5-q6} \\
Q_7&=& \frac{3}{2}(\bar{s}b)_{V-A} \sum_{q'} e_{q'} (\bar{q'}q')_{V+A}, \ \
Q_8= \frac{3}{2} (\bar{s}_\alpha b_{\beta})_{V-A} \sum_{q'} e_{q'}
(\bar{q'}_{\beta}q'_{\alpha})_{V+A}, \label{q7-q8} \\
Q_9&=& \frac{3}{2}(\bar{s}b)_{V-A} \sum_{q'} e_{q'} (\bar{q'}q')_{V-A}, \ \
Q_{10} = \frac{3}{2} (\bar{s}_\alpha b_{\beta})_{V-A} \sum_{q'} e_{q'}
(\bar{q'}_{\beta}q'_{\alpha})_{V-A}, \label{q9-q10} \\
Q_{g}&=& \frac{g_s}{8\pi^2}m_b \bar{s}_\alpha \sigma^{\mu \nu}
(1+ \gamma_5)T^a_{\alpha \beta} b_{\beta} G^a_{\mu \nu}
\label{q8}
\eeq
where $\alpha$ and $\beta$ are the $SU(3)$ color indices,
$T^a_{\alpha \beta}$ ( $a=1,...,8$) are the Gell-Mann matrices.
The sum over $q'$ runs over the quark fields that are active at the scale
$\mu=O(m_b)$, i.e., $q'\in \{u,d,s,c,b\}$.
$Q_1$ and $Q_2$ are current-current operators, $Q_{3,4,5,6}$
and $Q_{7,8,9,10}$ are QCD  and electroweak penguin operators, and
$Q_{g}$ is the chromo-magnetic dipole
(CMD) operator. Following Ref.\cite{ali9804}, we also neglect
the effects of the electro-magnetic penguin operator $Q_{7\gamma}$,
and do not consider the effects of the weak annihilation and exchange
diagrams.

In the SM, the Wilson coefficients $C_1(M_W),
\cdots, C_{10}(\mw)$ at NLO level and $C_{g}(\mw)$ at leading order (LO)
have been defined, for example, in Refs.\cite{buchalla96a,buras98}.
The explicit expressions of the coefficients in the naive dimensional
regularization (NDR) scheme  can also be found easily in
Refs.\cite{buchalla96a,buras98}

\subsection{Contributions of the charged-Higgs penguin diagrams}

\begin{figure}%fig.1
\begin{center}
\begin{picture}(400,440)(0,0)
\put(-60,-140) {\epsfxsize200mm\epsfbox{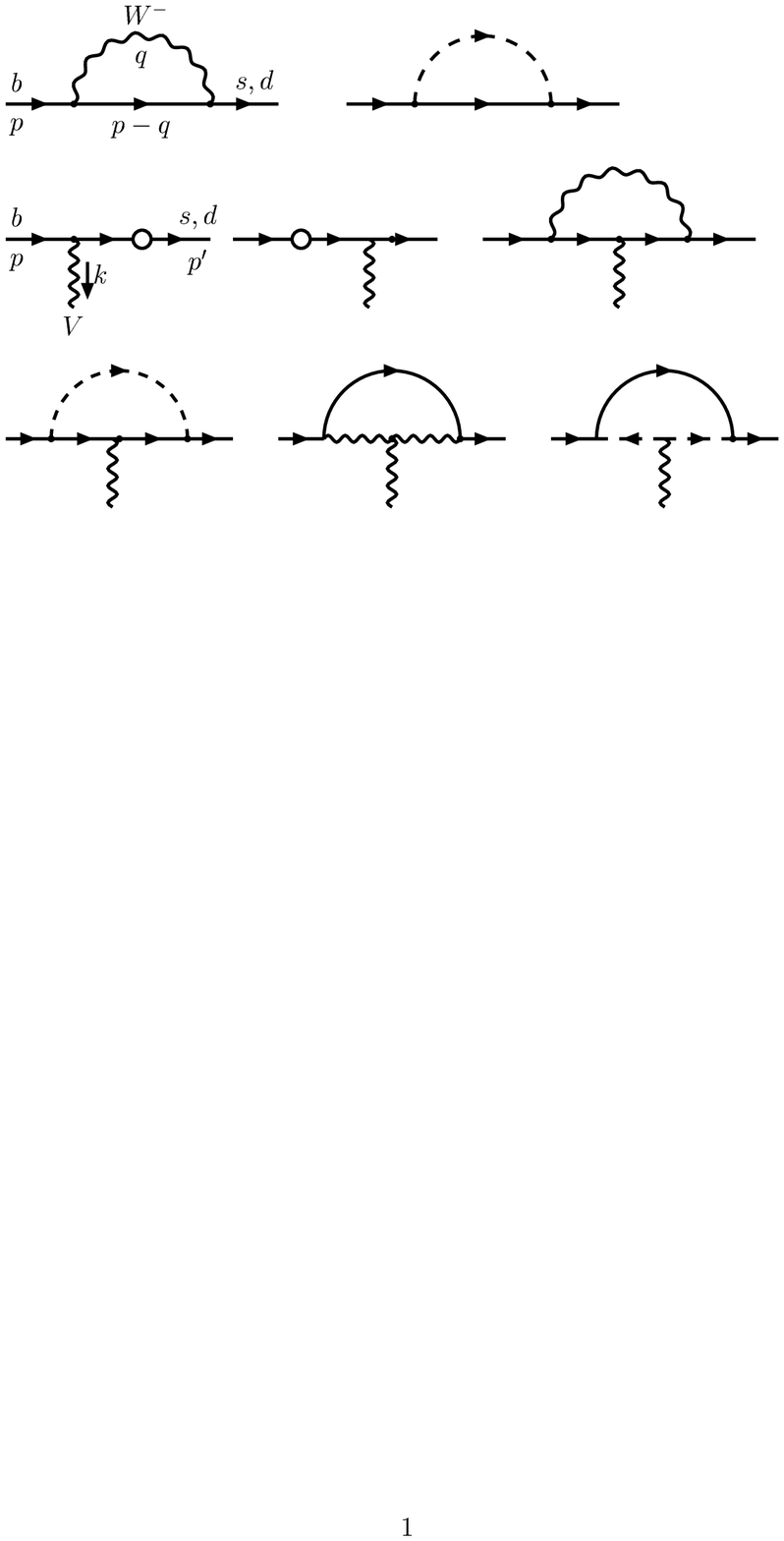}}
\end{picture}
\end{center}
\vspace{-80pt}
\caption{Typical one-loop Feynman diagrams for the quark level decays
$b \to (s, d) V^*$ ($V=\gamma, Z^0, g$),  with $W^{\pm}$ (internal wave lines)
and charged-Higgs exchanges (internal dashed lines) in the SM and two-Higgs-doublet
models. The internal quarks are the upper type quark $u, c$ and $t$.}
\label{fig:fig1}
\end{figure}

For the charmless hadronic decays of B meson under consideration, the new physics
will manifest itself by modifying the corresponding Inami-Lim
functions\cite{inami81}
$C_0(x), D_0(x), E_0(x)$ and $E'_0(x)$ which determine the coefficients
$C_3(\mw), \ldots, C_{10}(\mw)$ and $C_{g}(\mw)$ in the SM.
These modifications, in turn, will change for example the standard
model predictions for the branching ratios of decays $B \to h_1 h_2$.
The new strong and electroweak penguin diagrams can be obtained from the
corresponding penguin diagrams in the SM by replacing the internal $W^{\pm}$
lines with the charged-Higgs $H^\pm$ lines, as shown in \fig{fig:fig1}.
In the analytical calculations of those penguin diagrams, we use
the dimensional regularization to regulate all the ultraviolet divergence
in the virtual loop corrections and adopt the $\overline{MS}$
renormalization scheme. It is easy to show that all the ultraviolet
divergence is canceled after summing up all Feynman diagrams.

By  evaluating analytically the new $Z^0$-, $\gamma$- and gluonic penguin diagrams
induced by the exchanges of charged-Higgs boson
$H^\pm$ in the model III, we find the new $C_0$, $D_0$, $E_0$ and $E'_0$
functions
\beq
C_0^{III} &=& \frac{-x_t}{16} \left[ \frac{y_t}{1-y_t} + \frac{y_t}{(1-y_t)^2}\ln[y_t]
\right ]\cdot |\lambda_{tt}|^2~, \label{eq:c0m3}\\
D_0^{III} &=& -\frac{1}{3} H(y_t)|\lambda_{tt}|^2~,
\label{eq:d0m3}\\
E_0^{III} &=& -\frac{1}{2}I(y_t)|\lambda_{tt}|^2~,\label{eq:e0m3}\\
{E'_0}^{III}&=& \frac{1}{6}J(y_t)|\lambda_{tt}|^2
 - K(y_t) |\lambda_{tt} \lambda_{bb}| e^{i\theta}~,\label{eq:e0pm3}
\eeq
with
\beq
H(y)&=& \frac{38 y - 79 y^2 + 47 y^3}{72(1-y)^3}
+ \frac{4y -6y^2 + 3y^4}{12 (1-y)^4} \ln[y]~,\label{eq:hhy}\\
I(y)&=& \frac{16y -29y^2 +7 y^3}{36(1-y)^3} + \frac{2y- 3y^2}{6(1-y)^4}\log[y]~,
\label{eq:iiy} \\
J(y)&=& \frac{2y + 5y^2 - y^3}{4(1-y)^3} + \frac{3y^2}{2(1-y)^4}\log[y]~,
\label{eq:jjy} \\
K(y)&=& \frac{-3y + y^2}{4(1-y)^2} - \frac{y}{2(1-y)^3}\log[y]~,
\label{eq:kky}
\eeq
where $x_t=m_t^2/\mw^2$, $y_t=m_t^2/\mhp^2$, and the small terms
proportional to $m_b^2/m_t^2$ have been neglected.

In models I and II,  one can find the corresponding functions $C_0$,
$D_0$,$E_0$ and $E'_0$ by evaluating the new strong and electroweak
penguin diagrams in the same way as in model III
\beq
C_0^{I} &=& C_0^{II} = \frac{-x_t}{8 \tan^2{\beta}} \left[ \frac{y_t}{1-y_t}
    + \frac{y_t}{(1-y_t)^2}\ln[y_t] \right ]~, \label{eq:c0m2} \\
D_0^{I} &=& D_0^{II} =  -\frac{2}{3\tan^2{\beta}} H(y_t)~, \label{eq:d0m2}\\
E_0^{I} &=& E_0^{II} -\frac{1}{\tan^2{\beta}}I(y_t)|~, \label{eq:e0m2}\\
{E'_0}^{I}&=& \frac{1}{3 \tan^2{\beta}}\left [ J(y_t) - 6 K(y_t)\right]~,
\label{eq:e0pm1}\\
{E'_0}^{II}&=& \frac{1}{3 \tan^2{\beta}}J(y_t) + 2 K(y_t)~,\label{eq:e0pm2}
\eeq
where $y_t=m_t^2/\mhp^2$.

We combine the SM part and the new physics part of the corresponding
functions to define the functions at the scale $\mu =\mw$ as follows
\beq
F_0(M_W)&=& F_0^{SM} + F_0^{NP}, \label{eq:cdeep}
\eeq
where $F_0\in \{ C_0, D_0, E_0, E'_0 \}$. The explicit expressions
of the functions $C_0$, $D_0$, $E_0$ and $E'_0$ in the SM can be
found, for example,  in Ref.\cite{buras98}.

Since the heavy new particles appeared in the 2HDM's have been integrated out
at the scale $\mw$, the QCD running of the Wilson coefficients $C_i(\mw)$ down
to the scale $\mu=O(m_b)$ after including the new physics contributions will
be the same as in the SM. By using QCD renormalization group equations
\cite{buchalla96a,buras98}, it is
straightforward to run Wilson coefficients $C_i(\mw)$ from the scale $\mu =0( \mw)$
down to the lower scale $\mu =O(m_b)$. Working consistently  to the NLO precision,
the Wilson coefficients $C_i$ for $i=1,\ldots,10$ are needed in NLO precision,
while it is sufficient to use the leading logarithmic value for $C_{g}$:
\beq
{\bf C}(\mu)&=&U(\mu, \mw) {\bf C}(\mw), \label{eq:cmu}\\
C_g(\mu)&=& \eta^{14/23}C_g(\mw) + \sum_{i=1}^{8} \bar{h}_i
\eta^{a_i}, \label{eq:cgmu}
\eeq
where ${\bf C}(\mw)=( C_1(\mw), \ldots, C_{10}(\mw))^{T}$, $U(\mu,
\mw)$ is the five-flavor $10\times 10$ evolution matrix at NLO level
as defined in Ref.\cite{buchalla96a}, $\eta=\alpha_s(\mw)/\alpha_s(\mu)$,
and the constants $\bar{h}_i$ and $a_i$ can also be found in
Ref.\cite{buchalla96a}.

At the NLO level, the Wilson coefficients are usually renormalization scheme(RS)
dependent.
In the NDR scheme, by using the input parameters as given in Appendix
and Eq.(\ref{eq:lm3}), and setting $\mhp=200$ GeV, $\theta=0^0$ ,
$\tan{\beta}=2$ and
$\mu=2.5$ GeV,  we find the Wilson coefficients $C_g^{eff}(\mu)=C_g + C_5$
and $C_i(\mu)$ with $i=1,\ldots,10$ in the SM and models I, II and III, and
list them in \tab{cimu25}. From  the numerical results as listed in
\tab{cimu25}, one can easily see that:
\begin{itemize}
\item
The values of $C_i(\mu)$ ( $i=1,\ldots,10$ ) in models I, II and III
are almost identical with those in the SM. Only the coefficient $C_g^{eff}$
in models II and III are  clearly different from that in the SM.

\item
It is the coefficient $C_g^{eff}$ partially induced by the new
gluonic penguin diagrams which dominate the total new physics
corrections to the decay processes under study.

\end{itemize}

\subsection{The effective Wilson coefficients}

We know that the unphysical RS dependence of Wilson coefficients
will be cancelled by the corresponding dependence
in the matrix elements of the operators in ${\cal H}_{eff}$, as shown
explicitly in Refs.\cite{buras98,fleischer93}.
Very recently, Cheng {\it et al.} \cite{cheng99a} studied and resolved
the so-called gauge and infrared problems \cite{bs98} of generalized
factorization approach \footnote{The reliability of the generalized
factorization approach is improved by this progress.}. They
found that the gauge invariance is maintained
under radiative corrections by working in the physical
on-mass-shell scheme, while the infrared divergence in radiative
corrections should be isolated using the dimensional
regularization  and the resultant infrared poles are absorbed into
the universal meson wave functions \cite{cheng99a}.

The one-loop matrix elements can be rewritten in terms of the tree-level
matrix elements of the effective operators\cite{ali9804}
\beq
\langle s q'\bar q'\vert {\cal H}_{eff}\vert b \rangle =
\sum_{i,j} C_i^{eff}(\mu)
\langle sq'\bar q'\vert O_j \vert b\rangle ^{\rm tree}.
\label{eq:m1}
\eeq
where $C_i^{eff}(\mu)$ ($i=1,\ldots,10$) are the effective Wilson
coefficients. In the NDR scheme and for $SU(3)_C$, the effective
Wilson coefficients $C_i^{eff}$ can be written as \cite{ali9804,chen99},
\beq
C_i^{eff} &=& \left [ 1 + \frac{\alpha_s}{4\pi} \, \left( r_V^T +
 \gamma_{V}^T \log \frac{m_b}{\mu}\right) \right ]_{ij} \, C_j
 +\frac{\alpha_s}{24\pi} \, A_i' \left (C_t + C_p + C_g \right)
+ \frac{\alpha_{ew}}{8\pi}\, B_i' C_e ~, \label{eq:wceff}
\eeq
where $A_i'=(0,0,-1,3,-1,3,0,0,0,0)^T$, $B_i'=(0,0,0,0,0,0,1,0,1,0)^T$, the
matrices  $\hat{r}_V$ and $\gamma_V$ contain the process-independent
contributions from the vertex diagrams. Like Ref.\cite{chen99}, we
here include vertex corrections to $C_7 -C_{10}$~\footnote{Numerically, such
corrections are negligibly small.}. The anomalous dimension matrix
$\gamma_V$ has been given explicitly, for example, in Eq.(2.17) of
Ref.\cite{chen99}. Note that the correct value of
the element $(\hat{r}_{NDR})_{66}$ and  $(\hat{r}_{NDR})_{88}$ should be  17
instead of 1 as pointed out in Ref.\cite{cheng00a}, $\hat{r}_V$ in the NDR
scheme takes the form
\begin{equation}
\hat{r}_V^{NDR} = \left(
\begin{array}{ccccc ccccc}
3 & -9 & 0 & 0 & 0 & 0& 0 & 0 & 0 & 0 \\
-9 & 3 & 0 & 0 & 0 & 0& 0 & 0 & 0 & 0 \\
0 & 0  & 3 &-9 & 0 & 0& 0 & 0 & 0 & 0 \\
0 & 0  &-9 & 3 & 0 & 0& 0 & 0 & 0 & 0 \\
0 & 0  & 0 & 0 & -1& 3& 0 & 0 & 0 & 0 \\
0 & 0  & 0 & 0 & -3&17& 0 & 0 & 0 & 0 \\
0 & 0  & 0 & 0 & 0& 0&-1 & 3 & 0 & 0 \\
0 & 0  & 0 & 0 & 0& 0&-3 &17 & 0 & 0 \\
0 & 0  & 0 & 0 & 0& 0& 0 & 0 & 3 &-9 \\
0 & 0  & 0 & 0 & 0& 0& 0 & 0 &-9 & 3 \\
\end{array} \right) \quad .
\end{equation}

The function $C_t$, $C_p$, and $C_g$ describe the penguin-type corrections
to the operators $Q_{1,2}$, $Q_{3,\ldots,6}$, and the tree-level diagram
of the operator $Q_{g}$ respectively. We here follow the procedure of
Ref.\cite{ali98} to include $C_g$ in (\ref{eq:wceff}). The effective
Wilson coefficients $C_i^{eff}$ in Eq.(\ref{eq:wceff}) are now scheme
and scale independent in NLO precision, and also gauge invariant
and infrared safe.
The explicit expressions of functions $C_t$, $C_p$, and $C_g$ in the NDR scheme
have been given, for example, in Refs.\cite{ali9804,chen99}
\beq
C_t &=& \left [ \frac{2}{3} + {\lambda_u\over\lambda_t}G(m_u)
    +{\lambda_c\over\lambda_t} G(m_c) \right ] C_1,\label{cct} \\
C_p &=& \left [\frac{4}{3} - G(m_q) - G(m_b) \right ] C_3
    + \sum_{i=u,d,s,c,b} \left [ \frac{2}{3}- G(m_i) \right ] (C_4+C_6),  \label{ccp}\\
C_e &=& {8\over9}\left [ \frac{2}{3} +  {\lambda_u\over\lambda_t} G(m_u)
    + {\lambda_c\over\lambda_t} G(m_c) \right ] (C_1+3C_2),\label{cce}\\
C_g &=& -{2m_b\over \sqrt{< k^2>}}C^{\rm eff}_{g}, \label{ccg}
\eeq
with $\lambda_{q'}\equiv V_{q'b}V_{q'q}^*$. The function $G(m)$
is of the form \cite{abel98}
\beq
G(m)&=& \frac{10}{9}-\frac{2}{3}\ln[\frac{m^2}{\mu^2}] + \frac{2\mu^2}{3 m^2}
-\frac{2(1 + 2 z )}{3z } g(z)
\eeq
where $z=k^2/(4m^2)$, and
\beq
g(z) = \left \{\begin{array}{ll}
\sqrt{\frac{1-z}{z}} \arctan[\frac{z}{1-z}], & z < 1, \\
\sqrt{\frac{1-z}{4z}}\left [
\ln[\frac{\sqrt{z} + \sqrt{z-1}}{\sqrt{z}-\sqrt{z-1}}] - i \pi \right ], &
z >1. \\
\end{array} \right. \label{eq:gz}
\eeq
where $k$ is the momentum transferred by the virtual gluon, photon
or $Z$ to the $q^\prime \overline{q^\prime}$ quark pair in the inclusive three-body
decays $b \to q q^\prime \overline{q^\prime}$, and $m$ is the mass of internal
up-type quark in the penguin diagrams. For $k^2>4m^2$, an imaginary part of
$g(z)$ will appear because of the generation of a strong phase at the
$\bar{u}u$ and $\bar{c}c$ threshold  \cite{abel98,bss79,hou91}.

For the two-body exclusive B meson decays any information on $k^2$ is
lost in the factorization assumption, and it is not clear what ``relevant"
$k^2$ should be taken in numerical calculation. One usually  uses
the "physical" range for $k^2$:
$m_b^2/4 \stackrel{<}{\sim} k^2 \stackrel{<}{\sim} m_b^2/2$.
Following Refs.\cite{ali9804,chen99}, we also use $k^2=m_b^2/2$ in the numerical
calculation and will consider the $k^2$-dependence of branching ratios of
charmless  B meson decays for several typical decay channels.

%5702
\section{Decay amplitudes in the BSW model} \label{sec:bsw}

In numerical calculations, two sets of form factors at the zero momentum
transfer from the  Bauer, Stech and Wirbel (BSW) model \cite{bsw87}, as well
as Lattice QCD and Light-cone QCD sum rules (LQQSR)\cite{flynn97} will be used
respectively. Explicit values of these form factors can be found in
Ref.\cite{ali9804} and have also been given in Appendix.
Following Ref.\cite{ali9804}, the seventy six decay channels of $B_u$ and
$B_d$ mesons are classified into five classes according to their
$N^{eff}-$dependence:
\begin{itemize}

\item Class-I:  including four decay modes, $B^0 \to \pi^-\pi^+,
\rho^{\pm}\pi^{\mp}$ and $B^0 \to \rho^- K^+$, the large and $N^{eff}$
stable coefficient $a_1$ plays the major role.

\item
Class-II: including ten decay modes, for example $B^0 \to \pi^0 \pi^0$, and
the relevant coefficient for these decays is $a_2$ which shows a strong
$N^{eff}$-dependence.

\item
Class-III: including nine decay modes involving the interference of class-I
and class-II decays, such as the decays  $B^+ \to \pi^+ \etap$.

\item
Class-IV: including twenty two  $B \to h_1 h_2$ decay modes such as
$B \to K \etapp$ decays. The amplitudes of these decays involve one (or more) of
the dominant penguin coefficients $a_{4,6,9}$ with constructive interference
among them. The Class-IV decays are $N^{eff}$ stable.

\item
Class-V:  including twelve $B \to h_1 h_2$ decay modes, such as
$B \to \pi^0 \etapp$ and $B \to \phi K$ decays. Since the amplitudes of
these decays involve large and delicate cancellations due to
interference between strong $N^{eff}$-dependent coefficients $a_{3,5,7,10}$
and the dominant penguin coefficients $a_{4,6,9}$,
these decays are generally not stable against $N^{eff}$.

\end{itemize}

With the factorization ansatz \cite{bsw87,feynman65,ellis75}, the three-hadron
matrix elements or the decay amplitudes $<XY|H_{eff}|B>$ can be factorized
into a sum of products of two current matrix elements $<X|J_1^\mu|0>$ and
$<Y|J_{2\mu}|B>$ ( or $<Y|J_1^\mu|0>$ and $<X|J_{2\mu}|B>$). The explicit
expressions of the  matrix elements in terms of decay constants $(f_X, g_X)$ and
the Lorentz-scalar form factors $A_{0,1,2}(k^2)$ and $F_{0,1}(k^2)$ can be
found, for example, in Refs.\cite{bsw87,bijnens92,ali9804}.

In the B rest frame, the branching ratios of two-body B meson decays can
be written as
\beq
{\cal  B}(B \to X Y )=  \tau_{B}\,  \frac{|p|}{8\pi M_B^2}
|M(B\to XY)|^2\label{eq:brbpp}
\eeq
for $B \to P P$ decays, and
\beq
{\cal  B}(B \to X Y )=   \tau_{B} \, \frac{|p|^3}{8\pi M_V^2}
|M(B\to X Y )/(\epsilon \cdot p_{B})|^2\label{eq:brbpv}
\eeq
for $B \to P V$ decays. Here $\tau(B_u^-)=1.65\; ps$ and $\tau(B_d^0)=1.56\; ps$
\cite{pdg98}, $p_B$ is the four-momentum of the B
meson, $M_V$ and $\epsilon$ is the mass and polarization
vector of the produced light vector meson respectively,  and $|p|$
is the magnitude of momentum of particle X and Y in the B rest
frame
\beq
|p| =\frac{1}{2M_B}\sqrt{[M_B^2 -(M_X + M_Y)^2] [ M_B^2 -(M_X-M_Y)^2
]}~. \label{eq:pxy}
\eeq

For $B \to VV$ decays, one needs to evaluate the helicity matrix
elements $H_\lambda = <V_1(\lambda) V_2(\lambda)|H_{eff}|B)>$ with
$\lambda=0, \pm 1$. The branching ratio of the decay $B \to V_1 V_2$
is given in terms of $H_\lambda$ by
\beq
{\cal  B}(B \to V_1 V_2 ) &=&  \tau_{B}\,
\frac{|p|}{8\pi M_B^2}\left [ |H_0|^2 + |H_{+1}|^2 + |H_{-1}|^2
\right ] ~.
\eeq
The three independent
helicity amplitudes $H_0$, $H_{+1}$ and $H_{-1}$ can be expressed by three
invariant amplitudes $a, b, c$ defined by the decomposition
\beq
H_\lambda &=& i\epsilon^\mu(\lambda)\eta^\nu(\lambda)\left[
    a g_{\mu\nu}+\frac{b}{M_1 M_2}p_\mu
    p_\nu + \frac{ic}{M_1 M_2}\epsilon_{\mu\nu\alpha\beta}p_1^\alpha
    p^\beta \right] \label{eq:hl}
\eeq
where $p_{1,2}$ and $M_{1,2}$ are the four momentum and masses of $V_{1,2}$,
respectively. $p=p_1 + p_2 $ is the four-momentum of B meson, and
\beq
H_{\pm 1} &=& a \pm c \sqrt{x^2-1}, ~~~~
H_0 = -ax - b\left ( x^2-1 \right ) \label{eq:h01} \\
x &=& \frac{M_B^2-M_1^2-M_2^2}{2M_1 M_2}
\eeq

In the generalized factorization ansatz, the effective Wilson coefficients
$C_i^{eff}$ will appear in the decay amplitudes in the combinations:
\beq
a_{2i-1}\equiv C_{2i-1}^{eff} +\frac{{C}_{2i}^{eff}}{N^{eff}}, \ \
a_{2i}\equiv C_{2i}^{eff}     +\frac{{C}_{2i-1}^{eff}}{N^{eff}}, \ \ \
(i=1,\ldots,5) \label{eq:ai}
\eeq
where the effective number of colors $N^{eff}$ is treated as a free parameter
varying in the range of $2 \leq N^{eff} \leq \infty$, in order to model the
non-factorizable contribution to the hadronic matrix elements.
It is evident that  the reliability of generalized
factorization approach has been improved since the effective Wilson
coefficients $C_i^{eff}$ appeared in Eq.(\ref{eq:ai}) are now gauge
invariant and infrared safe. Although  $N^{eff}$ can in principle
vary from channel to channel, but in the energetic two-body hadronic B
meson decays, it is expected to be process insensitive as supported by
the data \cite{chen99}.  As argued in Ref.\cite{cheng98},
$N^{eff}(LL)$ induced by the $(V-A)(V-A)$ operators can be rather
different from $N^{eff}(LR)$ generated by  $(V-A)(V+A)$ operators.
In this paper, however, we will simply assume that $N^{eff}(LL)\equiv N^{eff}(LR)=
N^{eff}$ and consider the variation of $N^{eff}$ in the range
of $2 \leq N^{eff} \leq \infty$ since we here focus on the calculation
of new physics effects on the studied B meson decays induced by the new
penguin diagrams in the two-Higgs-doublet models.
For more details about the cases of
$N^{eff}(LL)\neq N^{eff}(LR)$, one can see for example Ref.\cite{chen99}.
We here will also not  consider the possible effects of final state
interaction (FSI) and the contributions from annihilation channels
although they may play a significant rule for some $B$ meson decays.

Using the input parameters as given in Appendix, and assuming $k^2=m_b^2/2$,
$\mhp=200$ GeV, $\theta=0^0$ and $\tan{\beta}=2$, the theoretical predictions
of effective
coefficients $a_i$ are calculated and displayed in \tab{ai:bd} and \tab{ai:bs}
for the transitions $b\to d$ ( $\bar{b} \to \bar{d}$ ) and
$b\to s$ ($\bar{b} \to \bar{s}$), respectively.  For coefficients $a_3,
\ldots, a_{10}$, the first, second and third entries in tables
(\ref{ai:bd},\ref{ai:bs}) refer to the values of $a_i$ in the SM and models
II and III, respectively. $a_i$ in model I are very similar with those in
the SM and hence was not given explicitly.

All branching ratios in the following three sections are the averages of the
branching ratios of $B$ and anti-$B$ decays. The ratio $\delta {\cal  B}$
describes the magnitude of new physics corrections on the SM
predictions of the decay ratios and is defined as
\beq
\delta {\cal  B} (B \to XY) = \frac{{\cal  B}(B \to XY)^{2HDM}
-{\cal  B}(B \to XY)^{SM}}{{\cal  B}(B \to XY)^{SM}} \label{eq:dbr}
\eeq

\section{$B \to P P$ decays}\label{sec:bpp}

Using formulae as given in last section, it is straightforward to find the decay
amplitudes of $B \to P P $ decays. As an example, we present here the decay
amplitude $M(B^- \to \pi^- \pi^0)=<\pi^-\pi^0|H_{eff}|B_u^->$,
\beq
M(B^- \to \pi^- \pi^0)&=& \frac{G_F}{2}
\left \{ V_{ub}V_{ud}^* \left ( a_1 M^{\pi^-\pi^0}_{uud} + a_2 M^{\pi^-\pi^0}_{duu}
\right )\right. \non
 &&  \left. - V_{tb} V_{td}^* \left [ \left ( a_4+a_{10}+ (a_6+a_8) R_1 \right )
 M^{\pi^-\pi^0}_{duu} \right. \right. \non
&& \left. \left. - \left ( a_4 + \frac{3}{2}(  a_7 -a_9 )
 -\frac{a_{10}}{2}  + (a_6-\frac{a_8}{2})R_2 \right ) M^{\pi^-\pi^0}_{uud}
\right ]\right\} \label{eq:bpipi0}
\eeq
with
\beq
R_1 &=&\frac{2 m_{\pi^-}^2}{(m_b-m_u) (m_u + m_d)},\label{eq:r1}\\
R_2 &=&\frac{m_{\pi^0}^2}{m_d (m_b -m_d)}, \label{eq:r2}\\
M^{\pi^-\pi^0}_{uud} &=& -i(m_B^2-m_{\pi^-}^2)f_{\pi}F^{B \to \pi}_0(m_{\pi^0}^2),\\
M^{\pi^-\pi^0}_{duu} &=& -i(m_B^2-m_{\pi^0}^2)f_{\pi}F^{B\to \pi}_0(m_{\pi^-}^2)
\eeq
where $f_{\pi}$ is the decay constant of $\pi$ meson. The form factor
$F_0^{B\to \pi}(m^2)$ can be found in Appendix.
Under the approximations of setting $m_u =m_d$ and $m_{\pi^0}=m_{\pi^-}$, the decay
amplitude $M(B^- \to \pi^- \pi^0)$ in Eq.(\ref{eq:bpipi0}) will be reduced to the
form as given in Eq.(80) of Ref.\cite{ali9804}.
In the following numerical calculations, we use the decay amplitudes as given
in Appendix A of Ref.\cite{ali9804} directly without further discussions about
details of individual amplitude.

In tables \ref{bppsm}-\ref{bppm2}, we present the numerical results of the
branching ratios for the twenty $B \to P P $ decays in the framework of the SM
and models I, II and III by using the BSW and LQQSR form factors, respectively.
Theoretical predictions are made by using the central values of input parameters
as given in Eq.(\ref{eq:lm3}) and Appendix, and assuming $\mhp=200$ GeV, $\theta=0^0$,
$\tan{\beta}=2$ and $N^{eff}=2, 3, \infty$ in the generalized
factorization approach.
The $k^2$-dependence of the branching ratios is small in the range of
$k^2=m_b^2/2\pm 2\; GeV^2$ and hence the numerical results are given by fixing
$k^2=m_b^2/2$.

The currently available CLEO data \cite{cleo99,cleo9912,cleo2001} are
listed in the last column of table \ref{bppsm}. From the numerical results,
we see that:

\begin{itemize}
\item
For $B \to K \etap$ decays, the observed branching ratios are clearly much larger
than the SM predictions\cite{cleosum,cleo-t0020}. All other estimated branching ratios in \tab{bppsm}
are, however, consistent with the new CLEO, BaBar and Belle measurements or
upper limits.

\item
In model III, the new physics corrections to most class-II, IV and V decay
channels can be rather large and  insensitive to the variations of the mass
$\mhp$ and the color number $N^{eff}$: from $~20\%$ to $~90\%$ $w.r.t$ the SM
predictions for both cases of $\theta=0^\circ, 30^\circ$.
For tree-dominated decay modes
$B \to \pi^+ \pi^-, \pi^+ \pi^0, \pi^+ \etapp$, the new physics corrections
are small in size.

\item
In models I and II, however, the new physics corrections to all $B \to PP$ decay
modes  are small in size within the considered parameter space: less
than $3\%$ in model I, and $\approx (-20 - 0)\%$ in model II,
as shown in \tab{bppm2}. So small corrections will be masked by other
large theoretical uncertainties.

\item
In model III, the new gluonic penguins will contribute effectively through
the mixing of chromo-magnetic operator $Q_g$ with QCD penguin operators
$Q_3-Q_6$. The $C_g^{eff}$ will strongly dominate the new physics
contributions to all $B \to h_1 h_2$ decay modes.

\item
The central values of the branching ratios obtained by using the LQQSR form factors
will be increased by about $15 \%$ when compared with the results using the BSW
form factors, as can be seen from \tab{bppsm}. We therefore use the BSW form
factors only to calculate the new physics effects on the ratios
${\cal B}(B \to h_1 h_2)$ and treat the difference induced by using
different set of form factors as one kind of theoretical uncertainties.

\end{itemize}

\subsection{$B \to \pi \pi, \; K \pi$ decays }

There are so far seven measured branching ratios of $B \to PP$ decays: one
$B \to \pi^+ \pi^-$ decay, four $B \to K \pi$ and two $B \to K \etap$
decays\cite{cleo9912,cleo2001,babar2000,belle2000}:
\beq
{\cal B}(B \to \pi^+ \pi^-)&=& \left \{\begin{array}{ll}
( 4.3 ^{+1.6}_{-1.5} \pm 0.5 )\times 10^{-6} & {\rm [CLEO]}, \\
( 9.3 ^{+2.8\; +1.2 }_{-2.1\; -1.4} )\times 10^{-6} & {\rm [BaBar]},  \\
\end{array} \right. \label{eq:brexp01} \\
{\cal B}(B \to K^+ \pi^0)&=& \left \{\begin{array}{ll}
( 11.6 ^{+3.0\; +1.4}_{-2.7\; -1.3} )\times 10^{-6} & {\rm [CLEO]}, \\
( 18.8 ^{+5.5}_{-4.9} \pm 2.3 )\times 10^{-6} & {\rm [Belle]},  \\
\end{array} \right. \label{eq:brexp11} \\
{\cal B}(B \to K^+ \pi^-)&=& \left \{\begin{array}{ll}
( 17.2 ^{+2.5}_{-2.4} \pm 1.2)\times 10^{-6} & {\rm [CLEO]}, \\
( 12.5 ^{+3.0\; +1.3}_{-2.6\; -1.7} \pm 2.3 )\times 10^{-6} & {\rm [BaBar]},  \\
( 17.4 ^{+5.1}_{-4.6} \pm 3.4)\times 10^{-6} & {\rm [Belle]}, \\
\end{array} \right. \label{eq:brexp12} \\
{\cal B}(B \to K^0 \pi^+)&=&
( 18.2 ^{+4.6}_{-4.0} \pm 1.6)\times 10^{-6}\ \   {\rm [CLEO]},
\label{eq:brexp13}\\
{\cal B}(B \to K^0 \pi^0)&=& \left \{\begin{array}{ll}
( 14.6 ^{+5.9\; +2.4}_{-5.1\; -3.3} )\times 10^{-6} & {\rm [CLEO]}, \\
( 21  ^{+9.3\; +2.5}_{-7.8\; -2.3} )\times 10^{-6} & {\rm [Belle]},  \\
\end{array} \right. \label{eq:brexp14} \\
{\cal B}(B \to K^+ \etap )&=& \left \{\begin{array}{ll}
( 80 ^{+10}_{-9} \pm 7 )\times 10^{-6} & {\rm [CLEO]}, \\
( 62 \pm 18 \pm 8 )\times 10^{-6} & {\rm [BaBar]},  \\
\end{array} \right. \label{eq:brexp16} \\
{\cal B}(B \to K^0 \etap )&=&
( 89 ^{+18}_{-16} \pm 9 )\times 10^{-6} \ \  {\rm [CLEO]}.
\label{eq:brexp18}
\eeq
The measurements of CLEO, BaBar and Belle Collaborations are in good
agreement with each other within errors. These decays are
sensitive to the relevant form factors $F_0^{B\to \pi}, F_0^{B \to \eta},
F_0^{B\to \etap}, etc.,$ and to the value of $N^{eff}$.

\begin{figure}[t] %fig.2
\vspace{-60pt}
\begin{minipage}[]{0.96\textwidth}
\centerline{\epsfxsize=\textwidth \epsffile{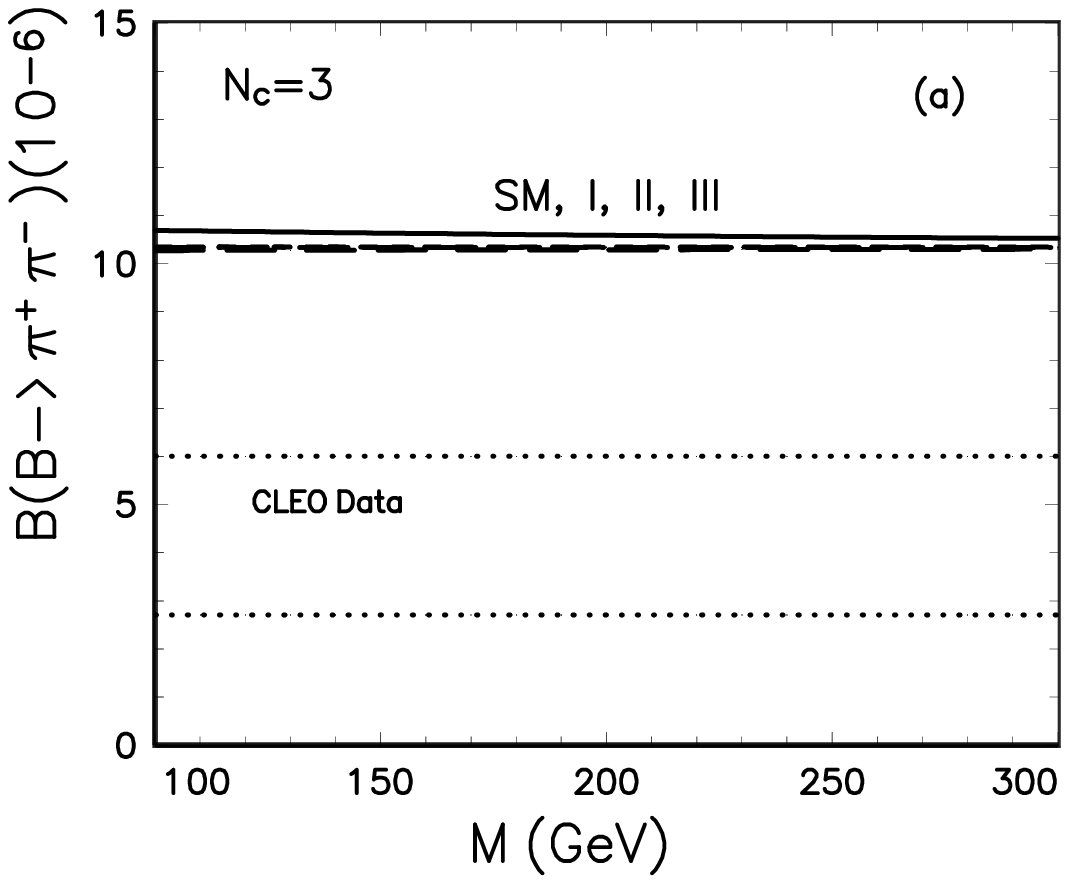}}
\vspace{-80pt}
\centerline{\epsfxsize=\textwidth \epsffile{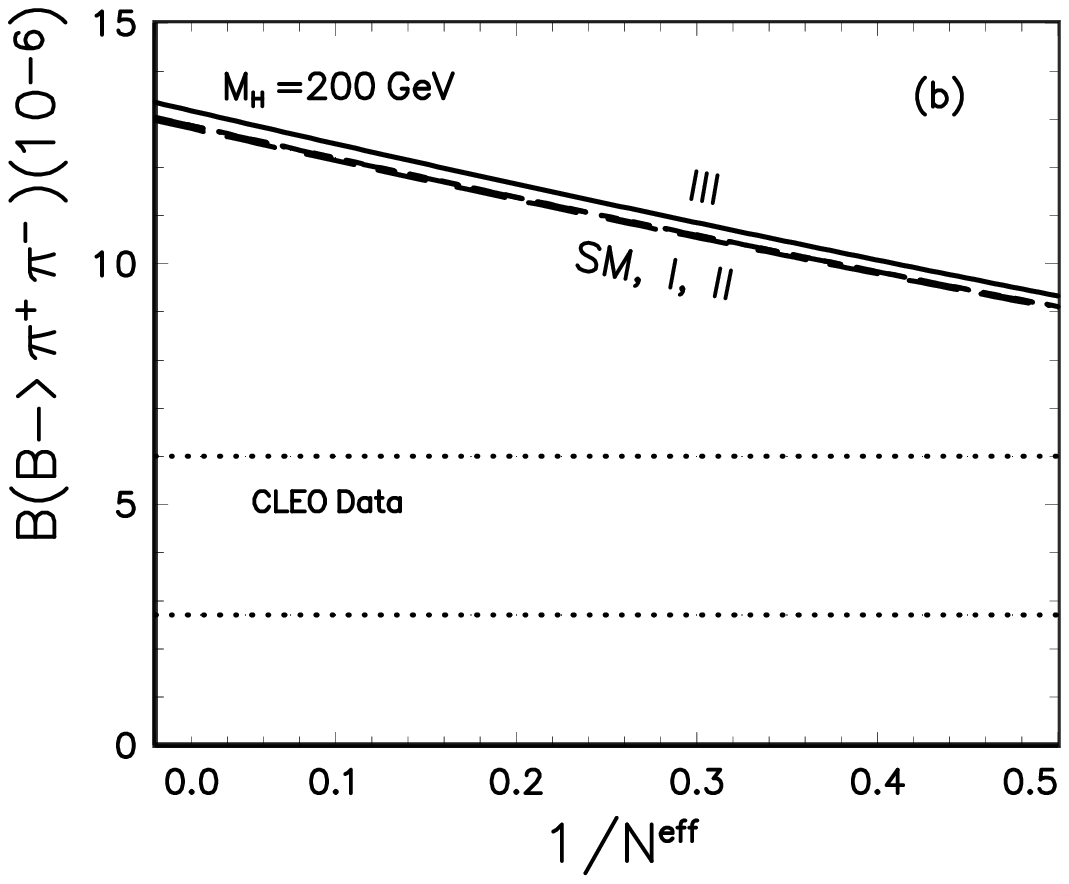}}
\vspace{-20pt}
\caption{Branching ratios ${\cal B}(B \to \pi^+ \pi^-)$ versus $\mhp$
and $1/N^{eff}$ in the SM and 2HDM's. For (a) and (b), we set $N^{eff}=3$ and
$\mhp=200$GeV, respectively. The four adjacent curves are the theoretical predictions in the SM and
models I,  II and III respectively. The band between two dots lines shows the
CLEO data with $1\sigma$ error: ${\cal B}(B \to \pi^+ \pi^-)
= (4.3 ^{+1.7}_{-1.6})\times 10^{-6}$.  }
\label{fig:fig2}
\end{minipage}
\end{figure}

As a Class-I decay channel, the $B^0 \to \pi^+  \pi^-$ decay is dominated
by the $b \to u$ tree diagram.  The band between two dots lines in
\fig{fig:fig2} shows the CLEO measurement. Since the new physics
corrections are very small in size, less than $3\%$ within
the considered parameter space, the four curves for the SM and 2HDM's are
close together and can not be separated clearly.
The theoretical predictions look higher than the CLEO measurement, but
they are still consistent with BaBar measurement because of very large
error of BaBar data. In fact the theoretical predictions for
${\cal B} (B \to \pi^+ \pi^-)$ in the SM and 2HDM's are  still consistent
with the CLEO data at the $2\sigma$ level if we consider currently
still large theoretical and experimental uncertainties.
On the other hand, if we take the average of CLEO and BaBar measurements,
${\cal B}(B_d^0\to \pi^+ \pi^-)=(5.5 \pm 1.5) \times 10^{-6}$, as
the experimental result, then the constraint on $F_0^{B\to \pi}(0)$ from the data
will be $F_0^{B\to \pi}(0)=0.25 \pm 0.03$ by setting $A=0.2205$, $\lambda=0.81$;
$\rho=0.12$, $\eta=0.34$, $N^{eff}=3$, and by neglecting FSI also.

\begin{figure}[t] %fig.3
\vspace{-80pt}
\begin{minipage}[]{0.96\textwidth}
\centerline{\epsfxsize=\textwidth \epsffile{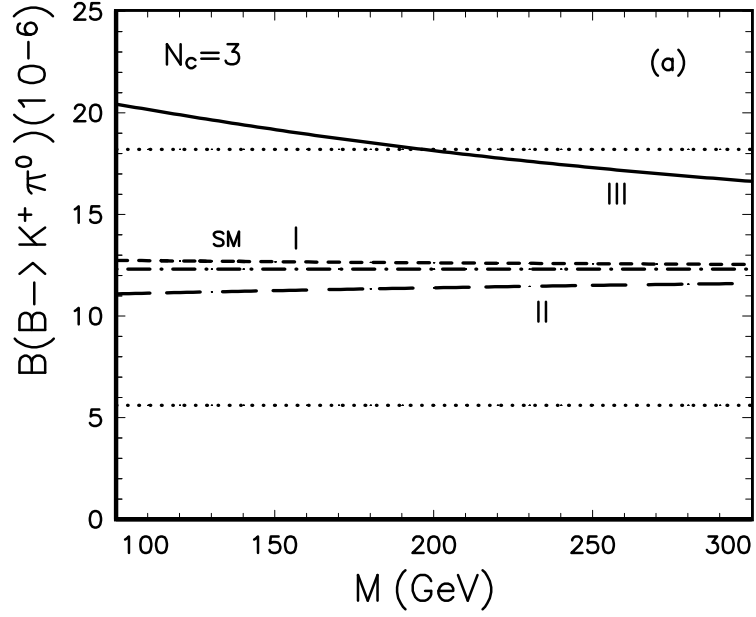}}
\vspace{-60pt}
\centerline{\epsfxsize=\textwidth \epsffile{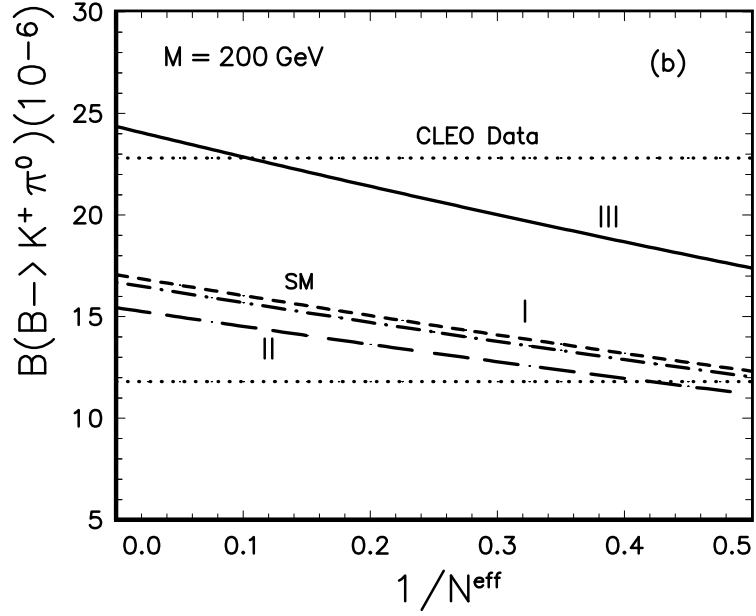}}
\vspace{-30pt}
\caption{Branching ratios ${\cal B}(B \to K^+ \pi^0)$ versus $\mhp$
and $1/N^{eff}$ in the SM and 2HDM's. For (a) and (b), we set $N^{eff}=3$ and
$\mhp=200$GeV, respectively. The dot-dashed, short-dashed,
long-dashed and solid curve correspond to the theoretical predictions in
the SM and models I, II and  III, respectively. The theoretical uncertainties
are not shown here.  The band between two dots lines shows the
CLEO data with $2\sigma$ errors: ${\cal B}(B \to K^+ \pi^0)=
(11.6 ^{+6.6}_{-6.0})\times 10^{-6}$.}
\label{fig:fig3}
\end{minipage}
\end{figure}

\begin{figure}[t] %fig.4
\vspace{-60pt}
\begin{minipage}[]{0.96\textwidth}
\centerline{\epsfxsize=\textwidth \epsffile{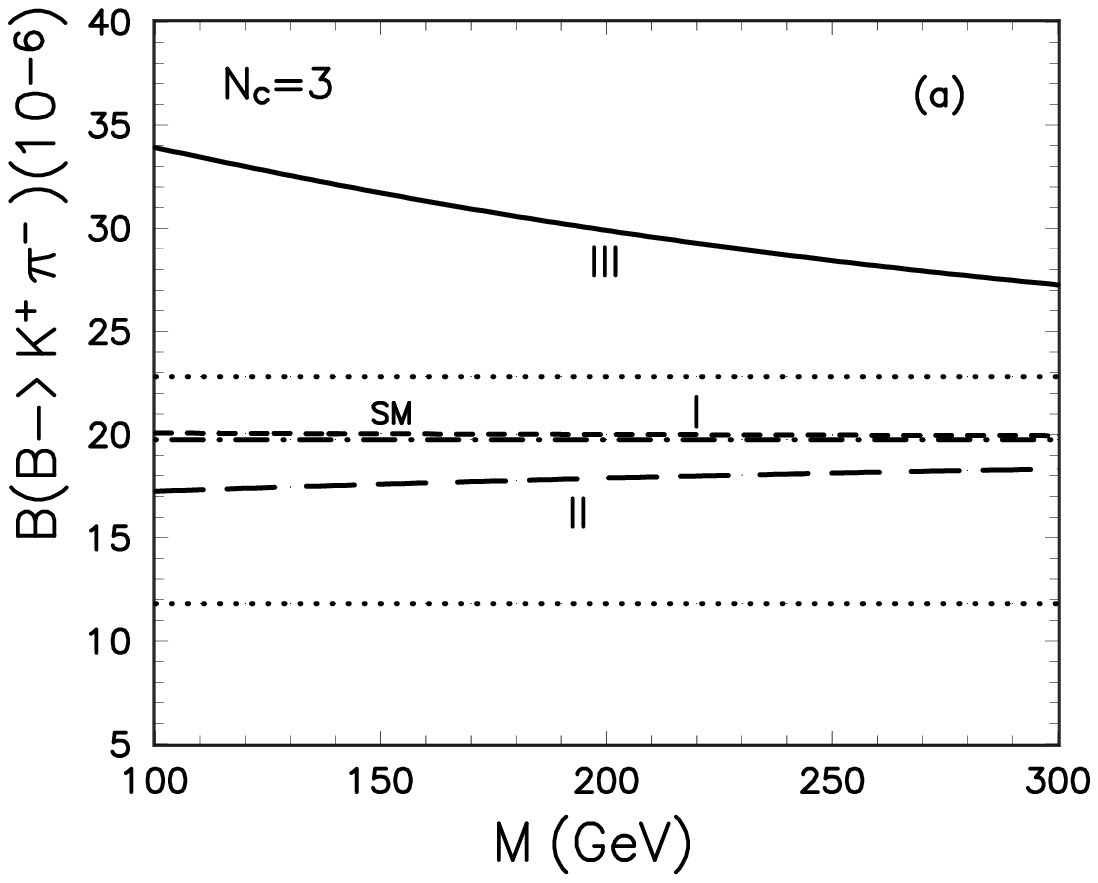}}
\vspace{-60pt}
\centerline{\epsfxsize=\textwidth \epsffile{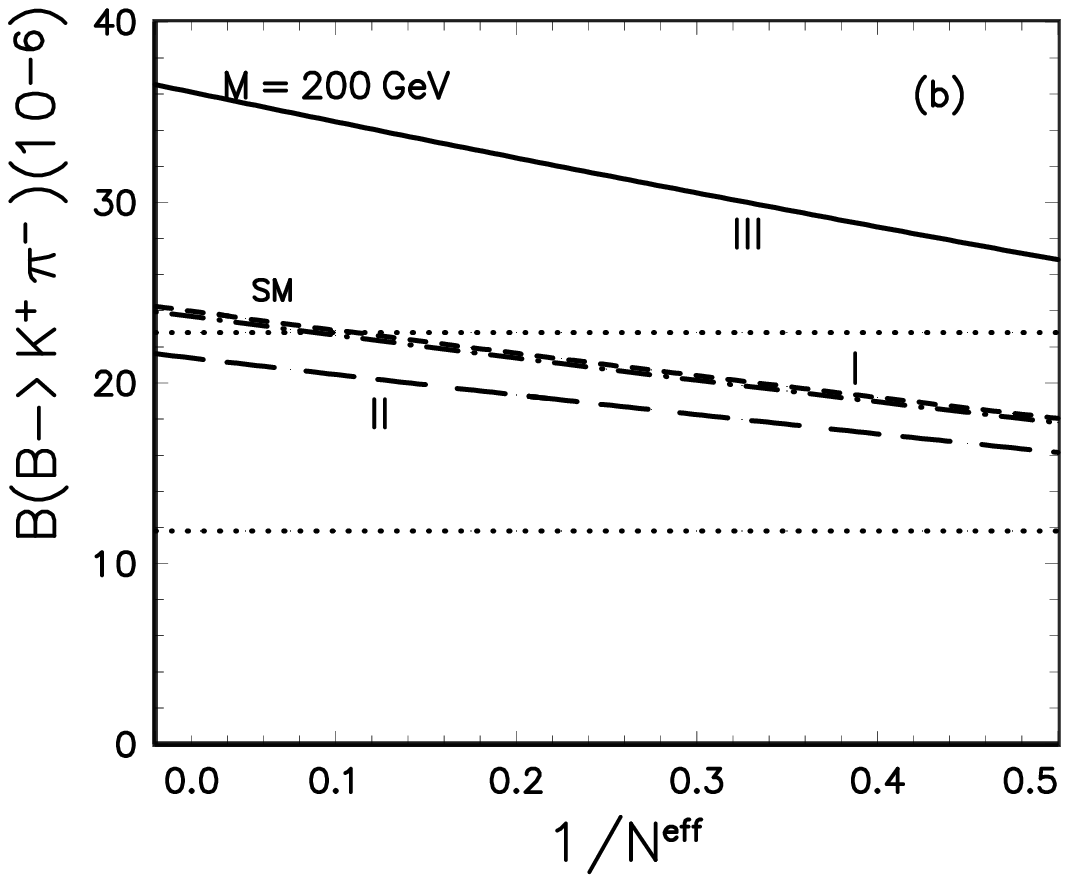}}
\vspace{-20pt}
\caption{Same as \fig{fig:fig3}, but for the decay $B \to K^+ \pi^-$.
 The dots band corresponds to the CLEO data with $2\sigma$ errors:
 ${\cal B}(B \to K^+ \pi^-)= (17.2 ^{+5.6}_{-5.4})\times 10^{-6}$.}
\label{fig:fig4}
\end{minipage}
\end{figure}

\begin{figure}[t]%fig.5
\vspace{-40pt}
\begin{minipage}[]{0.96\textwidth}
\centerline{\epsfxsize=\textwidth \epsffile{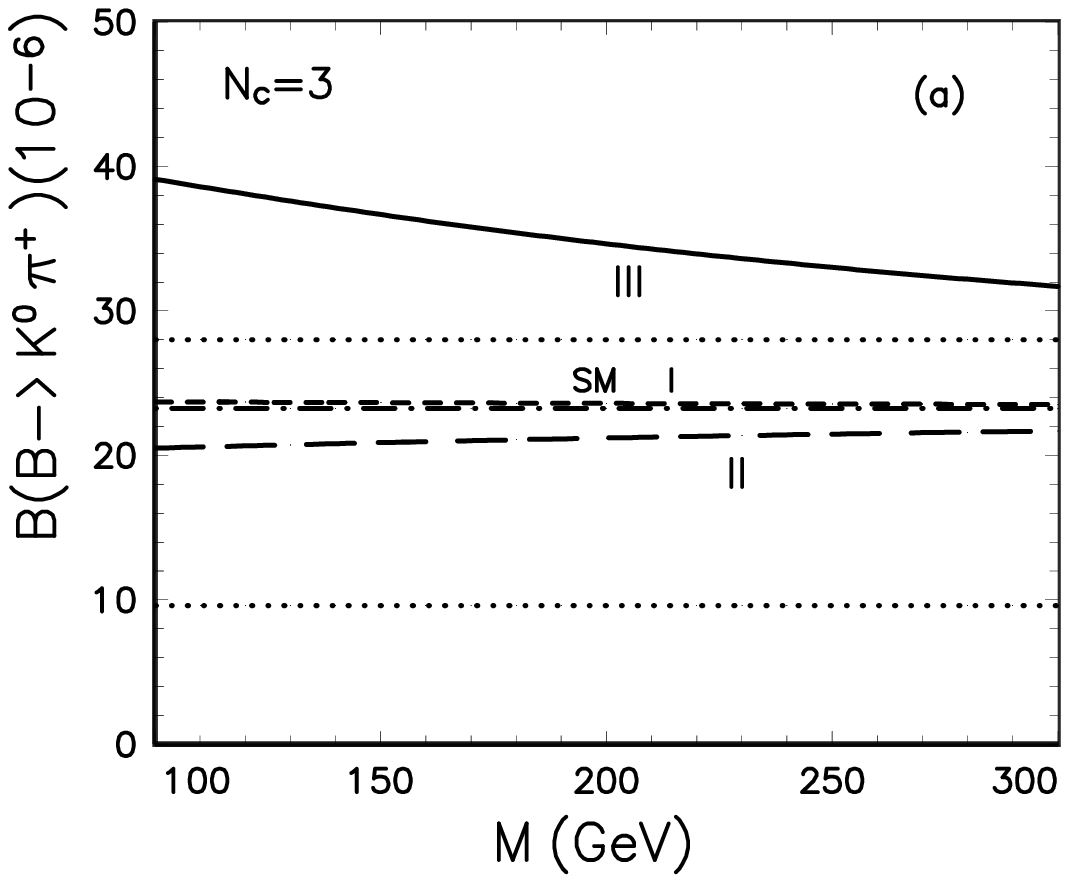}}
\vspace{-60pt}
\centerline{\epsfxsize=\textwidth \epsffile{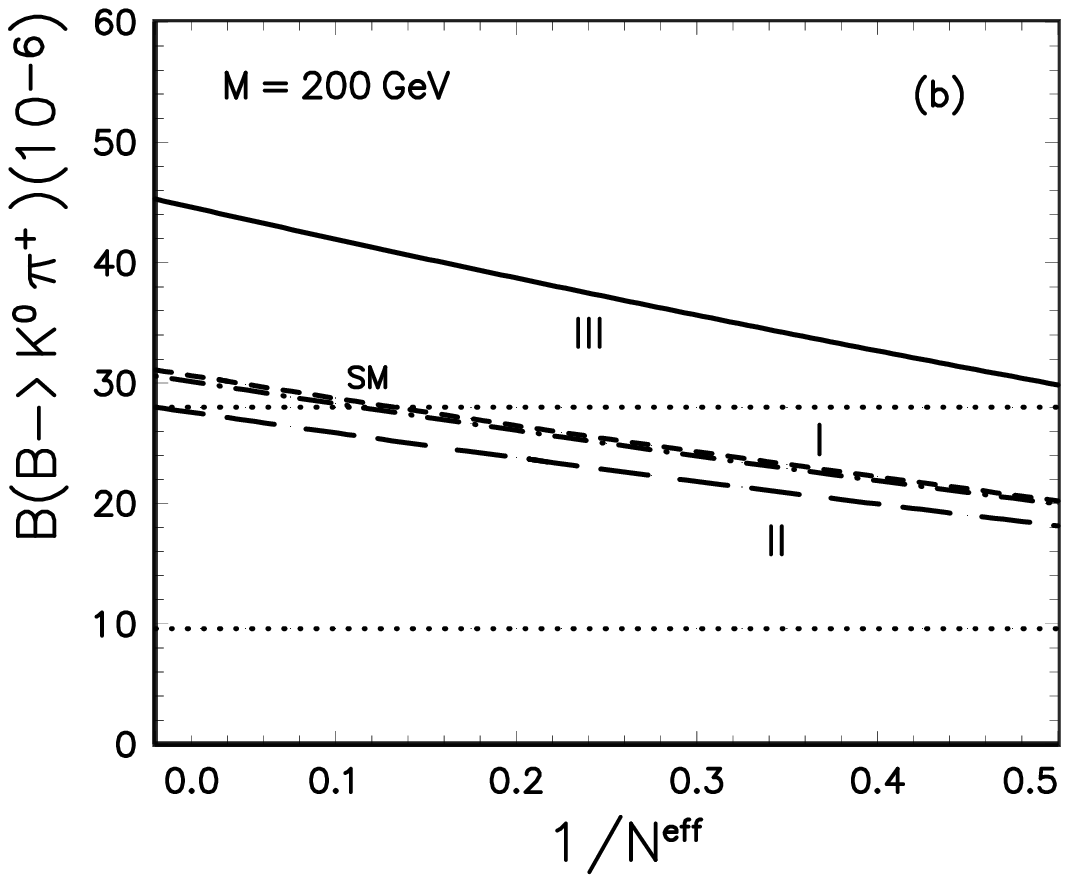}}
\vspace{-20pt}
\caption{Same as \fig{fig:fig3}, but for the decay $B \to K^0 \pi^+$.
The dots band corresponds to the CLEO data with $2\sigma$ errors:
${\cal B}(B \to K^0 \pi^+)=(18.2 ^{+9.8}_{-8.6})\times 10^{-6}$.}
\label{fig:fig5}
\end{minipage}
\end{figure}

\begin{figure}[t]%fig.6
\vspace{-40pt}
\begin{minipage}[]{0.96\textwidth}
\centerline{\epsfxsize=\textwidth \epsffile{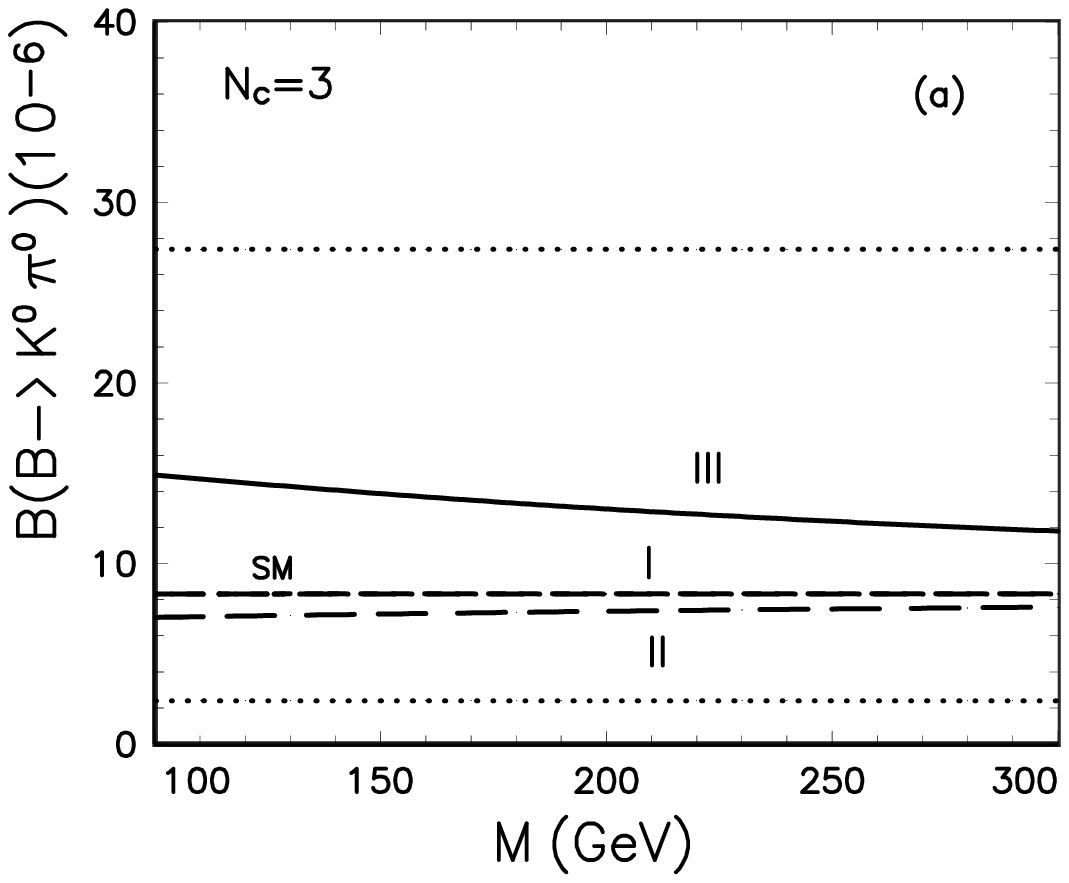}}
\vspace{-40pt}
\centerline{\epsfxsize=\textwidth \epsffile{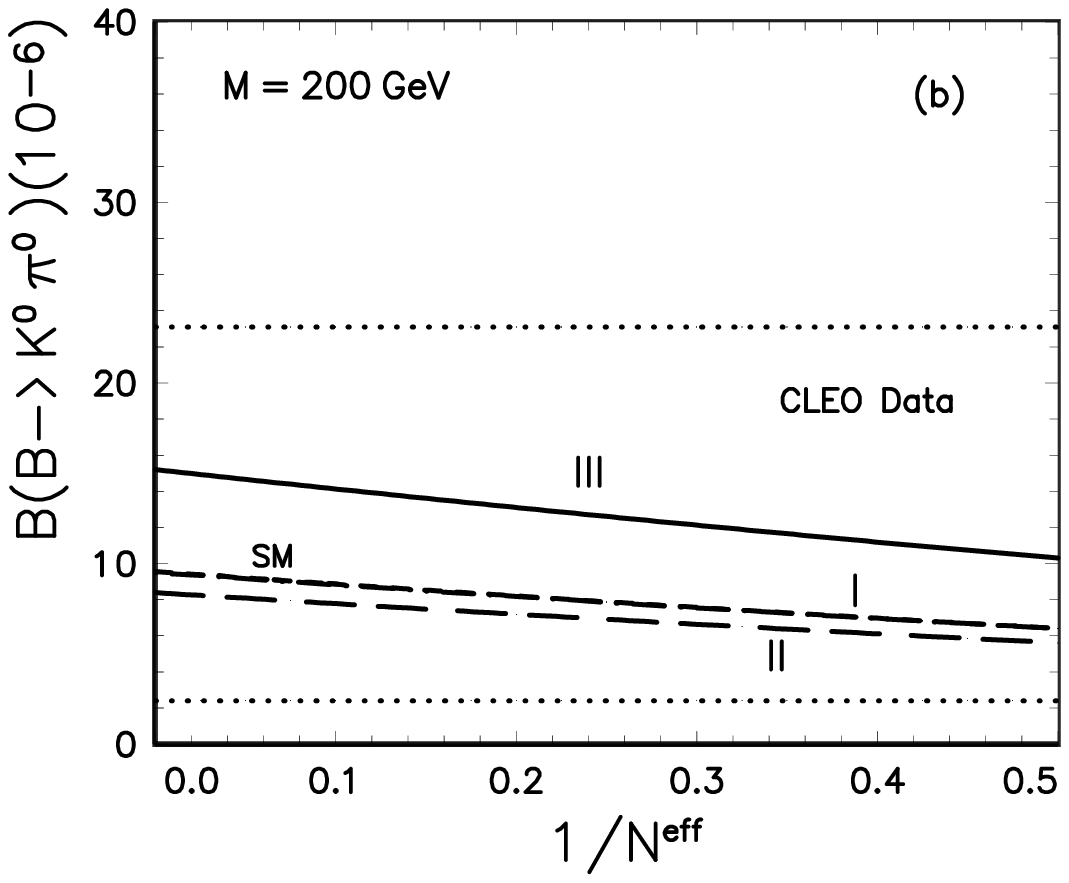}}
\vspace{-20pt}
\caption{ Same as \fig{fig:fig3}, but for the decay $B \to K^0 \pi^0$.
 The dots band corresponds to the CLEO data with $1\sigma$ error:
 ${\cal B}(B \to K^0 \pi^0)=(14.6 ^{+6.4}_{-6.1})\times 10^{-6}$.}
\label{fig:fig6}
\end{minipage}
\end{figure}

In the SM, the four Class-IV decays $B \to K \pi$ are dominated by the
$b\to s g$ gluonic penguin diagrams, with additional contributions from
$b \to u$ tree and electroweak penguin diagrams. Measurements of
$B\to K \pi$ decays are particularly important to measure the angle $\gamma$.
In model III, the new physics enhancements to the branching ratios
${\cal B}(B \to K \pi)$ are significant, $\sim (50 - 60)\%$, and show a moderate
dependence on the variations of other parameters, as illustrated in
figures (\ref{fig:fig3}-\ref{fig:fig6}). In models I and II,
however, the new  physics corrections are always very small in size.

For the decays  $B \to K^+ \pi^-, K^0
\pi^+$, the theoretical predictions in model III are higher than the CLEO
data as shown in \fig{fig:fig4} and \fig{fig:fig5}, but they are still
consistent with the CLEO data at the $2\sigma$ level if we consider
currently still large theoretical uncertainties.
As a simple illustration of effects of the theoretical uncertainties, we
recalculate the branching ratios of $B \to K^+\pi^-$ and $K^0\pi^+$ decays
by using $F_0^{B\to \pi}(0)=0.25$ instead of the ordinary BSW value
$F_0^{B\to \pi}(0)=0.33$ while keeping all other input parameters unchanged,
and find numerically that
\beq
{\cal B}(B \to K^+ \pi^-)&=& \left \{\begin{array}{ll}
( 11.3 ^{+2.9\; +2.5 }_{-2.5\; -1.1 } )\times 10^{-6} & {\rm in \ \ SM}, \\
( 17.2 ^{+4.3\; +3.7 }_{-3.9\; -1.8 } )\times 10^{-6} & {\rm in \ \ Model\ \  III}, \\
\end{array} \right. \label{eq:brpp12} \\
{\cal B}(B \to K^0 \pi^+)&=& \left \{\begin{array}{ll}
( 13.3 ^{+3.4 \; +4.2 }_{-3.0\; -1.9} )\times 10^{-6} & {\rm in \ \ SM}, \\
( 19.9 ^{+5.0\;  +6.1 }_{-4.5\; -2.8} )\times 10^{-6} & {\rm in \ \ Model \ \ III},  \\
\end{array} \right. \label{eq:brpp13}
\eeq
for $N^{eff}=3$ and $\mhp=200$ GeV.  Here the first and second error correspond to
$F_0^{B\to \pi}(0)=0.25\pm 0.03$ and $2\leq N^{eff} \leq \infty$, respectively.
It is evident that the
theoretical predictions of the two ratios ${\cal B}(B \to K^+ \pi^-)$ and
${\cal B}(B \to K^0 \pi^+)$ in the SM and model III can lie within the CLEO data.

Figures (\ref{fig:fig3}-\ref{fig:fig6}) show the mass and $N^{eff}$-dependence of
the branching ratios for four $B \to K \pi$ decay modes in the SM and models I, II
and III, using the input parameters as given in Eq.(\ref{eq:lm3}) and Appendix,
and assuming $\theta=0^0$, $\tan{\beta}=2$, and
$k^2=m_b^2/2$. For Figs.(\ref{fig:fig3}a-\ref{fig:fig6}a), we set $N^{eff}=3$
and assume that $\mhp=100 - 300$GeV. For Figs.(\ref{fig:fig3}b-\ref{fig:fig6}b),
we set $\mhp=200$GeV, and assume that $1/N^{eff}=0 - 0.5$. In all four figures,
the band between two dots lines shows the corresponding CLEO measurements
with $2\sigma$ errors.
For $B \to K^0 \pi^0$ decay, the inclusion of new physics contribution
will improve the agreement between the data and theoretical
prediction, as illustrated in \fig{fig:fig6}.
For other three $B \to K \pi$ decays, the theoretical
predictions in the model III are still consistent with the data if the
theoretical uncertainties are taken into account.

\subsection{$B \to K \etapp $ decays and the new physics effects}

For $B^+ \to K^+ \eta$ and $B^0 \to K^0 \eta$ decay modes, the new physics
corrections are large (small) in model III (models I and II). The theoretical
predictions in the SM and 2HDM's are consistent with the new CLEO
upper limits.

For $B \to K \etap$ decay modes, the situation is very interesting now.
In 1997, CLEO firstly reported the unexpectedly large $B \to K \etap$
rates \cite{cleo98}, which is confirmed very recently by  CLEO with the
full CLEO II/II.V data sample of $19$ million produced B mesons
\cite{cleo9912,cleo-t0020}. The  $K \etap$ signal is large, stable and
has small error ($\sim 14\%$). Those measured ratios are clearly
much larger than the SM predictions as given in table (\ref{bppsm}).
In \cite{cheng00a}, Cheng and Yang considered various
possible enhancements to $K \etap$ decay modes in the framework of the SM
\footnote{As discussed in \cite{cheng00b}, $B \to K \etap$ decay may
get enhanced due to  (i) a small $m_s$ at the scale $m_b$, (ii)
the sizable $SU(3)$ breaking,
(iii) large $F_0^{B\to \etap}$, (iv) the $\etap$ charm content, and (v) constructive
interference in tree amplitudes. But these possible enhancements are partially
washed out by the anomaly effect in the matrix element of pseudoscalar densities
\cite{ali98,hou98}.}, but found that the net enhancement is not very large:
${\cal B}(B^{\pm} \to K^{\pm} \etap) =(40- 50)\times 10^{-6}$, which is smaller
than the CLEO data \footnote{Although this prediction is consistent with BaBar
measurement, one should note that the error of BaBar measurement is still
much larger than that of CLEO data. More statistics is clearly  required for
BaBar to make a definite  conclusion.}
At present, it is indeed difficult to explain the observed large rate for
$B \to K \etap $ \cite{cleo9912,cleo-t0020}.
This fact strongly suggests the requirement for additional contributions
unique to the $\etap $ meson in the framework of the SM, or from new physics
beyond the SM.

In models I and II, the new physics contributions are too small (or negative)
to provide the required enhancement. This feature remains unchanged within
the considered range of $\tan{\beta}=1 - 50$.

\begin{figure}[t]%fig.7
\vspace{-80pt}
\begin{minipage}[]{0.96\textwidth}
\centerline{\epsfxsize=\textwidth \epsffile{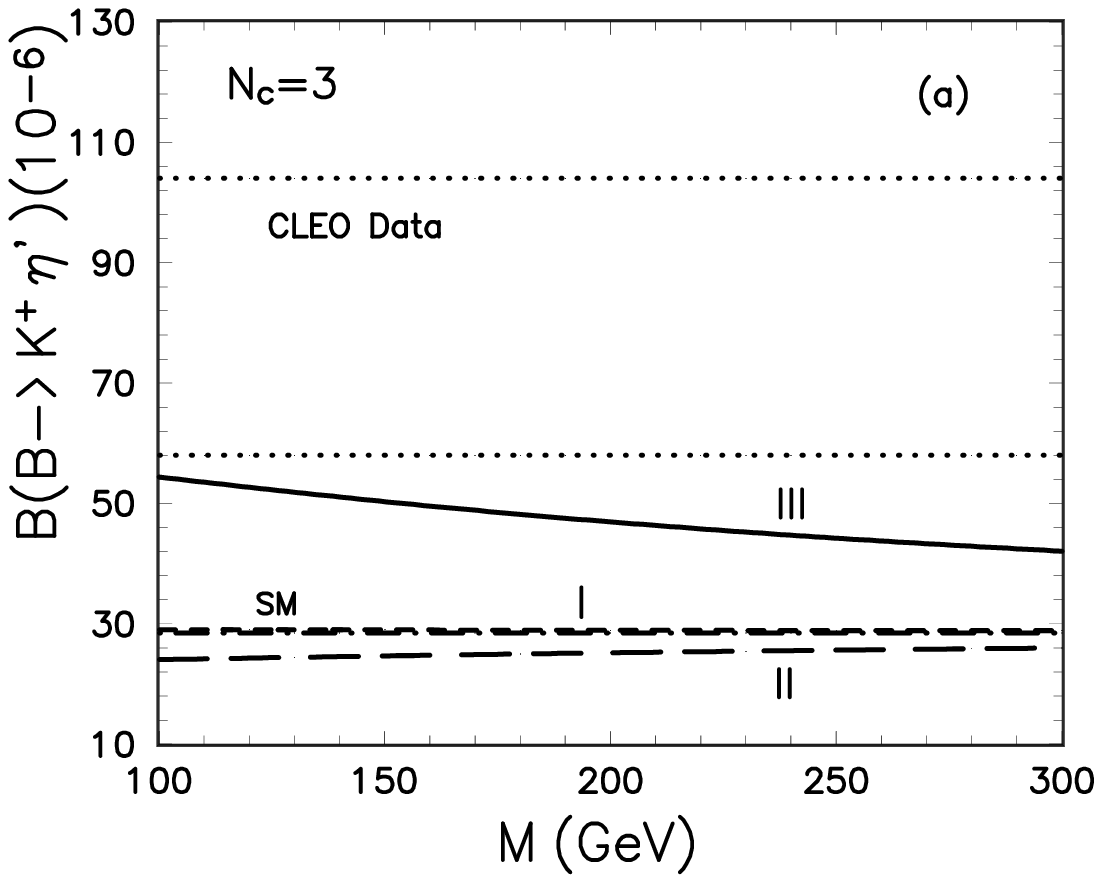}}
\vspace{-80pt}
\centerline{\epsfxsize=\textwidth \epsffile{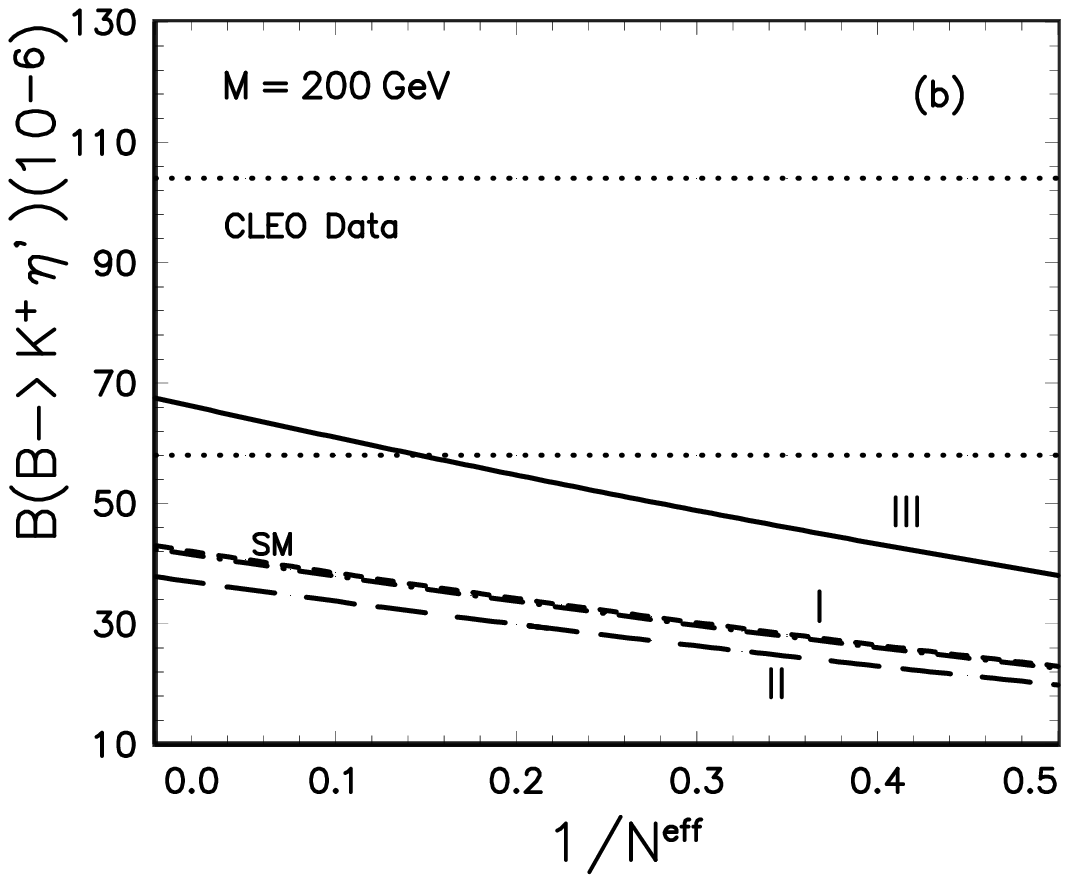}}
\vspace{-30pt}
\caption{Plots of ${\cal B}(B^+ \to K^+ \etap)$ {\it versus}
$\mhp$ and $1/N^{eff}$ in the SM and 2HDM's. For (a) and (b), we set
$N^{eff}=3$ and $\mhp=200$GeV, respectively.
 The dot-dashed curve and the closely adjacent short-dashed curve
refer to the theoretical predictions in the SM and model I; while
the long-dashed and solid curve correspond to the theoretical
predictions in models II and III, respectively.
The theoretical uncertainties are not shown here.
The dots band corresponds to the CLEO data with $2\sigma$ errors:
${\cal  B}(B^+ \to  K^+ \etap)=(80^{+24}_{-22})\times 10^{-6}$. }
\label{fig:fig7}
\end{minipage}
\end{figure}

\begin{figure}[t]%fig.8
\vspace{-40pt}
\begin{minipage}[]{0.96\textwidth}
\centerline{\epsfxsize=\textwidth \epsffile{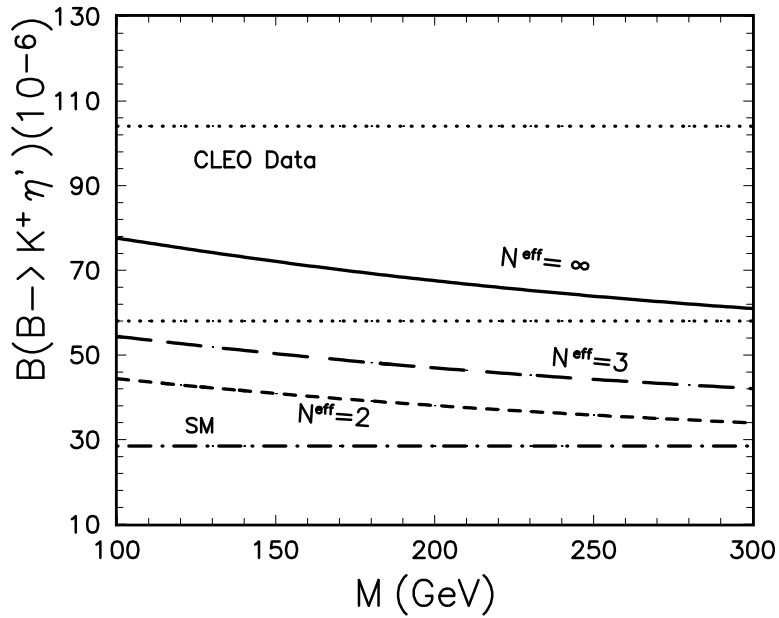}}
\caption{Plots of ${\cal B}(B^+ \to K^+ \etap)$ {\it versus} $\mhp$ in the
SM and model III. The dot-dashed line shows  the SM prediction with $N^{eff}=3$.
The short-dashed, long-dashed and solid curve correspond to model III
predictions for $N^{eff}=2,3,\infty$, respectively.
Other theoretical uncertainties are not shown here.
The dots band corresponds to the CLEO data with $2\sigma$ errors:
${\cal  B}(B^+ \to  K^+ \etap)=(80^{+24}_{-22})\times 10^{-6}$. }
\label{fig:fig8}
\end{minipage}
\end{figure}

\begin{figure}[t]%fig.9
\vspace{-40pt}
\begin{minipage}[]{0.96\textwidth}
\centerline{\epsfxsize=\textwidth \epsffile{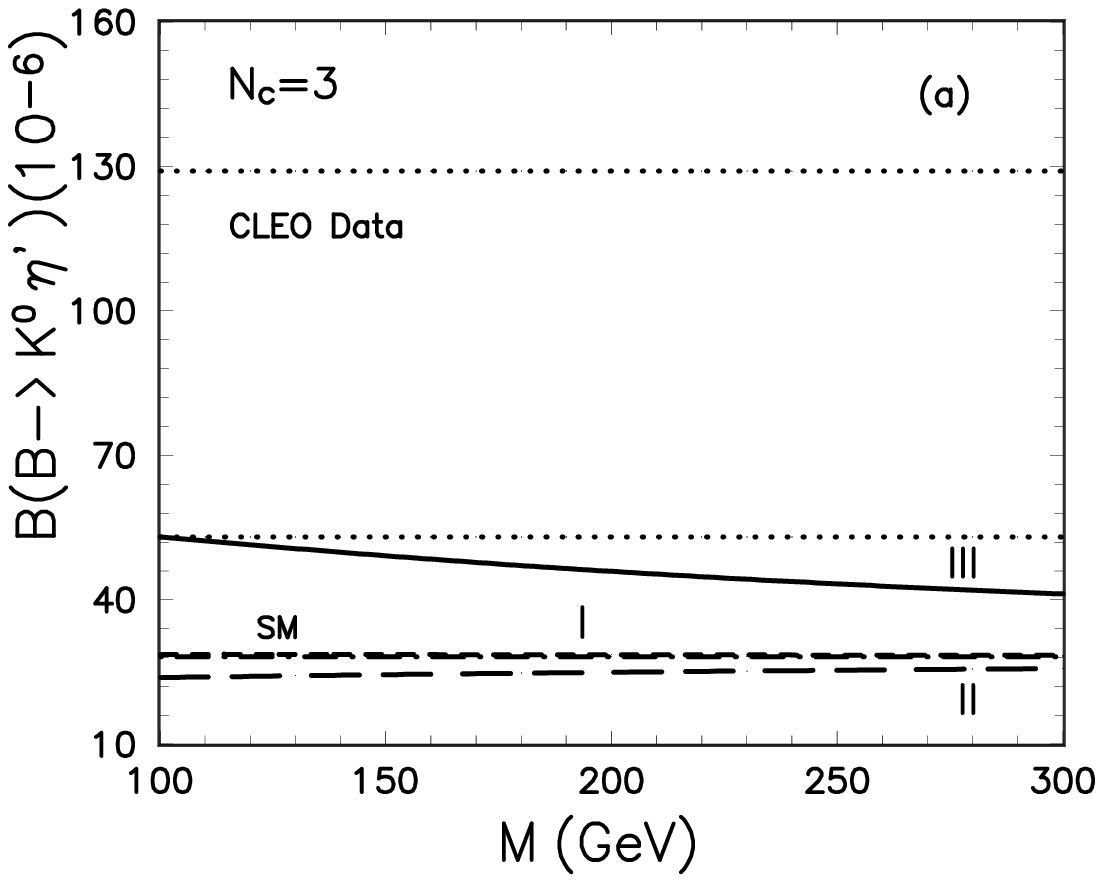}}
\vspace{-60pt}
\centerline{\epsfxsize=\textwidth \epsffile{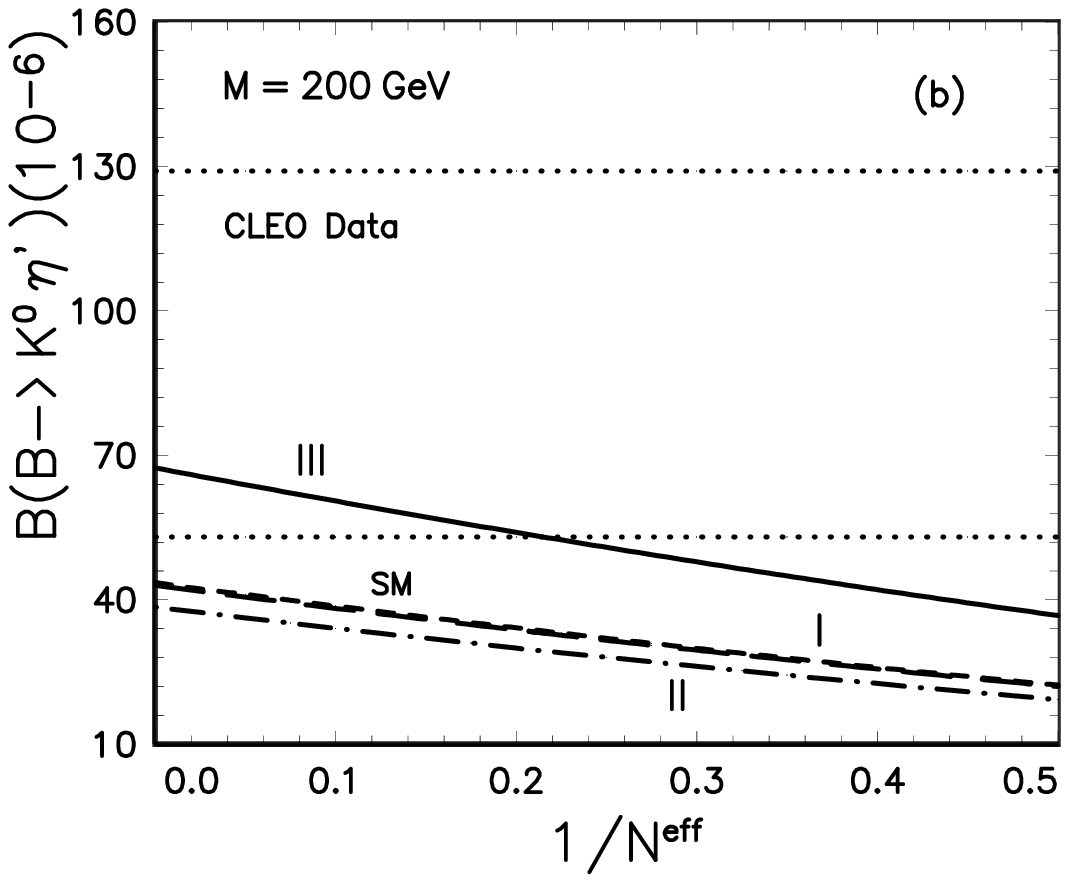}}
\vspace{-20pt}
\caption{Same as \fig{fig:fig7}, but for the decay $B \to K^0 \etap$.
The dots band corresponds to the CLEO data with $2\sigma$ errors:
${\cal  B}(B^0 \to  K^0 \etap)=(89^{+40}_{-36})\times 10^{-6}$.}
\label{fig:fig9}
\end{minipage}
\end{figure}

\begin{figure}[t]%fig.10
\vspace{-60pt}
\begin{minipage}[]{0.96\textwidth}
\centerline{\epsfxsize=\textwidth \epsffile{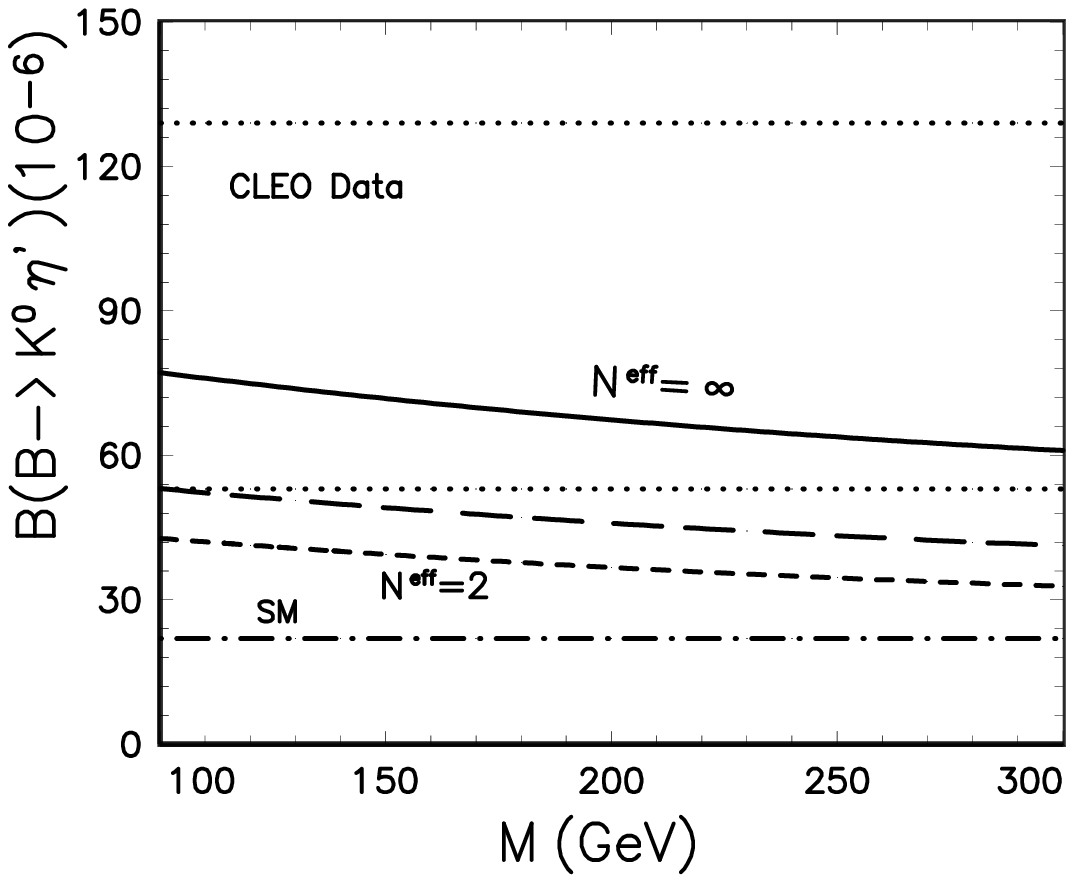}}
\caption{Same as \fig{fig:fig8}, but for the decay $B \to K^0 \etap$.
The dots band corresponds to the CLEO data with $2\sigma$ errors:
${\cal  B}(B^0 \to  K^0 \etap)=(89^{+40}_{-36})\times 10^{-6}$.}
\label{fig:fig10}
\end{minipage}
\end{figure}

In model III, however, the new
physics enhancements are significant, $\sim 60\%$, and have a moderate
dependence on $\mhp$ and $N^{eff}$, as illustrated  by the solid curves
in Figs.(\ref{fig:fig7}-\ref{fig:fig10}) where only the central values of
theoretical predictions in model III are shown. If we take into account
other theoretical uncertainties, the theoretical
predictions for ratios ${\cal B}(B \to K \etap )$ in model III will become
consistent with the CLEO data:
\beq
{\cal B}(B^+ \to K^+ \etap )&=& \left \{\begin{array}{ll}
( 69-92 )\times 10^{-6} & {\rm [CLEO] }~,  \\
( 20-52 )\times 10^{-6} & {\rm [SM] }~, \\
( 34-74 )\times 10^{-6} & {\rm [Model \ \ III]}~,  \\
\end{array} \right. \label{eq:brpp16}  \\
{\cal B}(B^0 \to K^0 \etap )&=& \left \{\begin{array}{ll}
( 71-109 )\times 10^{-6} & {\rm [CLEO] }~,  \\
( 19-53 )\times 10^{-6} & {\rm [SM]}~, \\
( 33-73 )\times 10^{-6} & {\rm [Model \ \ III]}~.  \\
\end{array} \right. \label{eq:brpp18}
\eeq
Here the major theoretical uncertainties induced by using different set of
form factors  and varying $k^2, \eta$ and $N^{eff}$ in the ranges of
$\delta k^2 =\pm 2 GeV^2$, $\delta \eta =\pm 0.08$ and
$N^{eff}=2-\infty$ have been taken into account.

Figures (\ref{fig:fig7},\ref{fig:fig9}) show the mass and $N^{eff}$
dependence of ${\cal B}(B \to K \etap)$ in the SM and 2HDM's. The upper
dots band shows the CLEO measurements with $2\sigma$ errors.
The short-dashed, dot-dashed, long-dashed and solid curve
refers to the theoretical predictions in the SM, models I, II and III,
respectively. As shown explicitly in Figs.(\ref{fig:fig8},\ref{fig:fig10}),
in which the short-dashed, long-dashed and solid curve corresponds to
the model III predictions for $N^{eff}=2,3,\infty$ respectively, the
theoretical predictions become now consistent with the CLEO
measurement due to the inclusion of new physics enhancement in
model III.

\section{$B \to P V$ decays}\label{sec:bpv}

In tables (\ref{bpvsm}-\ref{bpvm2}) we present the branching ratios
for the thirty seven $B \to P V$ decay modes involving $b\to d$ and  $b \to s$
transitions in the SM and models I, II and III by using the BSW form factors
and by employing generalized factorization approach.
Theoretical predictions are made by using the same input parameters as those for the
$B \to PP$ decays in last section.

For studied thirty seven $B \to PV$ decays, two general features are as
follows:

\begin{itemize}
\item
The theoretical predictions for those seven measured decay rates are consistent
with the CLEO data within $2\sigma$ errors. All other
estimated branching ratios in the SM and 2HDM's as given in tables
(\ref{bpvsm}-\ref{bpvm2}) are all consistent with the new CLEO upper limits.

\item
For most decay modes, the differences induced by using whether BSW or LQQSR
form factors are small, $\sim 15\% $. We therefore use the BSW form factors
only in the calculation of new physics effects.

\end{itemize}

There are so far seven measured branching ratios of $B \to PV$ decays.
For the first three decay modes, $B\to \rho^{\pm} \rho^{\mp}, \rho^0 \pi^+,
\omega \pi^+$, the new physics corrections are small in size, $< 5\%$, and
have a weak dependence on  $\mhp$ and  $N^{eff}$,
as shown in tables (\ref{bpvm3},\ref{bpvm2}). Consequently, the theoretical
predictions in the SM and models I, II and III agree well with CLEO
measurements.

Because of the appearance of very large ${\cal B}(B \to K \etap)$, the decay modes
$B \to K^* \etapp$ also draw more attentions now. Very recently, CLEO and
Belle reported
their first observation \cite{cleo99,cleo9912,belle2000}
of $B \to K^* \eta, K^{*+} \pi^-$ and $B \to K^{+}\phi $ decays:
\beq
{\cal B}(B^+ \to K^{*+} \eta )&=&(26.4^{+9.6}_{-8.2} \pm 3.3)\times 10^{-6},\\
{\cal B}(B^0 \to K^{*0} \eta )&=&(13.8^{+5.5}_{-4.6} \pm 1.6)\times 10^{-6},\\
{\cal B}(B^0 \to K^{*+} \pi^-)&=&(22^{+8 +4}_{-6 -5} )\times
10^{-6},\\
{\cal B}(B^+ \to K^{+} \phi )&=&(17.2^{+6.7}_{-5.4} \pm 1.8 )\times 10^{-6},
\eeq
while the theoretical predictions in the SM and model III are
\beq
{\cal B}(B^+ \to K^{*+} \eta )&=& \left \{\begin{array}{ll}
( 2-4 )\times 10^{-6} & {\rm [SM] }~, \\
( 2-5 )\times 10^{-6} & {\rm [Model \ \ III]}~,  \\
\end{array} \right. \label{eq:brpv26}  \\
{\cal B}(B^+ \to K^{*0} \eta )&=& \left \{\begin{array}{ll}
( 2-5 )\times 10^{-6} & {\rm [SM] }~, \\
( 3-6 )\times 10^{-6} & {\rm [Model \ \ III]}~,  \\
\end{array} \right. \label{eq:brpv28}  \\
{\cal B}(B^+ \to K^{*+} \pi^- )&=& \left \{\begin{array}{ll}
( 7-16)\times 10^{-6} & {\rm [SM] }~, \\
( 10-22)\times 10^{-6} & {\rm [Model \ \ III]}~,  \\
\end{array} \right. \label{eq:brpv30}  \\
{\cal B}(B^+ \to K^{+} \phi )&=& \left \{\begin{array}{ll}
( 0.5-28 )\times 10^{-6} & {\rm [SM] }~, \\
( 1-39 )\times 10^{-6} & {\rm [Model \ \ III]}~,  \\
\end{array} \right. \label{eq:brpv34}
\eeq
where the uncertainties induced by using the BSW or LQQSE form factors, and
setting $k^2=m_b^2/2 \pm 2 GeV^2$, $\eta=0.34 \pm 0.08$ and $N^{eff}=2-\infty$,
have been taken into account.
Although the central values of the theoretical
predictions in the SM are  much smaller than the corresponding central values
of the CLEO measurements, the theoretical predictions are still consistent with
the data within $2\sigma$ errors because current experimental error is still
large. Further improvement of experimental measurements about the decay modes
$B \to K^* \eta,\;  K^{*+} \pi^-$ will tell us whether there is any
discrepancy between the theory and experiments for these three decay modes.
At present, any positive contributions to the above three branching
ratios from new mechanisms in the SM or from new physics beyond the SM are
clearly preferred by the CLEO data.

\begin{figure}[t]%fig.11
\vspace{-60pt}
\begin{minipage}[]{0.96\textwidth}
\centerline{\epsfxsize=\textwidth \epsffile{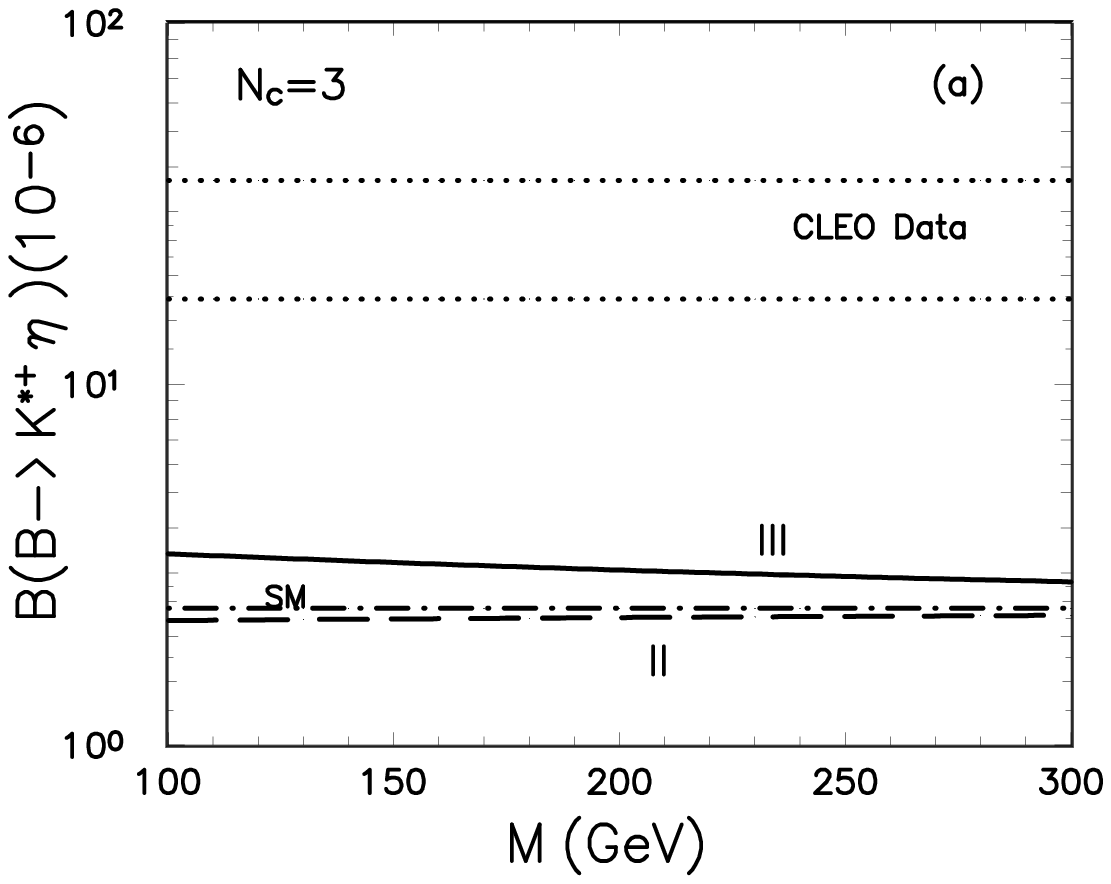}}
\vspace{-60pt}
\centerline{\epsfxsize=\textwidth \epsffile{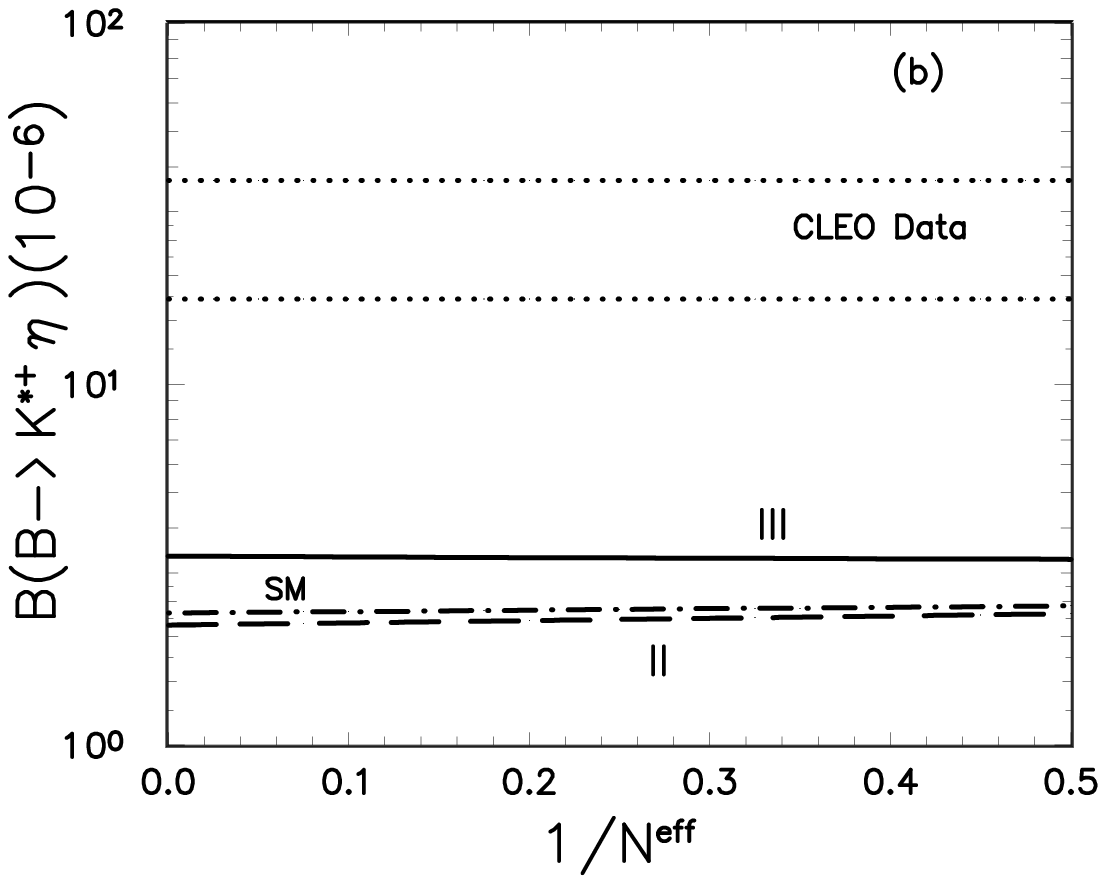}}
\vspace{-20pt}
\caption{ ${\cal B}(B \to K^{*+} \eta)$ versus $\mhp$ and $1/N^{eff}$
in the SM and 2HDM's.  For (a) and (b), we set $N^{eff}=3$ and $\mhp=200$GeV,
respectively. The dot-dashed, long-dashed and solid curve
show the theoretical predictions in the SM and models II and III, respectively.
The theoretical uncertainties are not shown here.
The dots band corresponds to the CLEO data with $1\sigma$ error:
${\cal  B}(B^+ \to  K^{*+} \eta) = ( 26.4^{+10.2}_{-8.8})\times 10^{-6}$.}
\label{fig:fig11}
\end{minipage}
\end{figure}

\begin{figure}[t]%fig.12
\vspace{-40pt}
\begin{minipage}[]{0.96\textwidth}
\centerline{\epsfxsize=\textwidth \epsffile{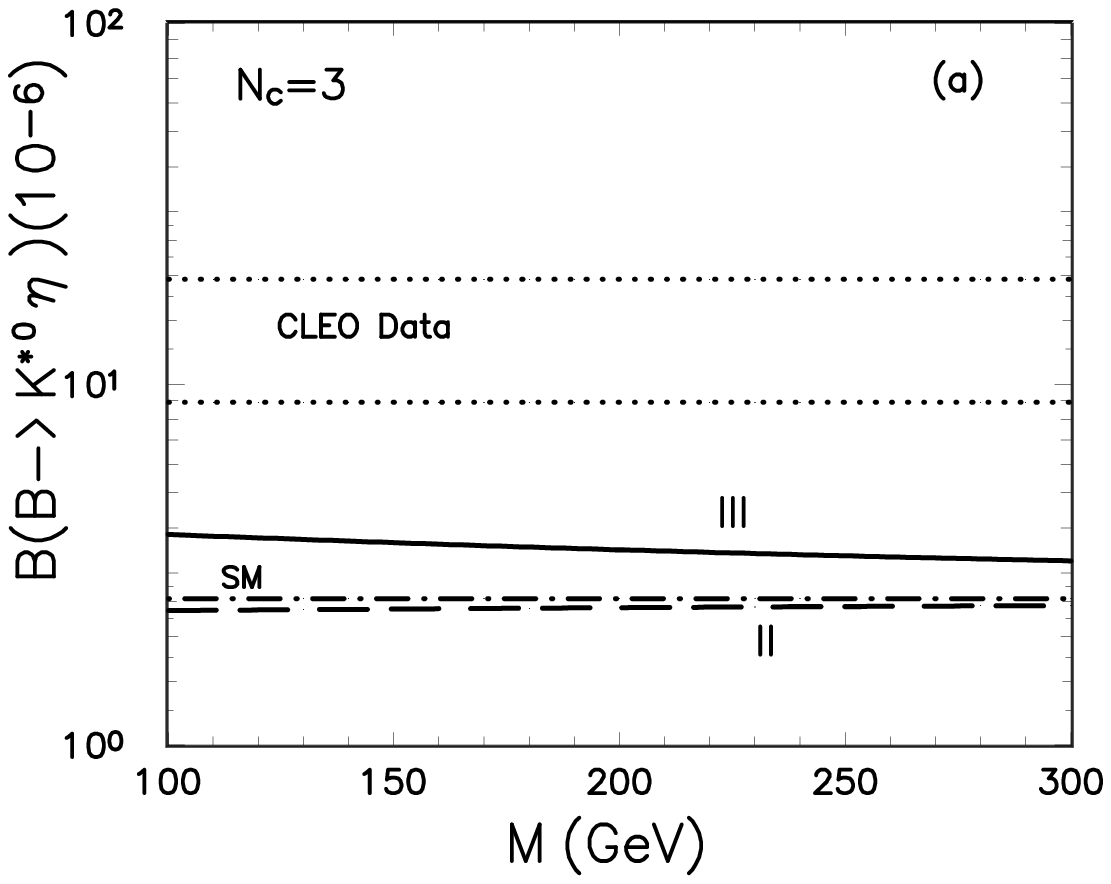}}
\vspace{-60pt}
\centerline{\epsfxsize=\textwidth \epsffile{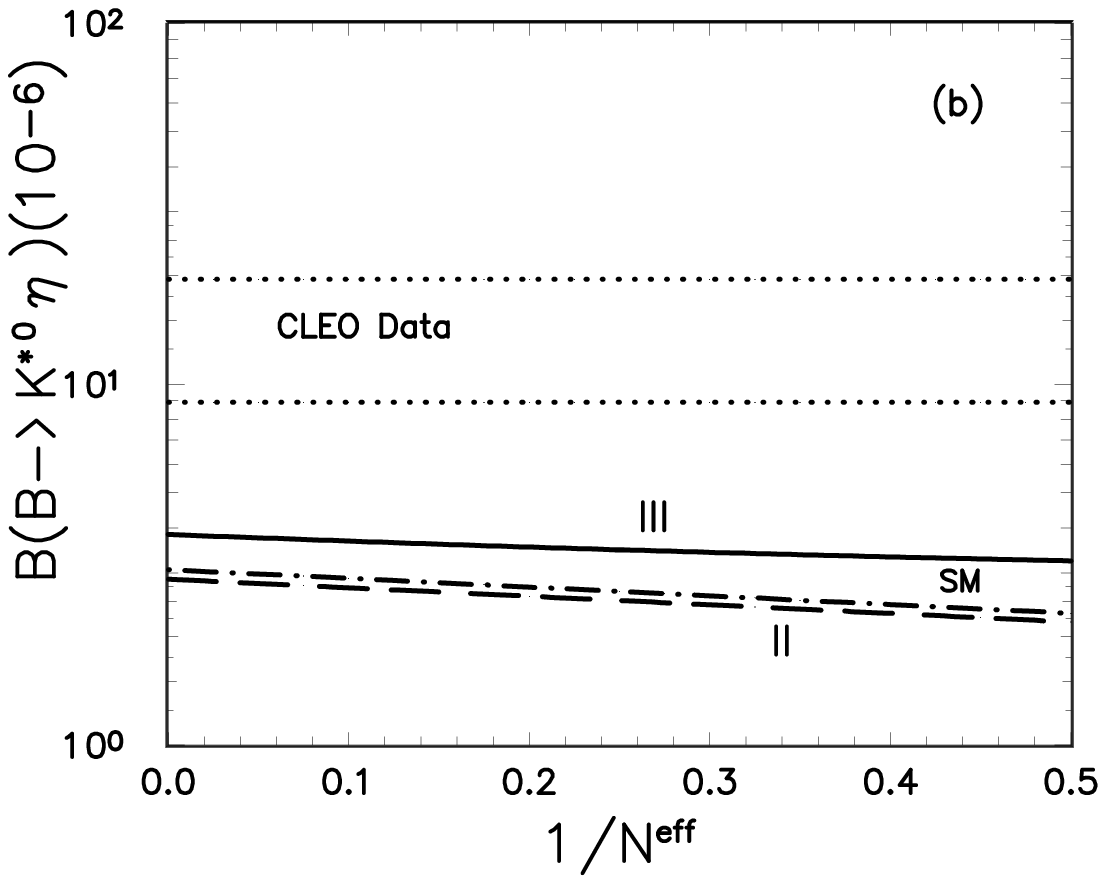}}
\vspace{-20pt}
\caption{Same as \fig{fig:fig11}, but for the decay $B \to K^{*0} \eta$.
The dots band corresponds to the CLEO data with $1\sigma$ error:
${\cal  B}(B^0 \to  K^{*0} \eta) = ( 13.8^{+5.7}_{-4.9})\times 10^{-6}$.}
\label{fig:fig12}
\end{minipage}
\end{figure}

\begin{figure}[t]%fig.13
\vspace{-40pt}
\begin{minipage}[]{0.96\textwidth}
\centerline{\epsfxsize=\textwidth \epsffile{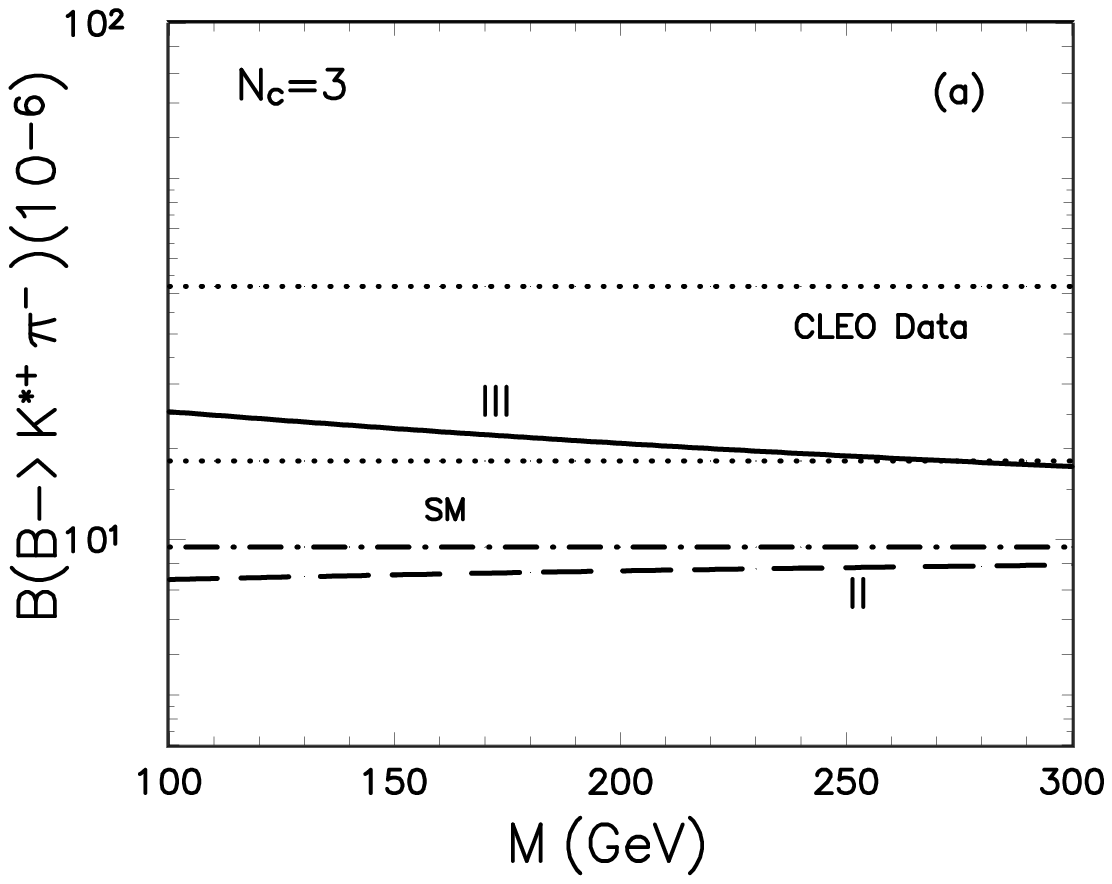}}
\vspace{-60pt}
\centerline{\epsfxsize=\textwidth \epsffile{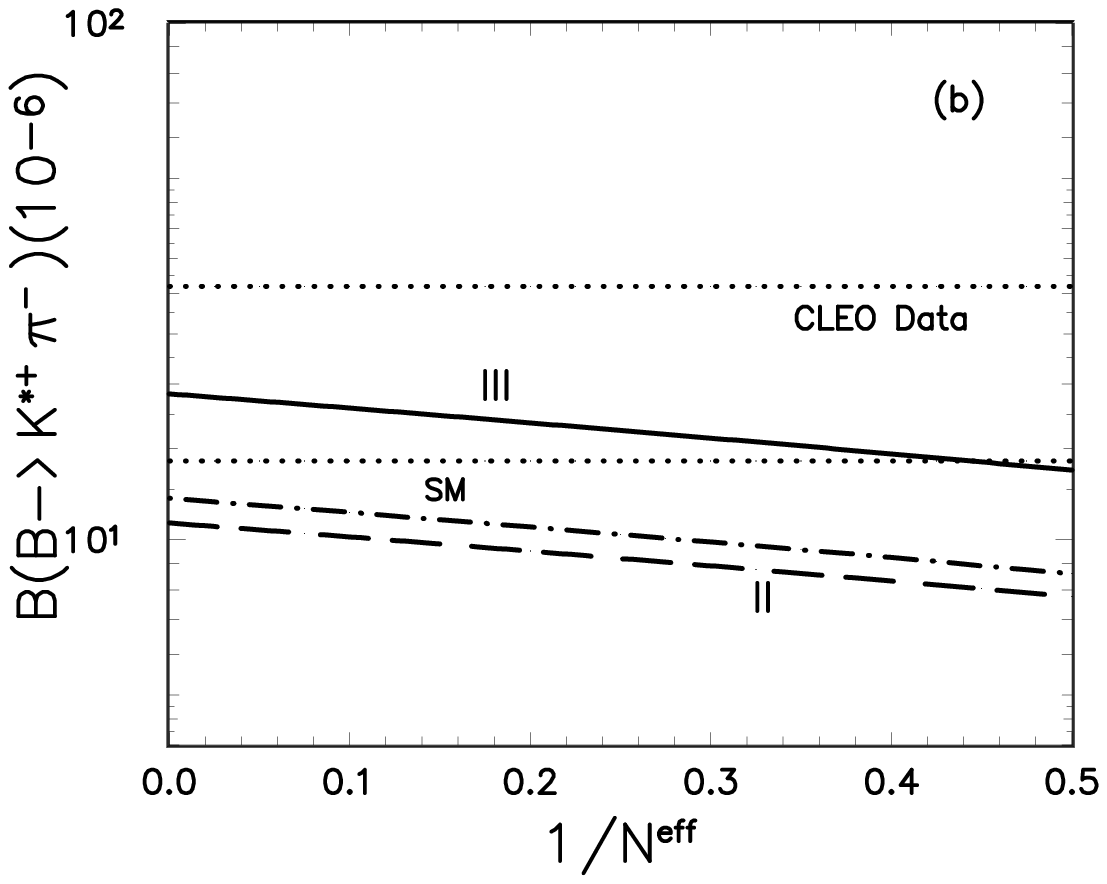}}
\vspace{-20pt}
\caption{Same as \fig{fig:fig11}, but for the decay $B \to K^{*+} \pi^-$.
The dots band corresponds to the CLEO data with $1\sigma$ error:
${\cal  B}(B^0 \to  K^{*+} \pi^-) = ( 22^{+8.9}_{-7.8})\times 10^{-6}$.}
\label{fig:fig13}
\end{minipage}
\end{figure}

\begin{figure}[t]%fig.14
\vspace{-40pt}
\begin{minipage}[]{0.96\textwidth}
\centerline{\epsfxsize=\textwidth \epsffile{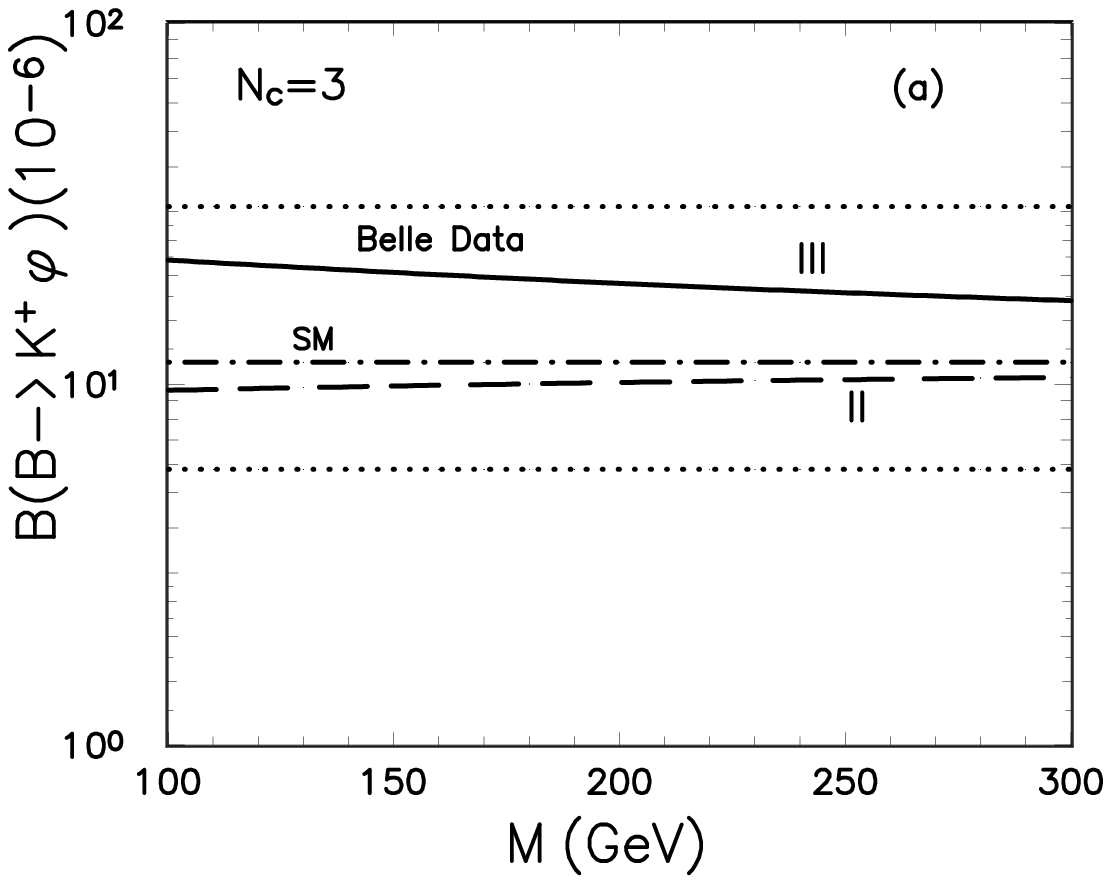}}
\vspace{-60pt}
\centerline{\epsfxsize=\textwidth \epsffile{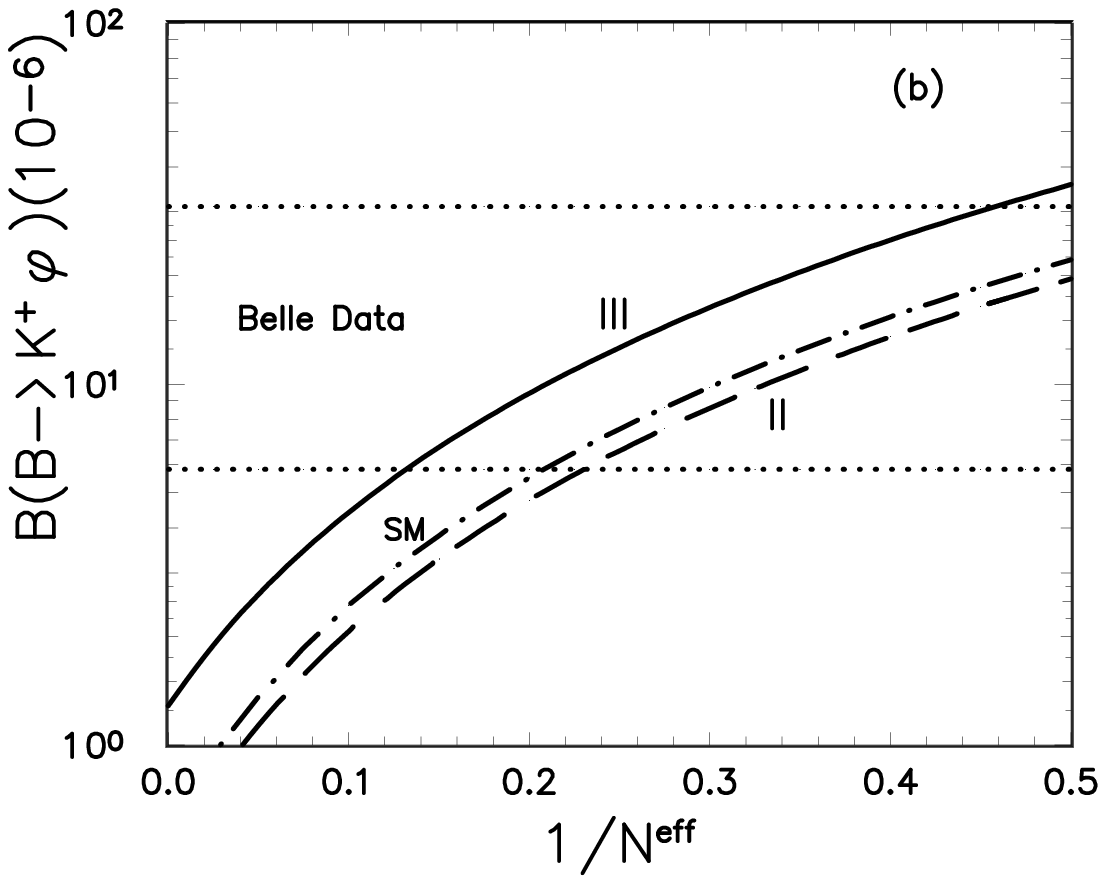}}
\vspace{-20pt}
\caption{Same as \fig{fig:fig11}, but for the decay $B \to K^{+} \phi$.
The dots band corresponds to the Belle data with $2\sigma$ errors:
${\cal  B}(B^+ \to  K^{+} \phi ) = ( 17.2^{+13.8}_{-11.4})\times 10^{-6}$.}
\label{fig:fig14}
\end{minipage}
\end{figure}

In models I and II, the new physics contributions are small in size:
from $-15\%$ to $20\%$ for most $B \to PV$  decay modes,
and have weak dependence on $\mhp$, $\tan{\beta}$ and $N^{eff}$, as shown in
table (\ref{bpvm2}), and illustrated in Figs.(\ref{fig:fig11}-\ref{fig:fig14})
where the long-dashed line shows the theoretical predictions in the model II
\footnote{Because the lines for the SM and model I are too close to be
separated clearly, we do not draw the line for model I in all four figures
for $B \to PV$ decays. }.
This feature remains unchanged within the considered range of $\tan{\beta}=1 - 50$.
When $\tan{\beta}$ becomes larger, the size of new physics corrections will become
even smaller.

In model III, however, the new physics contributions are significant, from $30\%$ to
$110\%$, and have also weak dependence on $\mhp$, $\theta $ and $N^{eff}$.
These new physics enhancements are very helpful to improve the agreement between
the theoretical predictions and the data, as shown
in Eqs.(\ref{eq:brpv26}-\ref{eq:brpv34}) and illustrated in
Figs.(\ref{fig:fig11}-\ref{fig:fig14}).

Figs.(\ref{fig:fig11}-\ref{fig:fig14}) show the mass and $N^{eff}$
dependence of the branching ratios for $B \to K^{*+} \eta, K^{*0}\eta,
K^{*+} \pi^-$ and $B \to K^+ \phi$ decays.
The dot-dashed line  is the SM prediction, while the long-dashed and
solid curve corresponds to the predictions in models II and III, respectively.
The theoretical uncertainties are not shown in these figures.
The dots band in Figs.(\ref{fig:fig11}-\ref{fig:fig13}) (Fig.\ref{fig:fig14})
corresponds to the CLEO data with $1\sigma$ ($2-\sigma$) error.

From Fig.(\ref{fig:fig11}), it can be seen that the CLEO measurement of
the ratio ${\cal B}(B^+ \to K^{*+} \eta )$ is  much larger than theoretical
predictions in the SM and 2HDM's. More positive contributions to this decay
mode are needed to improve the agreement between the data and theoretical
prediction. For $B \to K^{*0} \eta$ decay, the inclusion of new physics
contribution in the model III leads to a better agreement between data and
theory if we take into account still large theoretical
uncertainties. For $B \to K^{*+} \pi^-$ and $K^+ \phi$ decays, the
theoretical prediction becomes now consistent with the CLEO and Belle
measurements within $1\sigma$ error due to the large new physics
enhancement in  the model III.

For $B \to K^{*+}\etap$ and $K^{*0} \etap$ decays,  the new physics contributions
in model III are large in size, from $-77\%$ to
$200\%$, as shown in \tab{bpvm3}. But the theoretical predictions for these
two decay modes are $N^{eff}$-dependent and still far below the
current CLEO upper limits.

\section{$B \to V V$ decays}\label{sec:bvv}

Using the formulae as given in Sec. \ref{sec:bsw}, it is straightforward to
calculate the branching ratios of nineteen $B_{u,d} \to VV$ decays.
As an example, we show
here the calculation of the branching ratio for the Class-V decay
$B^- \to \rho^- \omega$ ($b \to d $ transition). We firstly find the explicit
expressions of the helicity amplitude
$H_\lambda=<\rho^-(\lambda) \omega (\lambda)|H_{eff}|B^->$,
and then compare this amplitude  with the standard form as defined in
Eq.(\ref{eq:hl}) to extract out the process dependent coefficients
$a, b$ and $c$
\beq
a&=& -\frac{1}{\sqrt{2}}\cdot \left [  f_1\, f_\omega M_\omega (M_B + M_\rho)
    A_1^{B \to \rho}(M_\omega^2)
    + f_2 \, f_\rho M_\rho (M_B+M_\omega)
    A_1^{B \to \omega}(M_\rho^2) \right ]~, \label{eq:hla}\\
b&=& f_1\cdot \frac{\sqrt 2f_\omega M_\omega^2M_\rho}{M_B+M_\rho}
    A_2^{B \to \rho}(M_\omega^2)
    + f_2 \cdot \frac{\sqrt 2f_\rho M_\rho^2M_\omega}{M_B+M_\omega}
    A_2^{B \to \omega}(M_\rho^2)~,  \label{eq:hlb}\\
c&=& f_1 \cdot \frac{\sqrt 2f_\omega M_\omega^2M_\rho}{M_B+M_\rho}
    V^{B \to \rho}(M_\omega^2)
    + f_2 \cdot \frac{\sqrt 2f_\rho M_\rho^2M_\omega}{M_B+M_\omega}
    V^{B \to \omega}(M_\rho^2)~,  \label{eq:hlc}
\eeq
with
\beq
f_1&=&  \frac{G_F}{\sqrt 2}\left[V_{ub}V_{ud}^* a_2 -V_{tb}V_{td}^*
\left ( 2a_3+a_4+2a_5+\frac{1}{2}a_7+\frac{1}{2}a_9-\frac{1}{2}a_{10}
\right ) \right]~, \label{eq:f1} \\
f_2&=&  \frac{G_F}{\sqrt 2}\left[V_{ub}V_{ud}^* a_1 -V_{tb}V_{td}^*
\left ( a_4+ a_{10} \right ) \right]~, \label{eq:f2}
\eeq
where the coefficients $a_{1,\ldots,10}$ have been defined in Eq.(\ref{eq:ai}),
the form factors and other input parameters can be found in Appendix.
With these coefficients $a, b$ and $c$, the branching ratio
${\cal  B}(B^- \to \rho^- \omega)$ can finally be written as
\beq
{\cal  B}(B^- \to \rho^- \omega) &=&  \tau_{B_u^-}
\frac{|p|}{8\pi M_B^2}\left ( |H_0|^2 + |H_{+1}|^2 + |H_{-1}|^2 \right
)~,
\eeq
where $|p|$ and $H_{1,0,-1}$ have been given in Eqs.(\ref{eq:pxy}) and
(\ref{eq:h01}).

In tables (\ref{bvvsm}-\ref{bvvm2}) we present the branching ratios
for the nineteen $B \to V V$ decay modes involving $b\to d$ and  $b \to s$
transitions in the SM and models I, II and III. Theoretical predictions are
made by using the same input parameters as those for the $B \to PP, PV$
decays in last two sections.

For $B \to VV$ decay modes, the differences induced by using whether BSW or
LQQSR form factors are around ten percent in the SM and models I, II and III.
We therefore show the numerical results obtained by using the BSW form
factors only for the cases of models I, II and III.
For all nineteen $B \to VV$ decays under study, the theoretical predictions
in the SM and 2HDM's are still under or far away from the current CLEO upper
limits, as can be seen from tables (\ref{bvvsm}-\ref{bvvm2}).

In models I and II, the new physics contributions to $B \to VV$ decays are small
in size:  from $-15\%$ to $\sim 10\%$ as shown in \tab{bvvm2}, and therefore
will be masked by other large theoretical uncertainties. This feature
remains unchanged within the considered range of $\tan{\beta}=1 - 50$.
When $\tan{\beta}$ becomes larger, the size of new physics corrections will become
smaller.

\begin{figure}[t]%fig.15
\vspace{-60pt}
\begin{minipage}[]{0.96\textwidth}
\centerline{\epsfxsize=\textwidth \epsffile{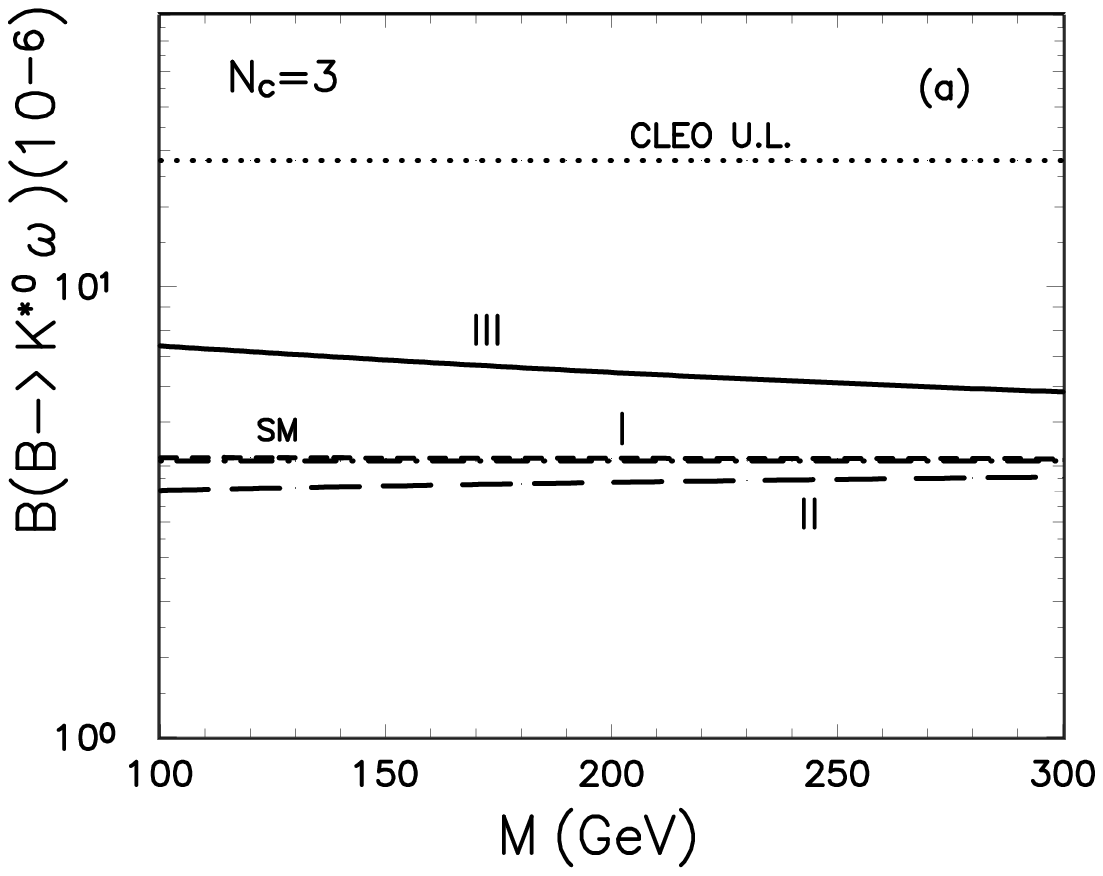}}
\vspace{-60pt}
\centerline{\epsfxsize=\textwidth \epsffile{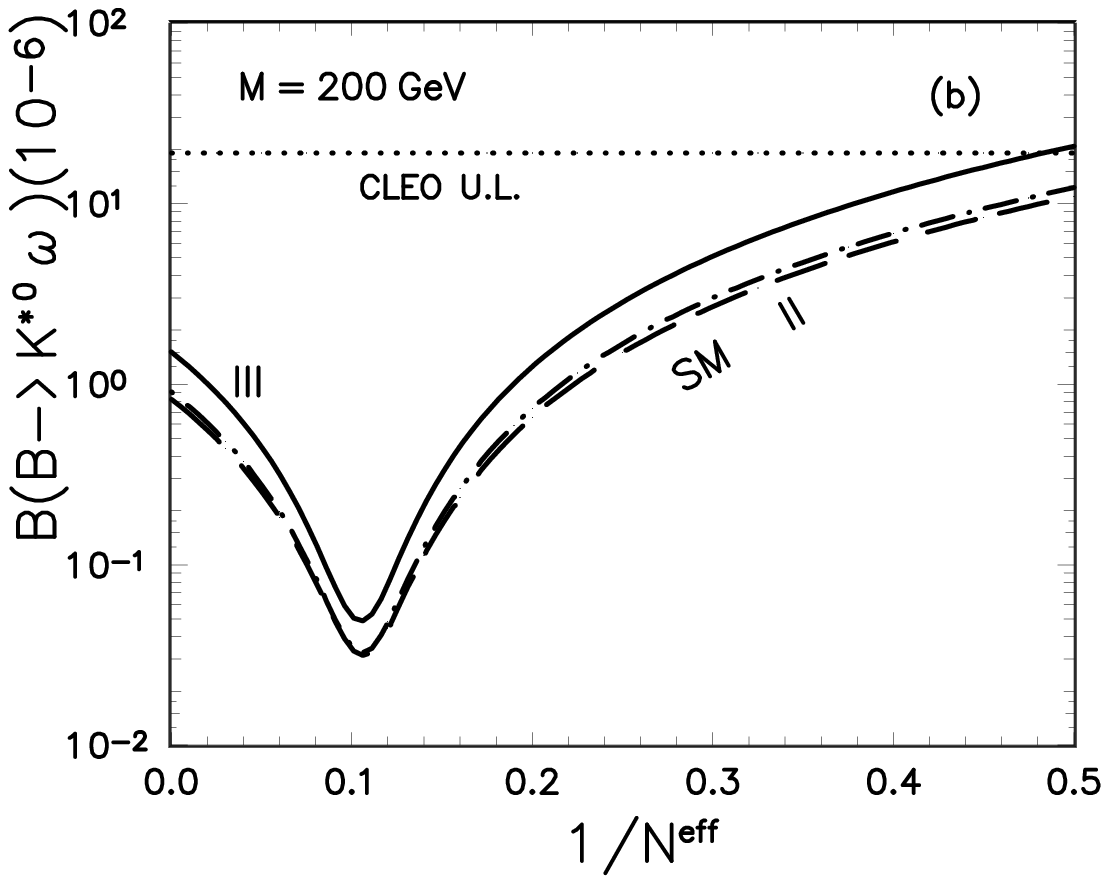}}
\vspace{-20pt}
\caption{
${\cal B}(B \to K^{*0} \omega)$ versus $\mhp$ and $N^{eff}$
in the SM and 2HDM's.  For (a) and (b), we set $N^{eff}=3$ and
$\mhp=200$GeV, respectively. The upper dots line shows the CLEO
upper limit: ${\cal B}(B \to K^{*0} \omega)\le 19\times 10^{-6}$.
The dot-dashed, long-dashed and solid curve correspond to the
theoretical prediction in the SM and models II and III, respectively.
The theoretical uncertainties are not shown here. }
\label{fig:fig15}
\end{minipage}
\end{figure}

\begin{figure}[t]%fig.16
\vspace{-40pt}
\begin{minipage}[]{0.96\textwidth}
\centerline{\epsfxsize=\textwidth \epsffile{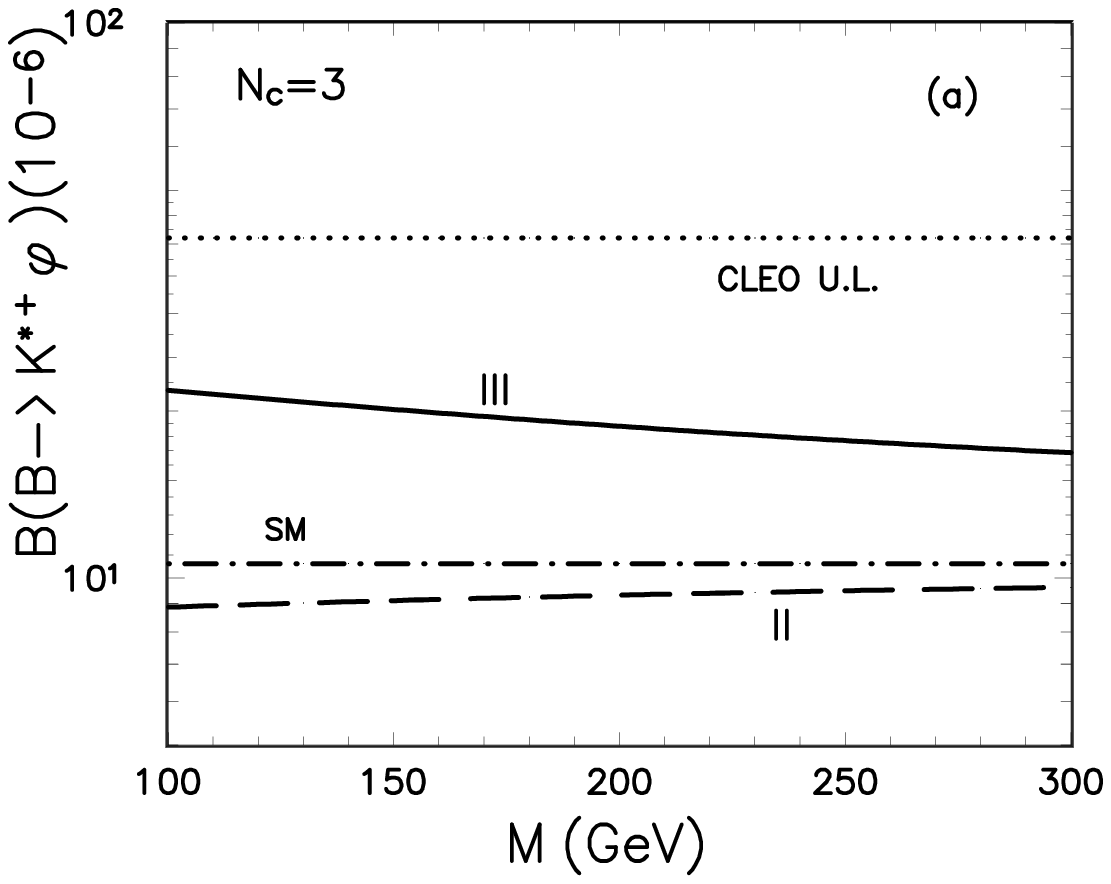}}
\vspace{-40pt}
\centerline{\epsfxsize=\textwidth \epsffile{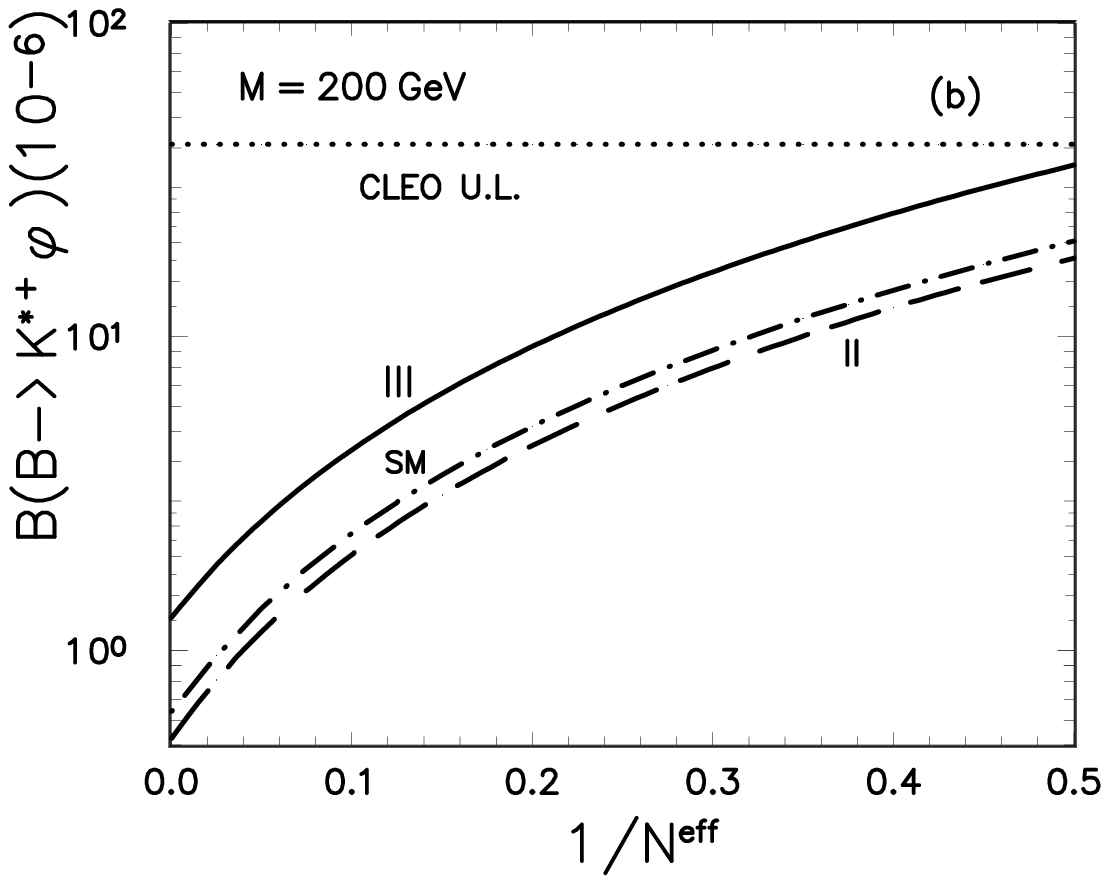}}
\vspace{-20pt}
\caption{Same as \fig{fig:fig15}, but for the decay $B \to K^{*+} \phi$.
The upper dots line shows the CLEO
upper limit: ${\cal B}(B \to K^{*+} \phi)\le 41\times 10^{-6}$.}
\label{fig:fig16}
\end{minipage}
\end{figure}

\begin{figure}[t]%fig.17
\vspace{-40pt}
\begin{minipage}[]{0.96\textwidth}
\centerline{\epsfxsize=\textwidth \epsffile{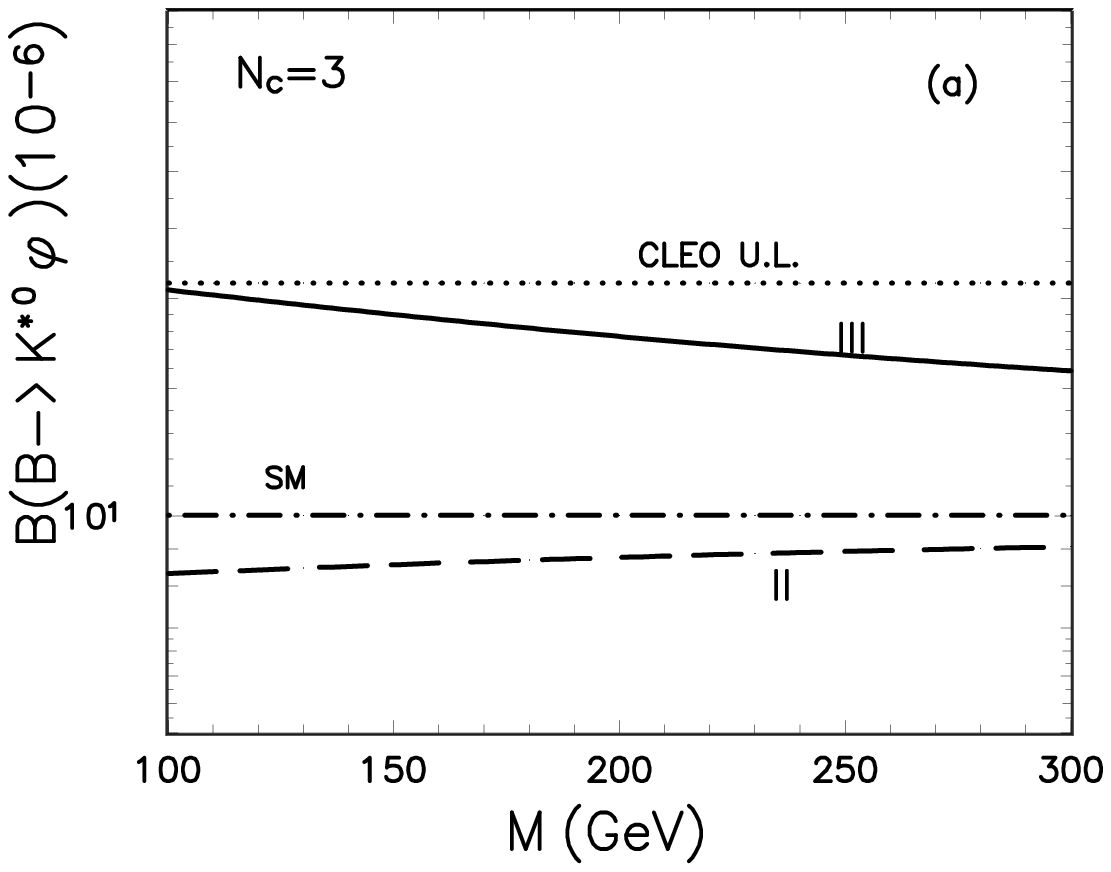}}
\vspace{-40pt}
\centerline{\epsfxsize=\textwidth \epsffile{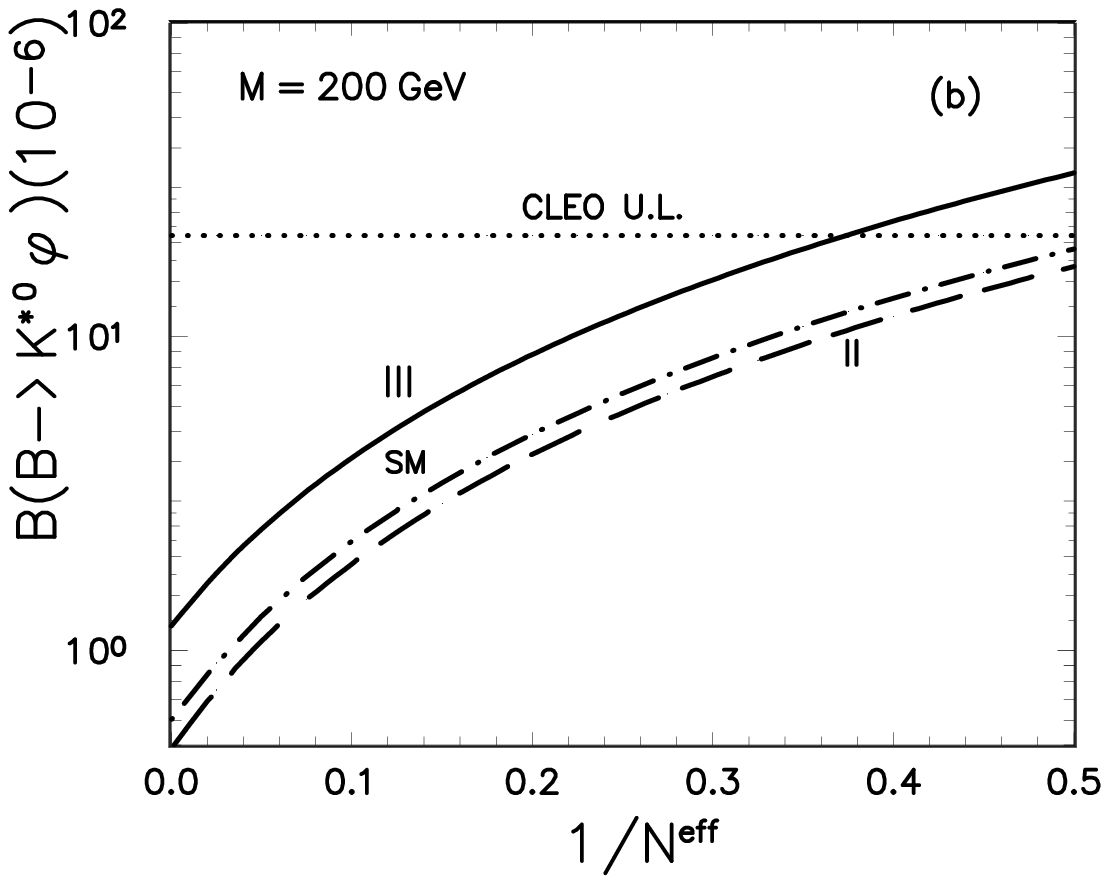}}
\vspace{-20pt}
\caption{Same as \fig{fig:fig15}, but for the decay $B \to K^{*0} \phi$.
The upper dots line shows the CLEO
upper limit: ${\cal B}(B \to K^{*0} \phi)\le 21\times 10^{-6}$. }
\label{fig:fig17}
\end{minipage}
\end{figure}

In model III, however, the new physics contributions to different channels
are varying greatly: from $-11\%$ to $\sim 110\%$, assuming $\mhp=200$GeV,
$N^{eff}=2-\infty$ and $\theta=0^\circ-30^\circ$.
For decay modes $B \to K^{*0}\omega,
K^{*+}\phi, K^{*0}\phi$, for example, the new physics enhancements are significant:
$\sim (60 -110)\%$. And hence the theoretical predictions in model III are
close to or slightly surpass the current CLEO upper limits, as illustrated in
Figs.(\ref{fig:fig15}-\ref{fig:fig17}) where the upper dots line shows the
corresponding CLEO upper limits at $90\% C.L.$.  These decay modes
will be observed soon.

Figures (\ref{fig:fig15}-\ref{fig:fig17}) show the mass and $N^{eff}$ dependence
of the ratios ${\cal B}(B \to K^{*0} \omega )$ and ${\cal B}(B \to K^{*} \phi )$.
The dot-dashed line is the SM prediction, while the long-dashed and solid curve
corresponds to the predictions in models II and III, respectively.
As the Class-V decays, these three decays show strong $N^{eff}$ dependence as
illustrated in Figs.(\ref{fig:fig15}-\ref{fig:fig17}).

\section{Summary and discussions}

In this paper, we calculated the branching ratios of two-body charmless hadronic
B meson decays $B_{u,d} \to PP, PV, VV$ in the SM and the general two-Higgs-doublet
models by employing the NLO effective Hamiltonian with the generalized
factorization.

In Sec. \ref{sec:2hdm}, with the help of previous works
\cite{atwood97,atwood96,aliev99,chao99,xiaonc}, we gave a brief review about the
2HDM's and studied corresponding experimental constraints on models I, II an III.
In Sec. \ref{sec:heff}, we evaluated analytically all new gluonic and
electroweak charged-Higgs
penguin diagrams and found the effective Wilson coefficients $C_i^{eff}$
in the SM and models I, II and III.
In Sec. \ref{sec:bsw}, we presented the formulae needed to calculate the
branching ratios ${\cal B} (B \to PP, PV, VV)$.

In sections \ref{sec:bpp}-\ref{sec:bvv}, we calculated the branching ratios
for seventy six $B  \to PP, PV, VV$ decays in the SM and models I, II and III,
presented the numerical results in tables (\ref{bppsm}-\ref{bvvm2}) and
displayed the $\mhp$ and $N^{eff}$-dependence for several phenomenologically
interesting decay modes in Figs.(\ref{fig:fig2}-\ref{fig:fig17}).

From the numerical results, we find following general features about the new
physics effects on the exclusive charmless hadronic $B \to PP, PV, VV$ decays
studied in this paper:

\begin{itemize}

\item[1]
The SM predictions for the B meson decay rates presented in this
paper agree well with those appeared in Refs.\cite{ali9804,chen99}.

\item[2]
The new physics effects due to new gluonic penguin diagrams strongly dominate
over those from the $\gamma-$ and $Z^0-$ penguin diagrams induced by exchanges of
charged-Higgs bosons appeared in models I, II and III.

\item[3]
For models I and II, the new physics contributions
to the decay rates ${\cal B}(B\to h_1 h_2)$ are always small in size:
from $-15\%$ to $20\%$ for most  decay modes.
So small contributions will be masked by other still large theoretical
uncertainties.

\item[4]
For model III, however, the new
physics enhancements to penguin-dominated decay modes can be significant,
$\sim (30 -200)\%$, and therefore can be measured in high precision B experiments.
In general, the new physics contributions in model III are large (small) for penguin-
dominated (tree-dominated) B meson decay channels.

\item[5]
The uncertainties of the theoretical predictions for the branching ratios of
$B \to h_1 h_2$ decays induced by varying $k^2$, $\eta$, $\theta$,
$\tan{\beta}$, and $\mhp$ are varying from $\sim 10\%$ to $\sim 50\%$
within the range of $k^2 = m_b^2/2 \pm 2 GeV^2$, $\eta=0.34 \pm 0.08$,
$\theta=0^0 - 30^0$, $\tan{\beta}=1-50$, and $\mhp=200\pm 100$ GeV.
The dependence of decay rates on whether using the BSW or LQSSR form
factors are weak, $\sim 10\%$. The
$N^{eff}-$dependence of branching ratios, however, are varying greatly for
different decay modes.

\item[6]
For phenomenologically interesting $B \to K \etap$ decay modes, the
new physics enhancements are significant in model III:  $\sim (35 -70)\%$,
and have a moderate dependence on $\mhp$ and $N^{eff}$. The theoretical
predictions for ${\cal B}(B \to K \etap )$ therefore  turn to be
consistent with the CLEO data in model III, as illustrated in
Figs.(\ref{fig:fig8},\ref{fig:fig10}).
For other $B \to PP$ decays, the theoretical predictions are still
consistent with the measurements if one takes into account still large
theoretical and experimental uncertainties.

\item[7]
For penguin-dominated $B \to PV$ decays, the new physics contributions in
model III are significant, from $30\%$ to $60\%$, and have a weak or moderate
dependence on $\mhp$, $\theta $ and $N^{eff}$, as illustrated in the tables
(\ref{bpvm3}-\ref{bpvm3}) and figures (\ref{fig:fig11}-\ref{fig:fig14}).
The CLEO measurements of ${\cal B}(B \to K^{*+} \eta, K^{*0} \eta )$ are  much
larger than theoretical predictions in the SM and hence large new physics
enhancements in model III are indeed helpful to lead to or improve the
agreement between the data and theoretical predictions.

\item[8]
In model III, the new physics contributions to different $B \to VV$ decay
modes are varying greatly: from $-11\%$ to $\sim 110\%$.
For decay modes $B \to K^{*0}\omega, K^{*+}\phi, K^{*0}\phi$,
for example, the new physics enhancements are significant:
$\sim (60 -110)\%$, and hence the theoretical predictions in model III are
close to or slightly surpass the current CLEO upper limits. These
decay modes  will be observed soon.

\end{itemize}

\section*{ACKNOWLEDGMENTS}

The authors are very grateful to D.S. Du, C.D. L\"u, Y.D. Yang and M.Z.Yang for helpful
discussions. C.S. Li and K.T. Chao  acknowledge the support by the National Natural
Science Foundation of China,  the State Commission of Science
and technology of China,  and the Doctoral Program Foundation of Institution
of Higher Education. Z.J. Xiao acknowledges the support by the National
Natural Science Foundation of China under Grant No.19575015 and
10075013, the Excellent Young Teachers Program of Ministry of Education,
P.R.China,  and the Natural Science Foundation of Henan Province under
Grant No. 994050500.

\vspace{1cm}
%\newpage
\begin{appendix}
\section{Input parameters and form factors} \label{app:a}

In this appendix we present relevant input parameters. We use the same set
of input parameters for the quark masses, decay constants, Wolfenstein parameters
and form factors as Ref.\cite{ali9804}.

\begin{itemize}

\item
Input parameters of electroweak and strong coupling constant, gauge boson masses,
B meson masses, light meson masses,
$\cdots$,  are as follows (all masses in unit of GeV )\cite{ali9804,pdg98}
\beq
\alpha_{em}&=&1/128, \;  \alpha_s(M_Z)=0.118,\;  \sin^2\theta_W=0.23,\;
G_F=1.16639\times 10^{-5} (GeV)^{-2}, \non
M_Z&=&91.187, \;   M_W=80.41,\;
m_{B_d^0}=m_{B_u^\pm}=5.279,\;   m_{\pi^\pm}=0.140,\;\non
m_{\pi^0}&=&0.135,\;   m_{\eta}=0.547,\; m_{\etap}=0.958,\;
m_{\rho}=0.770,\;  m_{\omega}=0.782,\non
m_{\phi}&=&1.019,\; m_{K^\pm}=0.494,\;  m_{K^0}=0.498,\;  m_{K^{*\pm}}=0.892,\;
m_{K^{*0}}=0.896,\non
\tau(B_u^\pm)&=& 1.64 ps,\; \tau(B_d^0)= 1.56 ps, \label{masses}
\eeq

\item
For the elements of CKM matrix, we use Wolfenstein parametrization, and fix
the parameters $A, \lambda, \rho$ to their central values,
$A=0.81,\; \lambda=0.2205, \; \rho=0.12$ and varying $\eta$ in the range of
$\eta=0.34 \pm 0.08$.

\item
We first treat the internal quark masses in the loops in connection with
the function $G(m)$ as constituent masses,
\beq
m_b=4.88 GeV, \; m_c=1.5 GeV,\; m_s=0.5 GeV, \; m_u=m_d=0.2 GeV.
\label{con-mass}
\eeq
Secondly, we will use the current quark masses for $m_i$ ($i=u,d,s,c,b$)
which appear through the equation of motion when working out the hadronic
matrix elements. For $\mu=2.5 GeV$, one finds\cite{ali9804}
\beq
m_b=4.88 GeV, \; m_c=1.5 GeV, m_s=0.122 GeV, \; m_d=7.6 MeV,\; m_u=4.2 MeV.
\label{cur-mass}
\eeq
For the mass of heavy top quark we also use $m_t=\overline{m_t}(m_t)=168 GeV$.

\item
For the decay constants of light mesons, the following values will be used in the
numerical calculations (in the units of MeV):
\beq
&&f_{\pi}=133,\; f_{K}=158,\; f_{K^*}=214, \; f_{\rho}=210, \;
f_{\omega}=195, \; f_{\phi}=233, \; \non
&&f^u_{\eta}=f^d_{\eta}=78,\; f^u_{\etap}=f^d_{\etap}=68,\;
f^c_{\eta}=-0.9,\; f^c_{\etap}=-0.23,\non
&&f^s_{\eta}=-113,\; f^s_{\etap}=141.
\label{fpis}
\eeq
where $f^u_{\etapp}$ and $f^s_{\etapp}$ have been defined in the two-angle-mixing
formalism with $\theta_0=-9.1^\circ$ and $\theta_8 =-22.2^\circ$\cite{fk97}
For more details about the mixings between $\eta$ and $\etap$,
one can see \cite{fk97,ali98}.

\item
The form factors at the zero momentum transfer in the BSW model
\cite{bsw87} have been collected in Table 2 of Ref.\cite{ali9804}. For the
convenience of the reader we list them here:
\beq
&&F_0^{B\to \pi}(0)=0.33,\; F_0^{B\to K}(0)=0.38,\;
F_0^{B\to \eta}(0)=0.145,\; F_0^{B\to \etap}(0)=0.135,\non
&&A_{0,1,2}^{B\to \rho}(0)=A_{0,1,2}^{B\to \omega}(0)=0.28,\;
A_{0}^{B\to K^*}(0)=0.32, \;A_{1,2}^{B\to K^*}(0)=0.33, \non
&&V^{B\to \rho}(0)=V^{B\to \omega}(0)=0.33, \; V^{B\to K^*}(0)=0.37,
\label{eq:bsw-f}
\eeq

\item
In the LQQSR approach, the form factors at zero momentum transfer being used in our
numerical calculations are,
\beq
&&F_0^{B\to \pi}(0)=0.36,\; F_0^{B\to K}(0)=0.41,\;
F_0^{B\to \eta}(0)=0.16,\; F_0^{B\to \etap}(0)=0.145,\non
&&\{ A_0,A_1,A_2,V \}( B\to \rho ) =\{0.30,0.27,0.26,0.35 \},\non
&&\{ A_0,A_1,A_2,V \}( B\to K^* )  =\{0.39,0.35,0.34,0.48 \},\non
&&\{ A_0,A_1,A_2,V \}( B\to \omega)=\{0.30,0.27,0.26,0.35 \}.
\label{eq:lqqsr-f}
\eeq

\item
The form factors $F_{0,1}(k^2),$ $A_{0,1,2}(k^2)$ and $V(k^2)$ were defined
in Ref.\cite{bsw87} as
\beq
F_0(k^2)&=& \frac{F_0(0)}{1-k^2/m^2(0^+)},\ \
F_1(k^2)=   \frac{F_1(0)}{1-k^2/m^2(1^-)}, \non
A_0(k^2)&=& \frac{A_0(0)}{1-k^2/m^2(0^-)}, \ \
A_1(k^2)=   \frac{A_1(0)}{1-k^2/m^2(1^+)},  \non
A_2(k^2)&=& \frac{A_2(0)}{1-k^2/m^2(1^+)}, \ \
V(k^2) =    \frac{V(0)}{1-k^2/m^2(1^-)}.
\eeq

\item
The pole masses being used to evaluate the $k^2$-dependence of form factors
are
\beq
\{ m(0^-),m(1^-),m(1^+),m(0^+) \}&=& \{ 5.2789, 5.3248,5.37,5.73 \},
\eeq
for $\bar{u}b$ and $ \bar{d}b$ currents, and
\beq
\{ m(0^-),m(1^-),m(1^+),m(0^+)\}&=& \{5.3693, 5.41,5.82,5.89\},
\eeq
for $\bar{s}b $ currents.
\end{itemize}

\end{appendix}

%5702

%\listoffigures

\newpage
\begin{table}[t]
\begin{center}
\caption{Wilson coefficients $C_i(\mu)$ and $C_g^{eff}(\mu)$ in the SM
and models I, II and III at the scale $\mu=2.5$ GeV, with $\mhp=200$
GeV,  $\tan{\beta}=2$ and $\theta=0^0, 30^0$.}
\label{cimu25}
\vspace{0.2cm}
\begin{tabular}{l|c|c|c|c|c}  \hline
& SM  & Model I   & Model II     & Model III: $\ \ \theta=0^0$  & Model III: $\ \ \theta=30^0$  \\ \hline
$C_1      $&     $  1.1245$ &     $  1.1245 $ &  $  1.1245$  &  $  1.1245$&  $  1.1245$\\
$C_2      $&     $ -0.2662$ &     $ -0.2662 $ &  $ -0.2662$  &  $ -0.2662$&  $ -0.2662$\\
$C_3      $&     $  0.0186$ &     $  0.0187 $ &  $  0.0187$  &  $  0.0186$&  $  0.0186$\\
$C_4      $&     $ -0.0458$ &     $ -0.0458 $ &  $ -0.0458$  &  $ -0.0458$&  $ -0.0458$\\
$C_5      $&     $  0.0113$ &     $  0.0113 $ &  $  0.0113$  &  $  0.0113$&  $  0.0113$\\
$C_6      $&     $ -0.0587$ &     $ -0.0585 $ &  $ -0.0585$  &  $ -0.0587$&  $ -0.0587$\\
$C_7      $&     $  0.0006$ &     $  0.0006 $ &  $  0.0006$  &  $  0.0006$&  $  0.0006$\\
$C_8      $&     $  0.0007$ &     $  0.0007 $ &  $  0.0007$  &  $  0.0007$&  $  0.0007$\\
$C_9      $&     $ -0.0095$ &     $ -0.0099 $ &  $ -0.0099$  &  $ -0.0096$&  $ -0.0096$\\
$C_{10}   $&     $  0.0026$ &     $  0.0027 $ &  $  0.0027$  &  $  0.0026$&  $  0.0026$\\
$C_g^{eff}$&     $ -0.1527$ &     $ -0.1321 $ &  $ -0.2487$  &  $  0.3364$&  $  0.2708+0.2448i$\\
\hline
\end{tabular}\end{center}
\end{table}

\begin{table}[t]
\begin{center}
\caption{Numerical values of $a_i$ for the transitions $b \to d$ [$\bar{b}
\to \bar{d}$ ].  The first, second and third entries for $a_3, \ldots, a_{10}$
refer to the values of $a_i$ in the SM and models II and III, respectively.
All entries for $a_3, \ldots, a_{10}$ should be multiplied with $10^{-4}$. }
\label{ai:bd}
\vspace{0.2cm}
\begin{tabular}{|l|c|c|c|} \hline
        & $N^{eff}=2$                 & $N^{eff}=3$ & $N^{eff}=\infty$  \\ \hline
 $a_1$  &$0.995 \;[0.995]$          &$1.061\;[1.061]$           & $1.192  \;[1.192]$     \\
 $a_2$  &$0.201 \;[0.201]$          &$0.003\;[0.003]$           & $-0.395 \;[-0.395]$   \\\hline
 $a_3$  &$-16-7i\;[-25-23i]$        &$77    \;[77]$             & $261+13i\;[280+47i]$ \\
        &$-10-7i\;[-19-23i]$        &$77    \;[77]$             & $252+13i\;[271+47i]$ \\
        &$-40-7i\;[-49-23i]$        &$77    \;[77]$             & $310+13i\;[329+47i]$        \\ \hline
 $a_4$  &$-423-33i\;[-470-117i]$    &$-467-35i\;[-517-125i]$    & $-554-39i\;[-610-141i]$ \\
        &$-398-33i\;[-445-117i]$    &$-440-35i\;[-490-125i]$    & $-524-39i\;[-581-141i]$\\
        &$-546-33i\;[-592-117i]$    &$-597-35i\;[-648-125i]$    & $-701-39i\;[-757-141i]$        \\ \hline
 $a_5$  &$-193-7i \;[-202-23i]$     &$-71    \;[-71]$           & $171+ 13i\;[190 + 47i]$ \\
        &$-187-7i \;[-196-24i]$     &$-71    \;[-71]$           & $161+ 13i\;[180 + 47i]$ \\
        &$-217-7i \;[-226-23i]$     &$-71    \;[-71]$           & $220+ 13i\;[239 + 47i]$        \\ \hline
 $a_6$  &$-642-33i\;[-689-117i]$    &$-671-35i\;[-721-125i]$    & $-728-39i\;[-784-141i]$ \\
        &$-616-33i\;[-663-117i]$    &$-642-35i\;[-693-125i]$    & $-696-39i\;[-752-141i]$  \\
        &$-764-33i\;[-811-117i]$    &$-801-35i\;[-851-125i]$    & $-874-39i\;[-931-141i]$        \\  \hline
 $a_7$  &$8.1-0.9i\;[7.7-1.7i]$     &$6.8-0.9i\;[6.4-1.7i]$     & $4.3-0.9i\;[3.9-1.7i]$  \\
        &$9.3-0.9i \;[8.9-1.7i]$    &$8.0-0.9i\;[7.5-1.7i]$     & $5.3-0.9i\;[4.9-1.7i]$   \\
        &$8.3-0.9i \;[7.9-1.7i]$    &$7.0-0.9i\;[6.6-1.7i]$     & $4.5-0.9i\;[4.1-1.7i]$        \\ \hline
 $a_8$  &$9.7-0.5i  \;[9.5-0.8i]$   &$9.0-0.3i  \;[8.8-0.6i]$   & $7.5     \;[7.5]$ \\
        &$11-0.5i   \;[11-0.8i]$    &$9.9-0.3i   \;[9.7-0.6i]$  & $8.1    \;[8.1]$  \\
        &$9.9-0.5i   \;[9.7-0.8i]$  &$9.1-0.3i  \;[9.0-0.6i]$   & $7.6    \;[7.6]$         \\ \hline
 $a_9$  &$-84-0.9i\;[-84-1.7i]$     &$-90-0.9i\;[-90-1.7i]$     & $-102-0.9i\;[-102-1.7i]$  \\
        &$-87-0.9i \;[-87-1.7i]$    &$-93-0.9i\;[-94-1.7i]$     & $-106-0.9i\;[-106-1.7i]$ \\
        &$-84-0.9i \;[-85-1.7i]$    &$-90-0.9i\;[-91-1.7i]$     & $-103-0.9i\;[-103-1.7i]$        \\ \hline
$a_{10}$&$-14-0.5i\;[-15-0.8i]$     &$2.6-0.3i\;[2.5-0.6i]$     & $37     \;[37]$   \\
        &$-15-0.5i  \;[-15-0.8i]$   &$2.8-0.3i\;[2.7-0.6i]$     & $38       \;[38]$   \\
        &$-15-0.5i  \;[-15-0.8i]$   &$2.7-0.3i\;[2.5-0.6i]$     & $37       \;[37]$   \\
\hline
\end{tabular}\end{center}
\end{table}

\begin{table}[t]
\begin{center}
\caption{Same as \tab{ai:bd} but for $b \to s$ [$\bar{b} \to \bar{s}$ ]
transitions. }
\label{ai:bs}
\vspace{0.2cm}
\begin{tabular}{l|c|c|c} \hline
       & $N^{eff}=2$              & $N^{eff}=3$ & $N^{eff}=\infty$  \\ \hline
 $a_1$ &$0.995  \;[0.995]$          &$1.061\;[1.061]$            &$1.192  \;[1.192]$       \\
 $a_2$ &$0.201  \;[0.201]$          &$0.026\;[0.026]$            &$-0.395 \;[-0.395]$ \\ \hline
 $a_3$ &$-21-14i\;[-19-14i]$        &$77   \;[77]$               &$272+29i\;[269+29i]$   \\
       &$-15-14i\;[-14-14i]$        &$77   \;[77]$               &$262+29i\;[260+29i]$    \\
       &$-45-14i\;[-44-14i]$        &$77   \;[77]$               &$320+29i\;[318+29i]$    \\ \hline
 $a_4$ &$-449-72i\;[-442-72i]$      &$-494-77i\;[-487-77i]$      &$-585-86i\;[-576-86i]$\\
       &$-424-72i\;[-417-72i]$      &$-468-77i\;[-460-77i]$      &$-555-87i\;[-547-87i]$\\
       &$-571-72i\;[-564-72i]$      &$-625-77i\;[-617-77i]$      &$-732-86i\;[-723-86i]$   \\ \hline
 $a_5$ &$-198-14i\;[-196-14i]$      &$-71     \;[-71]$           &$181+ 29i\;[179 + 29i]$  \\
       &$-192-14i\;[-191-14i]$      &$-71     \;[-71]$           &$172+ 29i\;[169 + 29i]$   \\
       &$-222-14i\;[-221-14i]$      &$-71     \;[-71]$           &$230+ 29i\;[228 + 29i]$\\\hline
 $a_6$ &$-667-72i \;[-660-72i]$     &$-698-77i\;[-691-77i]$      &$-758-86i\;[-750-86i]$  \\
       &$-641-72i \;[-635-72i]$     &$-670-77i\;[-663-77i]$      &$-727-87i\;[-719-87i]$ \\
       &$-790-72i \;[-783-72i]$     &$-828-77i\;[-821-77i]$      &$-905-86i\;[-897-87i]$ \\ \hline
 $a_7$ &$7.9-1.3i \;[7.9-1.3i]$     &$6.6-1.3i\;[6.7-1.3i]$      &$4.1-1.3i\;[4.2-1.3i]$   \\
       &$9.1-1.3i\;[9.2-1.3i]$      &$7.7-1.3i \;[7.8-1.3i]$     &$5.0-1.3i \;[5.1-1.3i]$  \\
       &$8.1-1.3i\;[8.2-1.3i]$      &$6.8-1.3i \;[6.9-1.3i]$     &$4.3-1.3i \;[4.3-1.3i]$  \\\hline
 $a_8$ &$9.6-0.6i  \;[9.6-0.6i]$    &$8.9-0.4i  \;[8.9-0.4i]$    &$7.5       \;[7.5]$    \\
       &$10.6-0.6i \;[10.6-0.6i]$   &$9.8-0.4i  \;[9.8-0.4i]$    &$8.1      \;[8.1]$   \\
       &$9.8-0.6i   \;[9.8-0.6i]$   &$9.1-0.4i  \;[9.1-0.4i]$    &$7.6      \;[7.6]$   \\\hline
 $a_9$ &$-84-1.3i\;[-84-1.3i]$      &$-90-1.3i\;[-90-1.3i]$      &$-102-1.3i\;[-102-1.3i]$     \\
       &$-87-1.3i \;[-87-1.3i]$     &$-94-1.3i \;[-94-1.3i]$     &$-106-1.3i \;[-106-1.3i]$    \\
       &$-85-1.3i \;[-84-1.3i]$     &$-91-1.3i \;[-91-1.3i]$     &$-103-1.3i \;[-103-1.3i]$    \\\hline
$a_{10}$&$-15-0.6i\;[-14-0.6i]$     &$ 2.6-0.4i \;[2.6-0.4i]$   &$37\;[37]$ \\
        &$-15-0.6i  \;[-15-0.6i]$   &$ 2.8-0.4i \;[2.8-0.4i]$   &$38\;[38]$  \\
        &$-15-0.6i  \;[-15-0.6i]$   &$ 2.6-0.4i \;[2.6-0.4i]$   &$37\;[37]$  \\
\hline
\end{tabular}\end{center}
\end{table}

\begin{table}[t]
\begin{center}
\caption{${\cal B}(B\to PP)$ (in units of $10^{-6}$) in the SM using the
BSW [LQSSR] form factors, with $k^2=m_b^2/2$ and $N^{eff}=2,\; 3,\; \infty$.
The last column
shows the CLEO measurements and upper limits at $90\% C.L.$ [22-25]. }
\label{bppsm}
\vspace{0.2cm}
\begin{tabular} {l|l|c|c|c|l}  \hline
Channel & Class & $N^{eff}=2$& $N^{eff}=3$ & $N^{eff}=\infty$ &Data  \\ \hline
$B^0 \to \pi^+ \pi^-$        & I  & $ 9.10\;[10.8]$ &$10.3 \;[12.3]$&$13.0 \;[15.5]$& $ 4.3^{+1.6}_{-1.5} \pm 0.5$ \\
$B^0 \to \pi^0 \pi^0$        & II & $ 0.28\;[0.33]$ &$0.15 \;[0.18]$&$0.92 \;[1.09]$& $<9.3$ \\
$ B^+ \to \pi^+ \pi^0$       & III& $ 6.41\;[7.62]$ &$5.06 \;[6.02]$&$2.85 \;[3.39]$& $ < 12.7 $  \\ \hline
$ B^0 \to \eta \eta$         & II & $ 0.14\;[0.17]$ &$0.10 \;[0.13]$&$0.29 \;[0.36]$& $<18$  \\
$ B^0 \to \eta \eta^\prime$  & II & $ 0.14\;[0.17]$ &$0.08 \;[0.09]$&$0.38 \;[0.45]$& $<27$ \\
$ B^0 \to \eta' \eta^\prime$ & II & $ 0.04\;[0.05]$ &$0.01 \;[0.01]$&$0.13 \;[0.15]$& $< 47 $  \\ \hline
$ B^+ \to \pi^+ \eta$        & III& $ 3.51\;[4.25]$ &$2.78 \;[3.37]$&$1.75 \;[2.13]$& $<5.7$ \\
$ B^+ \to \pi^+ \eta^\prime$ & III& $ 2.49\;[2.90$ &$1.88 \;[2.17]$&$1.02 \;[1.17]$& $<12$ \\
$ B^0 \to \pi^0 \eta$        &  V & $ 0.26\;[0.31]$ &$0.29 \;[0.35]$&$0.39 \;[0.47]$& $<2.9 $  \\
$ B^0 \to \pi^0 \eta^\prime$ &  V & $ 0.06\;[0.07]$ &$0.08 \;[0.09]$&$0.14 \;[0.17]$& $<5.7$ \\ \hline
$B^+ \to K^+ \pi^0$          & IV & $ 12.0\;[14.3]$ &$13.5 \;[16.0]$&$16.7 \;[19.8]$& $ 11.6^{+3.0 +1.4}_{-2.7 -1.3}$  \\
$B^0 \to K^+ \pi^-$          & IV & $ 17.8\;[21.2]$ &$19.8 \;[23.5]$&$24.0 \;[28.5]$& $17.2^{+2.5}_{-2.4} \pm 1.2$ \\
$B^+ \to K^0 \pi^+$          & IV & $ 19.9\;[23.7]$ &$23.2 \;[27.7]$&$30.6 \;[36.4]$& $18.2^{+4.6}_{-4.0}\pm 1.6 $\\
$B^0 \to K^0 \pi^0$          & IV & $ 7.27\;[8.68]$ &$8.31 \;[9.92]$&$10.7 \;[12.7]$& $ 14.6 ^{+5.9\; +2.4}_{-5.1\; -3.3}$ \\ \hline
$ B^+ \to  K^+ \eta$         & IV & $ 3.91\;[4.37]$ &$4.56 \;[5.10]$&$6.07 \;[6.80]$& $ <6.9$  \\
$ B^+ \to  K^+ \eta^\prime$  & IV & $ 22.6\;[26.2]$ &$28.5 \;[33.1]$&$42.4 \;[49.2]$& $ 80^{+10}_{-9}\pm 7 $ \\
$B^0 \to K^0 \eta$           & IV & $ 3.22\;[3.57]$ &$3.63 \;[4.02]$&$4.58 \;[5.07]$& $ <9.3 $ \\
$B^0 \to K^0 \eta^\prime$    & IV & $ 21.9\;[25.5]$ &$28.2 \;[32.7]$&$43.0 \;[49.9]$& $ 89^{+18}_{-16}\pm 9 $  \\ \hline
$ B^+ \to K^+ \bar K^0$      & IV & $ 1.16\;[1.35]$ &$1.35 \;[1.58]$&$1.78 \;[2.07]$& $<5.1 $ \\
$ B^0 \to K^0\bar K^0$       & IV & $ 1.10\;[1.28]$ &$1.28 \;[1.49]$&$1.68 \;[1.96]$& $ <17$  \\
\hline
\end{tabular}\end{center}
\end{table}

\begin{table}[t]
\begin{center}
\caption{ ${\cal B}(B\to PP)$ (in units of $10^{-6}$) in model III using the
BSW form factors, with $k^2=m_b^2/2$, $N^{eff}=2,\; 3,\; \infty$, $\mhp=200$GeV and
$\theta=0^0, 30^0$, respectively. }
\label{bppm3}
\vspace{0.2cm}
\begin{tabular} {l|c|c|c| c|c|c| c|c|c| c|c|c}  \hline
 &  \multicolumn{3}{c|}{$\theta=0^0$ }& \multicolumn{3}{c|}{$\delta {\cal  B}\;
 [\%]$}& \multicolumn{3}{c|}{$\theta=30^0$}&
 \multicolumn{3}{c}{$\delta {\cal  B}\; [\%]$}   \\ \cline{1-13}
Channel &  $2$& $3$ & $\infty$ & $2$& $3$ & $\infty$
                & $2$& $3$ & $\infty$& $2$ & $3$ & $\infty$  \\ \hline
$B^0 \to \pi^+ \pi^-$        & $9.33 $ &$10.6  $&$13.3 $&$2.5$&$2.5$&$2.4$&$8.83$ &$10.0$&$12.6$&$-3.0$ &$ -3.1$&$-3.1$ \\
$B^0 \to \pi^0 \pi^0$        & $0.36 $ &$0.25  $&$1.03 $&$30$&$61$&$13$&$0.39$ &$0.23$&$0.92$&$ 40$ &$  52$&$-0.5$   \\
$ B^+ \to \pi^+ \pi^0$       & $6.41 $ &$5.06  $&$2.85 $&$0.0$&$0.0$&$0.0$&$6.41$ &$5.06$&$2.85$&$ 0.0$ &$  0.0$&$ 0.0$  \\ \hline
$ B^0 \to \eta \eta$         & $0.18 $ &$0.15  $&$0.36 $&$29$&$47$&$21$&$0.16$ &$0.14$&$0.38$&$ 15$ &$  39$&$ 29$   \\
$ B^0 \to \eta \eta^\prime$  & $0.19 $ &$0.13  $&$0.46 $&$29$&$68$&$20$&$0.16$ &$0.12$&$0.506$&$ 9.6$ &$  57$&$ 30$  \\
$ B^0 \to \eta' \eta^\prime$ & $0.05 $ &$0.02  $&$0.15 $&$20$&$127$&$17$&$0.04$ &$0.02$&$0.16$&$-5.4$ &$  103$&$ 30$  \\ \hline
$ B^+ \to \pi^+ \eta$        & $3.82 $ &$3.13  $&$2.20 $&$8.7$&$13$&$26$&$3.48$ &$2.81$&$1.95$&$-1.1$ &$  1.2$&$ 12$  \\
$ B^+ \to \pi^+ \eta^\prime$ & $2.63 $ &$2.05  $&$1.27 $&$5.4$&$9.0$&$24$&$2.37$ &$1.81$&$1.09$&$-4.8$ &$ -3.5$&$ 6.5$   \\
$ B^0 \to \pi^0 \eta$        & $0.39 $ &$0.44  $&$0.59 $&$50$&$51$&$49$&$0.36$ &$0.42$&$0.57$&$ 39$ &$  43$&$ 46$    \\
$ B^0 \to \pi^0 \eta^\prime$ & $0.11 $ &$0.14  $&$0.25 $&$92$&$91$&$72$&$0.10$ &$0.13$&$0.25$&$ 65$ &$  76$&$ 70$    \\ \hline
$B^+ \to K^+ \pi^0$          & $17.4 $ &$19.6  $&$24.4 $&$45$&$45$&$46$&$17.2$ &$19.3$&$23.8$&$ 43$ &$  43$&$ 43$    \\
$B^0 \to K^+ \pi^-$          & $26.8 $ &$29.9  $&$36.5 $&$51$&$51$&$53$&$26.5$ &$29.5$&$36.1$&$ 49$ &$  49$&$ 51$    \\
$B^+ \to K^0 \pi^+$          & $29.8 $ &$34.6  $&$45.3 $&$50$&$49$&$48$&$28.6$ &$33.3$&$43.5$&$ 44$ &$  43$&$ 42$    \\
$B^0 \to K^0 \pi^0$          & $11.4 $ &$13.0  $&$16.7 $&$57$&$57$&$56$&$10.8$ &$12.5$&$16.1$&$ 49$ &$  50$&$ 51$    \\ \hline
$ B^+ \to  K^+ \eta$         & $5.69 $ &$6.63  $&$8.78 $&$45$&$45$&$45$&$5.30$ &$6.22$&$8.34$&$ 36$ &$  36$&$ 37$    \\
$ B^+ \to  K^+ \eta^\prime$  & $38.0 $ &$46.9  $&$67.5 $&$68$&$65$&$59$&$36.8$ &$45.2$&$64.9$&$ 63$ &$  59$&$ 53$    \\
$B^0 \to K^0 \eta$           & $4.86 $ &$5.50  $&$6.93 $&$51$&$51$&$51$&$4.63$ &$5.27$&$6.73$&$ 44$ &$  45$&$ 47$     \\
$B^0 \to K^0 \eta^\prime$    & $36.7 $ &$45.9  $&$67.3 $&$67$&$63$&$57$&$35.1$ &$43.8$&$64.2$&$ 60$ &$  56$&$ 49$    \\ \hline
$ B^+ \to K^+ \bar K^0$      & $1.73 $ &$2.01  $&$2.62 $&$49$&$48$&$47$&$1.64$ &$1.91$&$2.49$&$ 41$ &$  41$&$ 40$    \\
$ B^0 \to K^0\bar K^0$       & $1.64 $ &$1.90  $&$2.48 $&$49$&$48$&$47$&$1.55$ &$1.80$&$2.36$&$ 41$ &$  41$&$ 40$    \\
\hline
\end{tabular}\end{center}
\end{table}

\begin{table}[t]
\begin{center}
\caption{ ${\cal B}(B\to PP)$ (in units of $10^{-6}$) in models I and II using the
BSW form factors, with $k^2=m_b^2/2$, $N^{eff}=2,\; 3,\; \infty$, $\tan{\beta}=2$
and $\mhp=200$GeV. }
\label{bppm2}
\vspace{0.2cm}
\begin{tabular} {l|c|c|c| c|c|c| c|c|c| c|c|c}  \hline
 &  \multicolumn{3}{c|}{Model I }& \multicolumn{3}{c|}{$\delta {\cal  B}\;
 [\%]$}& \multicolumn{3}{c|}{Model II} &
 \multicolumn{3}{c}{$\delta {\cal  B}\; [\%]$}   \\ \cline{1-13}
Channel &  $2$& $3$ & $\infty$ & $2$& $3$ & $\infty$
                & $2$& $3$ & $\infty$& $2$ & $3$ & $\infty$  \\ \hline
$B^0 \to \pi^+ \pi^-$        & $9.11$ &$10.3 $&$13.0$  &$ 0.1$&$ 0.1$&$0.1$  & $9.1$ &$10.3$&$13.0$  &$-0.5$ &$ -0.4$&$ -0.4$  \\
$B^0 \to \pi^0 \pi^0$        & $0.28$ &$0.15 $&$0.92$  &$-0.1$&$-0.1$&$0.1$  & $0.3$ &$0.1$&$0.9$  &$-6.2$ &$ -12.6$&$ -2.5$  \\
$ B^+ \to \pi^+ \pi^0$       & $6.41$ &$5.06 $&$2.85$  &$ 0.0$&$ 0.0$&$0.0$  & $6.4$ &$5.1$&$2.9$  &$ 0.0$ &$  0.0$&$  0.0$   \\ \hline
$ B^0 \to \eta \eta$         & $0.14$ &$0.11 $&$0.30$  &$ 1.5$&$ 2.4$&$1.1$  & $0.1$ &$0.1$&$0.3$  &$-4.7$ &$ -7.5$&$ -3.4$   \\
$ B^0 \to \eta \eta^\prime$  & $0.14$ &$0.08 $&$0.38$  &$ 1.0$&$ 2.3$&$0.7$  & $0.1$ &$0.1$&$0.4$  &$-4.8$ &$ -11.4$&$ -3.5$   \\
$ B^0 \to \eta' \eta^\prime$ & $0.04$ &$0.01 $&$0.13$  &$ 0.3$&$ 2.0$&$0.2$  & $0.04$ &$0.01$&$0.1$  &$-2.7$ &$ -20.2$&$ -3.1$   \\ \hline
$ B^+ \to \pi^+ \eta$        & $3.53$ &$2.79 $&$1.77$  &$ 0.4$&$ 0.5$&$1.0$  & $3.5$ &$2.7$&$1.7$  &$-1.5$ &$ -2.1$&$ -4.40$   \\
$ B^+ \to \pi^+ \eta^\prime$ & $2.50$ &$1.88 $&$1.03$  &$ 0.1$&$ 0.2$&$0.4$  & $2.5$ &$1.9$&$1.0$  &$-0.9$ &$ -1.6$&$ -4.5$   \\
$ B^0 \to \pi^0 \eta$        & $0.26$ &$0.30 $&$0.40$  &$ 1.5$&$ 1.6$&$1.7$  & $0.2$ &$0.3$&$0.4$  &$-9.0$ &$ -9.1$&$ -8.6$   \\
$ B^0 \to \pi^0 \eta^\prime$ & $0.06$ &$0.08 $&$0.15$  &$ 0.9$&$ 0.9$&$0.7$  & $0.05$ &$0.1$&$0.1$  &$-15.9$ &$ -16.4$&$ -13.5$   \\ \hline
$B^+ \to K^+ \pi^0$          & $12.3$ &$13.8 $&$17.1$  &$ 2.4$&$ 2.3$&$2.2$  & $11.2$ &$12.5$&$15.4$  &$-7.1$ &$ -7.3$&$ -7.5$   \\
$B^0 \to K^+ \pi^-$          & $18.0$ &$20.0 $&$24.2$  &$ 1.3$&$ 1.3$&$1.2$  & $16.2$ &$17.9$&$21.6$  &$-9.3$ &$ -9.4$&$ -9.7$   \\
$B^+ \to K^0 \pi^+$          & $20.2$ &$23.6 $&$31.1$  &$ 1.4$&$ 1.5$&$1.6$  & $18.1$ &$21.2$&$28.0$  &$-9.0$ &$ -8.8$&$ -8.5$  \\
$B^0 \to K^0 \pi^0$          & $7.28$ &$8.33 $&$10.7$  &$ 0.1$&$ 0.2$&$0.5$  & $6.4$ &$7.4$&$9.5$  &$-11.6$ &$ -11.5$&$ -11.1$   \\ \hline
$ B^+ \to  K^+ \eta$         & $3.91$ &$4.58 $&$6.12$ &$  0.0$&$ 0.3$&$0.8$  & $3.5$ &$4.1$&$5.6$  &$-9.6$ &$ -9.3$&$ -8.7$  \\
$ B^+ \to  K^+ \eta^\prime$  & $23.0$ &$28.9 $&$43.0$  &$ 1.6$&$ 1.5$&$1.4$  & $19.8$ &$25.2$&$37.8$  &$-12.1$ &$ -11.6$&$ -10.8$   \\
$B^0 \to K^0 \eta$           & $3.24$ &$3.66 $&$4.63$  &$ 0.6$&$ 0.8$&$1.1$  & $2.9$ &$3.3$&$4.1$  &$-9.9$ &$ -9.8$&$ -9.5$    \\
$B^0 \to K^0 \eta^\prime$    & $22.3$ &$28.5 $&$43.5$  &$ 1.4$&$ 1.4$&$1.3$  & $19.3$ &$24.9$&$38.5$  &$-12.2$ &$ -11.5$&$ -10.5$   \\ \hline
$ B^+ \to K^+ \bar K^0$      & $1.18$ &$1.37 $&$1.81$  &$ 1.4$&$ 1.5$&$1.6$  & $1.1$ &$1.2$&$1.6$  &$-8.9$ &$ -8.7$&$ -8.4$   \\
$ B^0 \to K^0\bar K^0$       & $1.11$ &$1.30 $&$1.71$  &$ 1.4$&$ 1.5$&$1.6$  & $1.0$ &$1.2$&$1.5$  &$-8.9$ &$ -8.7$&$ -8.4$   \\
\hline
\end{tabular}\end{center}
\end{table}

\begin{table}[t]
\begin{center}
\caption{$B\to PV$ branching ratios (in units of $10^{-6}$) using the BSW
[ LQQSR ] form factors in the SM, with  with $k^2=m_b^2/2$, $N^{eff}=2,\;
3,\; \infty$. The last column shows the CLEO measurements
and upper limits ($90\%$ C.L.) [22-25,27]. }
\label{bpvsm}
\begin{tabular} {l|l|c|c|c|l}  \hline
Channel & Class & $N^{eff}=2$& $N^{eff}=3$ & $N^{eff}=\infty$ &Data  \\ \hline
\parbox[c]{2.5cm}{$B^0 \to \rho^+  \pi^- $ \\
                $B^0 \to \rho^-  \pi^+ $}  &\parbox[l]{1cm}{ I\\I} &
\parbox[c]{2.5cm}{\centering$21.1\;[25.1]$\\
                          $5.7\;[6.5]$ }&
\parbox[c]{2.5cm}{\centering$24.0 \;[28.5]$ \\
                          $6.5 \;[7.4]$ }&
\parbox[c]{2.5cm}{\centering$30.3\;[36.0]$\\
                          $8.2 \;[9.4]$} & $ \}\; 27.6^{+8.4}_{-7.4}\pm 4.2$  \\
$ B^0 \to \rho^0 \pi^0$       &II &$0.49\;[0.58]$&$0.06\;[0.07]$&$2.05\; [2.41]$&$< 5.1$ \\
$ B^+ \to \rho^0 \pi^+ $      &III&$5.72\;[6.63]$&$3.46\;[3.97]$&$0.71\; [0.78]$&$10.4^{+3.3}_{-3.4}\pm 2.1$ \\
$ B^+ \to \rho^+ \pi^0$       &III&$13.5\;[16.0]$&$12.6\;[15.0]$&$10.9\; [13.1]$&$<43$\\ \hline
$ B^0 \to \rho^0 \eta $       &II &$0.01\;[0.02]$&$0.02\;[0.02]$&$0.06\; [0.08]$&$<10$  \\
$ B^0 \to \rho^0 \eta^\prime$ &II&$0.01\;[0.01]$&$0.002\;[0.003]$&$0.03\;[0.03]$&$<12$     \\
$ B^+ \to \rho^+ \eta $       &III&$5.44\;[6.57]$&$4.75\;[5.79]$&$3.54\; [4.38]$&$<15$     \\
$ B^+ \to \rho^+ \eta ^\prime$&III&$4.35\;[5.02]$&$3.81\;[4.40]$&$2.85\; [3.29]$&$<33$     \\ \hline
$ B^0 \to \omega \pi^0$       &II &$0.29\;[0.35]$&$0.08\;[0.09]$&$0.15\; [0.19]$&$<5.5$    \\
$ B^+ \to \omega \pi^+ $      &III&$6.32\;[7.35]$&$3.75\;[4.31]$&$0.78\; [0.85]$&$11.3^{+3.3}_{-2.9}\pm 1.4$ \\
$ B^0 \to \omega \eta $       &II &$0.32\;[0.38]$&$0.03\;[0.04]$&$0.82\; [0.98]$&$<12 $   \\
$ B^0 \to \omega \eta^\prime$ &II&$0.20\;[0.23]$&$0.001\;[0.002]$&$0.68\;[0.79]$&$<60$    \\ \hline
$ B^0 \to \phi \pi^0$         &V &$0.03\;[0.04]$&$0.002\;[0.002]$&$0.23\;[0.27]$&$<5.4$   \\
$ B^+ \to \phi \pi^+ $        &V &$0.06\;[0.08]$&$0.004\;[0.005]$&$0.49\;[0.58]$&$<4$     \\
$ B^0 \to \phi \eta $         &V &$0.01\;[0.01]$&$0.001\;[0.001]$&$0.09\;[0.10]$&$<9 $    \\
$ B^0 \to  \phi \eta^\prime $ &V &$0.01\;[0.01]$&$0.001\;[0.001]$&$0.07\;[0.08]$&$<31$   \\ \hline
$ B^+ \to  \bar K^{*0} K^+ $  &IV &$0.42\;[0.49]$&$0.53\;[0.61]$&$0.78\; [0.90]$&$<5.3$    \\
$ B^0 \to \bar{K}^{*0} K^0$   &IV &$0.40\;[0.46]$&$0.50\;[0.58]$&$0.73\; [0.89]$&$-$   \\
$ B^+ \to K^{*+}  \bar{K}^0$ &V &$0.005\;[0.007]$&$0.002\;[0.003]$&$0.001\;[0.001]$&$-$    \\
$ B^0 \to K^{*0} \bar{K}^0$&IV&$0.004\;[0.006]$&$0.002\;[0.003]$&$0.001\;[0.001]$&$< 12$  \\ \hline
$ B^0 \to \rho^0 K^0$        &IV  &$0.52\;[0.60]$&$0.53\;[0.62]$&$0.71\; [0.83]$&$< 27$ \\
$ B^+ \to \rho^0 K^+ $       &IV  &$0.39\;[0.46]$&$0.31\;[0.36]$&$0.31\; [0.36]$&$< 17$   \\
$ B^0 \to  \rho^- K^{+} $      &I &$0.54\;[0.62]$&$0.59\;[0.68]$&$0.70\; [0.81]$&$ <25$      \\
$ B^+ \to \rho^+ K^0$          &IV&$0.11\;[0.12]$&$0.05\;[0.05]$&$0.01\; [0.01]$&$< 48$      \\ \hline
$ B^+ \to K^{*+} \eta$         &IV&$2.43\;[3.12]$&$2.39\;[3.04]$&$2.32\; [2.89]$&$26.4^{+9.6}_{-8.2}\pm 3.3$  \\
$ B^+ \to K^{*+} \etap$       &III&$0.66\;[1.14]$&$0.36\;[0.61]$&$0.24\; [0.23]$&$<35$                        \\
$ B^0 \to K^{*0} \eta$         &IV&$2.32\;[2.98]$&$2.54\;[3.23]$&$3.06\; [3.82]$&$13.8^{+5.5}_{-4.6} \pm 1.6$ \\
$ B^0 \to K^{*0} \etap$        &V &$0.33\;[0.69]$&$0.09\;[0.23]$&$0.31\; [0.26]$&$<20$                        \\ \hline
$ B^0 \to K^{*+} \pi^-$        &IV&$8.59\;[10.2]$&$9.67\;[11.5]$&$12.0\; [14.3]$&$22 ^{+8\, +4}_{-6\, -5}$    \\
$ B^0 \to K^{*0} \pi^0$        &IV&$2.44\;[2.77]$&$3.02\;[3.43]$&$4.42\; [5.01]$&$<3.6$                       \\
$ B^+ \to K^{*+ }\pi^0$        &IV&$4.95\;[6.09]$&$5.55\;[6.84]$&$6.91\; [8.52]$&$<31$                        \\
$ B^+ \to  K^{*0} \pi^+ $      &IV&$7.35\;[8.75]$&$9.23\;[11.0]$&$13.6\; [16.2]$&$< 16$                       \\ \hline
$ B^+ \to \phi K^+ $           &V &$22.1\;[25.7]$&$11.5\;[13.4]$&$0.60\; [0.70]$&$17.2^{+6.7}_{-5.4}\pm 1.8 $                      \\
$ B^0 \to \phi K^0 $           &V &$20.9\;[24.3]$&$10.9\;[12.6]$&$0.57\; [0.66]$&$< 28$                       \\ \hline
$ B^0 \to \omega K^0 $         &V&$3.31\;[3.86]$&$0.002\;[0.003]$&$13.3\;[15.4]$&$< 21$                       \\
$ B^+  \to \omega K^+ $        &V &$3.53\;[4.11]$&$0.25\;[0.28]$&$16.5\; [19.2]$&$< 7.9$                      \\
\hline
\end{tabular}\end{center}
\end{table}

\begin{table}[t]
\begin{center}
\caption{$B\to PV$ branching ratios (in units of $10^{-6}$)
using the BSW form factors in model III, assuming $\mhp=200$GeV,
$\theta=0^0$ and  $N^{eff}=2,3,\infty$. }
\label{bpvm3}
\vspace{0.2cm}
\begin{tabular} {l|l|c|c|c| c|c|c| c|c|c } \hline
 & & \multicolumn{3}{c|}{SM }& \multicolumn{3}{c|}{ Model III }&
\multicolumn{3}{c}{$\delta {\cal B}\; [\%]$}  \\ \cline{3-11}
Channel & Class& $2$& $3$ & $\infty$ & $2$& $3$ & $\infty$
                & $2$& $3$ & $\infty$  \\ \hline
$ B^0 \to \rho^+  \pi^- $     &I  &$21.1$  &$24.0$  &$30.3$&  $21.2$  &$24.1$  &$30.5$&  $ 0.7$  &$ 0.7$  &$ 0.7$   \\
$ B^0 \to \rho^-  \pi^+ $     &I  &$5.70$  &$6.48$  &$8.19$&  $5.70$  &$6.48$  &$8.19$&  $ 0.0$  &$ 0.0$  &$ 0.0$ \\
$ B^0 \to \rho^0 \pi^0$       &II &$0.49$  &$0.06$  &$2.05$&  $0.54$  &$0.11$  &$2.12$&  $ 9.8$  &$ 99.6$  &$ 3.5$  \\
$ B^+ \to \rho^0 \pi^+ $      &III&$5.72$  &$3.46$  &$0.71$&  $5.79$  &$3.54$  &$0.81$&  $ 1.3$  &$ 2.3$  &$ 14.0$    \\
$ B^+ \to \rho^+ \pi^0$       &III&$13.5$  &$12.6$  &$10.9$&  $13.6$  &$12.7$  &$11.0$&  $ 0.4$  &$ 0.5$  &$ 0.7$    \\ \hline
$ B^0 \to \rho^0 \eta $       &II &$0.01$  &$0.02$  &$0.06$&  $0.03$  &$0.03$  &$0.08$&  $ 86.0$  &$ 100$  &$ 40.1$  \\
$ B^0 \to \rho^0 \eta^\prime$ &II &$0.01$  &$0.003$ &$0.03$&  $0.004$  &$0.001$  &$0.03$&  $-47.3$  &$-54.4$  &$ 18.1$    \\
$ B^+ \to \rho^+ \eta $       &III&$5.44$  &$4.75$  &$3.54$&  $5.46$  &$4.79$  &$3.59$&  $ 0.5$  &$ 0.7$  &$ 1.4$    \\
$ B^+ \to \rho^+ \eta ^\prime$&III&$4.35$  &$3.81$  &$2.85$&  $4.34$  &$3.81$  &$2.86$&  $-0.2$  &$-0.08$  &$ 0.4$ \\ \hline
$ B^0 \to \omega \pi^0$       &II &$0.29$  &$0.08$  &$0.15$&  $0.45$  &$0.14$  &$0.15$&  $ 54.4$  &$ 77.0$  &$ 0.8$   \\
$ B^+ \to \omega \pi^+ $      &III&$6.32$  &$3.75$  &$0.78$&  $6.63$  &$3.86$  &$0.79$&  $ 5.0$  &$ 3.1$  &$ 1.1$  \\
$ B^0 \to \omega \eta $       &II &$0.32$  &$0.03$  &$0.82$&  $0.37$  &$0.05$  &$0.83$&  $ 16.3$  &$ 67.1$  &$ 0.2$  \\
$ B^0 \to \omega \eta^\prime$ &II &$0.20$  &$0.001$ &$0.68$&  $0.22$  &$0.004$  &$0.69$&  $ 9.5$  &$ 155$  &$ 0.5$  \\ \hline
$ B^0 \to \phi \pi^0$         &V  &$0.03$  &$0.002$ &$0.23$&  $0.05$  &$0.002$  &$0.33$&  $ 59.1$  &$ 1.9$  &$ 42.3$  \\
$ B^+ \to \phi \pi^+ $        &V  &$0.06$  &$0.004$ &$0.49$&  $0.10$  &$0.004$  &$0.69$&  $ 59.1$  &$ 1.9$  &$ 42.3$  \\
$ B^0 \to \phi \eta $         &V  &$0.01$  &$0.001$ &$0.09$&  $0.02$  &$0.001$  &$0.12$&  $ 59.1$  &$ 1.9$  &$ 42.3$  \\
$ B^0 \to  \phi \eta^\prime $ &V  &$0.01$  &$0.001$ &$0.07$&  $0.01$  &$0.001$  &$0.10$&  $ 59.1$  &$ 1.9$  &$ 42.3$  \\ \hline
$ B^+ \to  \bar K^{*0} K^+ $  &IV &$0.42$  &$0.53$  &$0.78$&  $0.68$  &$0.83$  &$1.19$&  $ 61.0$  &$ 57.9$  &$ 53.3$ \\
$ B^0 \to \bar{K}^{*0} K^0$   &IV &$0.40$  &$0.50$  &$0.73$&  $0.64$  &$0.79$  &$1.12$&  $ 61.0$  &$ 57.9$  &$ 53.3$ \\
$ B^+ \to K^{*+}  \bar K^0$   &V  &$0.005$&$0.002$ &$0.001$&  $0.002$  &$0.001$  &$0.003$&  $-58.4$  &$-72.5$  &$ 256$  \\
$ B^0 \to K^{*0} \bar{ K}^0$  &IV &$0.004$&$0.002$ &$0.001$&  $0.002$  &$0.001$  &$0.003$&  $-58.4$  &$-72.5$  &$ 256$  \\ \hline
$ B^0 \to \rho^0 K^0$         &IV &$0.52$  &$0.53$  &$0.71$&  $0.43$  &$0.44$  &$0.60$&  $-17.1$  &$-18.3$  &$-16.5$    \\
$ B^+ \to \rho^0 K^+ $        &IV &$0.39$  &$0.31$  &$0.31$&  $0.43$  &$0.36$  &$0.40$&  $ 8.0$  &$ 16.4$  &$ 30.6$  \\
$ B^0 \to  \rho^- K^{+} $     &I  &$0.54$  &$0.59$  &$0.70$&  $0.47$  &$0.52$  &$0.62$&  $-13.1$  &$-12.7$  &$-11.8$     \\
$ B^+ \to \rho^+ K^0$         &IV &$0.11$  &$0.05$  &$0.01$&  $0.05$  &$0.01$  &$0.02$&  $-50.8$  &$-70.1$  &$ 363$  \\ \hline
$ B^+ \to K^{*+} \eta$        &IV &$2.43$  &$2.39$  &$2.32$&  $3.27$  &$3.29$  &$3.34$&  $ 34.4$  &$ 37.4$  &$ 43.5$ \\
$ B^+ \to K^{*+} \etap$       &III&$0.66$  &$0.36$  &$0.24$&  $0.31$  &$0.24$  &$0.65$&  $-52.2$  &$-34.2$  &$ 170$  \\
$ B^0 \to K^{*0} \eta$        & IV&$2.32$  &$2.54$  &$3.06$&  $3.15$  &$3.47$  &$4.20$&  $ 35.8$  &$ 36.5$  &$ 37.4$ \\
$ B^0 \to K^{*0} \etap$       &V  &$0.33$  &$0.09$  &$0.31$&  $0.08$  &$0.10$  &$0.96$&  $-77.3$  &$ 6.9$  &$ 204$ \\ \hline
$ B^0 \to K^{*+} \pi^-$       &IV &$8.59$  &$9.67$  &$12.0$&  $13.6$  &$15.4$  &$19.1$&  $ 58.6$  &$ 58.8$  &$ 59.3$ \\
$ B^0 \to K^{*0} \pi^0$       &IV &$2.44$  &$3.02$  &$4.42$&  $4.26$  &$5.18$  &$7.34$&  $ 74.9$  &$ 71.6$  &$ 66.0$ \\
$ B^+ \to K^{*+ }\pi^0$       &IV &$4.95$  &$5.55$  &$6.91$&  $7.42$  &$8.38$  &$10.5$&  $ 49.9$  &$ 50.9$  &$ 52.2$ \\
$ B^+ \to  K^{*0} \pi^+ $     &IV &$7.35$  &$9.23$  &$13.6$&  $11.9$  &$14.7$  &$21.0$&  $ 62.0$  &$ 58.8$  &$ 54.1$ \\ \hline
$ B^+ \to \phi K^+ $          &V  &$22.1$  &$11.5$  &$0.60$&  $35.7$  &$19.0$  &$1.29$&  $ 61.5$  &$ 65.3$  &$ 113$ \\
$ B^0 \to \phi K^0 $          &V  &$20.9$  &$10.9$  &$0.57$&  $33.7$  &$18.0$  &$1.21$&  $ 61.5$  &$ 65.3$  &$ 113$ \\ \hline
$ B^0 \to \omega K^0 $        &V  &$3.31$  &$0.002$ &$13.3$&  $5.33$  &$0.01$  &$19.4$&  $ 60.9$  &$ 175$  &$ 46.5$  \\
$ B^+  \to \omega K^+ $       &V  &$3.53$  &$0.25$  &$16.5$&  $5.57$  &$0.23$  &$23.6$&  $ 57.8$  &$-7.7$  &$ 42.9$  \\
\hline
\end{tabular}\end{center}
\end{table}

\begin{table}[t]
\begin{center}
\caption{$B\to PV$ branching ratios (in units of $10^{-6}$) using the BSW
form factors in models I and  II, assuming $\mhp=200$GeV, $tan{\beta}=2$
and $N^{eff}=2,3,\infty$. }
\label{bpvm2}
\vspace{0.2cm}
\begin{tabular} {l|c|c|c| c|c|c| c|c|c |c|c|c} \hline
 &  \multicolumn{3}{c|}{Model I  }& \multicolumn{3}{c|}{$\delta {\cal B}\; [\%]$ }&
 \multicolumn{3}{c|}{ Model II }
 &\multicolumn{3}{c}{$\delta {\cal B}\; [\%]$}  \\ \cline{2-13}
Channel & $2$& $3$ & $\infty$ &$2$& $3$ & $\infty$ & $2$& $3$ & $\infty$
                & $2$& $3$ & $\infty$   \\ \hline
$ B^0 \to \rho^+  \pi^- $     &$ 21.1$ &$24.0$ &$30.3$ &$ 0.0$ &$ 0.0$ &$ 0.0$ &$21.1$  &$23.9$ &$30.2$  &$ -0.1$ &$-0.1$  &$ -0.1$   \\
$ B^0 \to \rho^-  \pi^+ $     &$ 5.70$ &$6.48$ &$8.19$ &$-0.0$ &$-0.0$ &$ 0.0$ &$5.70$  &$6.48$ &$8.19$  &$  0.0$ &$ 0.0$  &$  0.0$   \\
$ B^0 \to \rho^0 \pi^0$       &$ 0.49$ &$0.06$ &$2.05$ &$-0.1$ &$-0.4$ &$ 0.0$ &$0.48$  &$0.05$ &$2.04$  &$ -1.9$ &$-18.9$  &$ -0.6$   \\
$ B^+ \to \rho^0 \pi^+ $      &$ 5.72$ &$3.46$ &$0.71$ &$-0.0$ &$-0.0$ &$ 0.1$ &$5.70$  &$3.44$ &$0.69$  &$ -0.3$ &$-0.5$  &$ -2.8$     \\
$ B^+ \to \rho^+ \pi^0$       &$ 13.5$ &$12.6$ &$10.9$ &$ 0.0$ &$ 0.0$ &$ 0.0$ &$13.5$  &$12.6$ &$10.9$  &$ -0.1$ &$-0.1$  &$ -0.1$     \\ \hline
$ B^0 \to \rho^0 \eta $       &$ 0.02$ &$0.02$ &$0.06$ &$ 2.9$ &$ 3.7$ &$ 1.6$ &$0.01$  &$0.01$ &$0.05$  &$ -12.8$ &$-15.2$  &$ -6.2$    \\
$ B^0 \to \rho^0 \eta^\prime$ &$ 0.01$ &$0.002$ &$0.03$ &$-0.2$&$-0.1$ &$ 0.0$ &$0.01$  &$0.003$ &$0.03$  &$  18.1$ &$ 39.1$  &$ -1.1$     \\
$ B^+ \to \rho^+ \eta $       &$ 5.44$ &$4.76$ &$3.54$ &$ 0.0$ &$ 0.0$ &$ 0.1$ &$5.43$  &$4.75$ &$3.53$  &$ -0.1$ &$-0.1$  &$ -0.2$     \\
$ B^+ \to \rho^+ \eta ^\prime$&$ 4.35$ &$3.81$ &$2.85$ &$-0.0$ &$-0.0$ &$ 0.0$ &$4.36$  &$3.81$ &$2.85$  &$  0.1$ &$ 0.1$  &$ -0.0$   \\ \hline
$ B^0 \to \omega \pi^0$       &$ 0.29$ &$0.08$ &$0.15$ &$ 0.2$ &$-0.4$ &$ 0.5$ &$0.26$  &$0.07$ &$0.15$  &$ -10.7$ &$-15.4$  &$  0.4$    \\
$ B^+ \to \omega \pi^+ $      &$ 6.33$ &$3.75$ &$0.78$ &$ 0.1$ &$ 0.1$ &$ 0.0$ &$6.26$  &$3.73$ &$0.78$  &$ -0.9$ &$-0.5$  &$ -0.2$   \\
$ B^0 \to \omega \eta $       &$ 0.32$ &$0.03$ &$0.82$ &$ 0.7$ &$ 3.6$ &$ 0.0$ &$0.31$  &$0.03$ &$0.82$  &$ -2.7$ &$-9.9$  &$ -0.1$   \\
$ B^0 \to \omega \eta^\prime$ &$ 0.209$ &$0.001$ &$0.68$&$ 0.2$ &$ 0.3$&$ 0.1$ &$0.20$  &$0.002$ &$0.68$  &$ -1.3$ &$ 8.7$  &$ -0.1$   \\ \hline
$ B^0 \to \phi \pi^0$         &$ 0.03$ &$0.002$ &$0.23$&$-0.8$ &$ 10.9$&$ 2.4$ &$0.03$  &$0.002$ &$0.21$  &$ -12.9$ &$ 10.9$  &$ -6.7$    \\
$ B^+ \to \phi \pi^+ $        &$ 0.06$ &$0.004$ &$0.50$&$-0.8$ &$ 10.9$&$ 2.4$ &$0.06$  &$0.005$ &$0.45$  &$ -12.9$ &$ 10.9$  &$ -6.7$    \\
$ B^0 \to \phi \eta $         &$ 0.01$ &$0.001$ &$0.09$&$-0.8$ &$ 10.9$&$ 2.4$ &$0.01$  &$0.001$ &$0.08$  &$ -12.9$ &$ 10.9$  &$ -6.7$    \\
$ B^0 \to  \phi \eta^\prime $ &$ 0.01$ &$0.001$ &$0.07$&$-0.8$ &$ 10.9$&$ 2.4$ &$0.01$  &$0.001$ &$0.06$  &$ -12.9$ &$ 10.9$  &$ -6.7$    \\ \hline
$ B^+ \to  \bar K^{*0} K^+ $  &$ 0.43$ &$0.54$ &$0.79$ &$ 1.8$ &$ 1.9$ &$ 2.1$ &$0.37$  &$0.47$ &$0.71$  &$ -10.7$ &$-10.0$  &$ -9.0$    \\
$ B^0 \to \bar{K}^{*0} K^0$   &$ 0.40$ &$0.51$ &$0.75$ &$ 1.8$ &$ 1.9$ &$ 2.1$ &$0.35$  &$0.45$ &$0.67$  &$ -10.7$ &$-10.0$  &$ -9.0$    \\
$ B^+ \to K^{*+}  \bar K^0$   &$ 0.005$ &$0.002$ &$0.001$&$-3.5$&$-7.1$&$13.5$ &$0.01$  &$0.002$ &$0.001$  &$  14.3$ &$ 20.9$  &$ -16.6$   \\
$ B^0 \to K^{*0} \bar{ K}^0$  &$ 0.004$ &$0.002$ &$0.001$&$-3.5$&$-7.1$&$13.5$ &$0.005$ &$0.002$ &$0.001$  &$  14.3$ &$ 20.9$  &$ -16.6$    \\ \hline
$ B^0 \to \rho^0 K^0$         &$ 0.53$ &$0.55$ &$0.74$ &$ 3.1$ &$ 3.3$ &$ 3.1$ &$0.56$  &$0.58$ &$0.77$  &$  7.6$ &$ 8.2$  &$  7.6$      \\
$ B^+ \to \rho^0 K^+ $        &$ 0.40$ &$0.32$ &$0.34$ &$ 2.4$ &$ 4.9$ &$ 9.0$ &$0.40$  &$0.31$ &$0.32$  &$  0.7$ &$ 1.4$  &$  2.4$     \\
$ B^0 \to  \rho^- K^{+} $     &$ 0.53$ &$0.58$ &$0.70$ &$-1.8$ &$-1.4$ &$-0.6$ &$0.55$  &$0.60$ &$0.72$  &$  1.6$ &$ 2.0$  &$  2.7$      \\
$ B^+ \to \rho^+ K^0$         &$ 0.10$ &$0.04$ &$0.005$ &$-3.0$ &$-6.3$&$11.6$ &$0.12$  &$0.05$ &$0.005$  &$  11.6$ &$ 16.7$  &$ -6.5$    \\ \hline
$ B^+ \to K^{*+} \eta$        &$ 2.49$ &$2.45$ &$2.36$ &$ 2.5$ &$ 2.3$ &$ 1.7$ &$2.31$  &$2.26$ &$2.15$  &$ -4.9$ &$-5.8$  &$ -7.5$   \\
$ B^+ \to K^{*+} \etap$       &$ 0.64$ &$0.35$ &$0.24$ &$-2.4$ &$-2.3$ &$ 2.0$ &$0.77$  &$0.43$ &$0.21$  &$  16.6$ &$ 19.5$  &$ -12.5$    \\
$ B^0 \to K^{*0} \eta$        &$ 2.37$ &$2.60$ &$3.13$ &$ 2.4$ &$ 2.3$ &$ 2.2$ &$2.19$  &$2.40$ &$2.88$  &$ -5.4$ &$-5.6$  &$ -5.9$   \\
$ B^0 \to K^{*0} \etap$       &$ 0.32$ &$0.09$ &$0.33$ &$-3.3$ &$-4.7$ &$ 5.0$ &$0.42$  &$0.13$ &$0.23$  &$  27.3$ &$ 42.6$  &$ -25.4$    \\ \hline
$ B^0 \to K^{*+} \pi^-$       &$ 8.76$ &$9.84$ &$12.2$ &$ 2.0$ &$ 1.8$ &$ 1.4$ &$7.74$  &$8.69$ &$10.8$  &$ -9.8$ &$-10.1$  &$ -10.5$    \\
$ B^0 \to K^{*0} \pi^0$       &$ 2.44$ &$3.03$ &$4.45$ &$ 0.1$ &$ 0.4$ &$ 0.8$ &$2.08$  &$2.60$ &$3.87$  &$ -14.7$ &$-13.9$  &$ -12.5$    \\
$ B^+ \to K^{*+ }\pi^0$       &$ 5.11$ &$5.72$ &$7.10$ &$ 3.2$ &$ 3.0$ &$ 2.7$ &$4.60$  &$5.14$ &$6.35$  &$ -7.2$ &$-7.5$  &$ -8.1$   \\
$ B^+ \to  K^{*0} \pi^+ $     &$ 7.48$ &$9.41$ &$13.9$ &$ 1.8$ &$ 1.9$ &$ 2.1$ &$6.55$  &$8.29$ &$12.4$  &$ -10.8$ &$-10.2$  &$ -9.2$   \\ \hline
$ B^+ \to \phi K^+ $          &$ 22.3$ &$11.6$ &$0.61$ &$ 1.1$ &$ 1.0$ &$ 0.7$ &$19.5$  &$10.1$ &$0.48$  &$ -11.5$ &$-12.2$  &$ -19.9$    \\
$ B^0 \to \phi K^0 $          &$ 21.1$ &$11.0$ &$0.57$ &$ 1.1$ &$ 1.0$ &$ 0.7$ &$18.5$  &$9.54$ &$0.46$  &$ -11.5$ &$-12.2$  &$ -19.9$    \\ \hline
$ B^0 \to \omega K^0 $        &$ 3.35$ &$0.002$ &$13.4$ &$ 1.2$&$ 1.1$ &$ 1.3$ &$2.94$  &$0.003$ &$12.1$  &$ -11.3$ &$ 25.7$  &$ -8.5$   \\
$ B^+  \to \omega K^+ $       &$ 3.59$ &$0.25$ &$16.7$ &$ 1.6$ &$-0.1$ &$ 1.4$ &$3.17$  &$0.25$ &$15.2$  &$ -10.1$ &$ 2.3$  &$ -7.7$   \\
\hline
\end{tabular}\end{center}
\end{table}

\begin{table}[t]
\begin{center}
\caption{$B\to V V$ branching ratios (in units of $10^{-6}$) using the BSW
[ LQQSR ] form factors in the SM,  with $k^2=m_b^2/2$, $N^{eff}=
2,\; 3,\; \infty$.  The last column
shows the CLEO upper limits ($90\%$ C.L.) [22-25].}
\label{bvvsm}
\vspace{0.2cm}
\begin{tabular} {l|l|c|c|c|l} \hline
Channel & Class & $N^{eff}=2$& $N^{eff}=3$ & $N^{eff}=\infty$ &Data  \\ \hline
$ B^0 \to \rho^+ \rho^-$       &I  &$17.8\;[19.8] $&$20.2\;[22.5] $&$25.5\;[28.4] $ &$<2200$ \\
$ B^0 \to \rho^0 \rho^0 $      &II &$0.39\;[0.44] $&$0.09\;[0.10] $&$1.56\;[1.73] $ &$<4.8$ \\
$ B^0 \to \omega \omega$       &II &$0.81\;[0.90] $&$0.15\;[0.17] $&$1.22\;[1.35] $ &$<19$   \\
$ B^+ \to \rho^+ \rho^0 $      &III&$12.8\;[14.3] $&$10.1\;[11.3] $&$5.69\;[6.33] $ &$<120$   \\
$ B^+ \to \rho^+ \omega$       &III&$15.7\;[17.4] $&$12.2\;[13.5] $&$6.69\;[7.45] $ &$<47$   \\
$ B^0 \to K^{*+} \rho^- $      &IV &$6.17\;[6.82] $&$6.95\;[7.68] $&$8.64\;[9.55] $ &$ -$    \\
$ B^0 \to K^{*0} \rho^0 $      &IV &$1.73\;[1.82] $&$2.01\;[2.11] $&$2.79\;[2.91] $ &$<16.1$    \\
$ B^+ \to K^{*+} \rho^0 $      &IV &$5.22\;[5.97] $&$5.97\;[6.82] $&$7.76\;[8.91] $ &$<52$    \\
$ B^+ \to K^{*0} \rho^+ $      &IV &$6.65\;[7.35] $&$8.36\;[9.24] $&$12.4\;[13.7] $ &$-$    \\
$ B^+ \to K^{*+} \bar{K}^{*0}$ &IV &$0.38\;[0.49] $&$0.48\;[0.61] $&$0.70\;[0.90] $ &$<62$    \\
$ B^0 \to K^{*0} \bar{K}^{*0}$ &IV &$0.36\;[0.47] $&$0.46\;[0.58] $&$0.67\;[0.86] $ &$<7.4$    \\
$ B^0 \to \rho^0 \omega$       &V  &$0.45\;[0.50] $&$0.24\;[0.27] $&$0.02\;[0.02] $ &$<11$     \\
$ B^0 \to K^{*0} \omega$       &V  &$13.5\;[16.1] $&$4.52\;[5.03] $&$1.04\;[1.79] $ &$<19$     \\
$ B^+ \to K^{*+} \omega$       &V  &$13.4\;[16.1] $&$3.94\;[4.39] $&$2.74\;[4.01] $ &$<52$     \\
$ B^+ \to K^{*+} \phi$         &V  &$21.8\;[27.8] $&$11.3\;[14.5] $&$0.60\;[0.76] $ &$<41$     \\
$ B^0 \to K^{*0} \phi$         &V  &$20.6\;[26.2] $&$10.7\;[13.6] $&$0.56\;[0.72] $ &$<21$     \\
$ B^+ \to \rho^+ \phi$         &V  &$0.06\;[0.07] $&$0.004\;[0.005] $&$0.47\;[0.52] $ &$<16$     \\
$ B^0 \to \rho^0 \phi$         &V  &$0.03\;[0.03] $&$0.001\;[0.002] $&$0.22\;[0.25] $ &$<13$     \\
$ B^0 \to \omega \phi$         &V  &$0.03\;[0.03] $&$0.001\;[0.002] $&$0.22\;[0.24] $ &$<21$     \\
\hline
\end{tabular}\end{center}
\end{table}

\begin{table}[t]
\begin{center}
\caption{$B\to V V$ branching ratios (in units of $10^{-6}$) using the BSW
form factors in model III, assuming $\mhp=200$ GeV, $N^{eff}=2,3,\infty$ and
$\theta=0^0, 30^0$. }
\label{bvvm3}
\vspace{0.2cm}
\begin{tabular} {l| c|c|c| c|c|c| c|c|c |c|c|c} \hline
 & \multicolumn{3}{c|}{$\theta=0^0$ }& \multicolumn{3}{c|}{$\delta {\cal  B}\; [\%]$}
 & \multicolumn{3}{c|}{$\theta=30^0$}&
   \multicolumn{3}{c }{$\delta{\cal B}\;[\%] $ }  \\ \cline{2-13}
Channel & $2$ & $3$ & $\infty$ & $2$& $3$ & $\infty$
                & $2$ & $3$ & $\infty$  & $2$ & $3$ & $\infty$   \\ \hline
$ B^0 \to \rho^+ \rho^-$       &$17.9$&$20.3$&$25.7$  &$ 0.7$&$ 0.7$&$ 0.7$  &$19.3$&$19.7$&$24.8$  &$-2.6$&$-2.6$&$-2.6$ \\
$ B^0 \to \rho^0 \rho^0 $      &$0.46$&$0.16$&$1.65$  &$ 16$&$ 79$&$ 5.7$  &$0.56$&$0.15$&$1.50$  &$ 28$&$ 67$&$-4.0$ \\
$ B^0 \to \omega \omega$       &$1.05$&$0.24$&$1.23$  &$ 29$&$ 54$&$ 1.0$  &$1.02$&$0.22$&$1.19$  &$ 14$&$ 45$&$-2.3$ \\
$ B^+ \to \rho^+ \rho^0 $      &$12.8$&$10.1$&$5.69$  &$ 0.0$&$ 0.0$&$ 0.0$  &$14.3$&$10.1$&$5.69$  &$ 0.0$&$ 0.0$&$ 0.0$ \\
$ B^+ \to \rho^+ \omega$       &$13.8$&$10.9$&$6.12$  &$-12$&$-11$&$-8.5$  &$15.2$&$10.9$&$6.12$  &$-13$&$-11$&$-8.5$ \\
$ B^0 \to K^{*+} \rho^- $      &$9.78$&$11.0$&$13.8$  &$ 59$&$ 59$&$ 59$  &$11.1$&$11.3$&$14.1$  &$ 62$&$ 63$&$ 63$ \\
$ B^0 \to K^{*0} \rho^0 $      &$3.18$&$3.70$&$4.98$  &$ 84$&$ 84$&$ 78$  &$3.14$&$3.50$&$4.90$  &$ 72$&$ 74$&$ 75$ \\
$ B^+ \to K^{*+} \rho^0 $      &$7.61$&$8.75$&$11.4$  &$ 46$&$ 47 $&$ 47$  &$8.79$&$8.80$&$11.3$  &$ 47$&$ 48$&$ 45$ \\
$ B^+ \to K^{*0} \rho^+ $      &$10.8$&$13.3$&$19.0$  &$ 62$&$ 59$&$ 54$  &$11.4$&$12.7$&$18.2$  &$ 55$&$ 52$&$ 48$ \\
$ B^+ \to K^{*+} \bar{K}^{*0}$ &$0.61$&$0.75$&$1.08$  &$ 61$&$ 58$&$ 53$  &$0.73$&$0.71$&$1.02$  &$ 51$&$ 49$&$ 45$ \\
$ B^0 \to K^{*0} \bar{K}^{*0}$ &$0.59$&$0.72$&$1.03$  &$ 61$&$ 58$&$ 53$  &$0.70$&$0.68$&$0.98$  &$ 51$&$ 49$&$ 45$ \\
$ B^0 \to \rho^0 \omega$       &$0.72$&$0.39$&$0.04$  &$ 61$&$ 65$&$ 79$  &$0.76$&$0.37$&$0.03$  &$ 53$&$ 54$&$ 51$  \\
$ B^0 \to K^{*0} \omega$       &$20.8$&$6.97$&$1.51$  &$ 53$&$ 54$&$ 45$  &$23.5$&$6.58$&$1.49$  &$ 46$&$ 46$&$ 44$  \\
$ B^+ \to K^{*+} \omega$       &$20.8$&$6.23$&$3.47$  &$ 55$&$ 58$&$ 27$  &$24.3$&$6.25$&$3.24$  &$ 51$&$ 59$&$ 18$  \\
$ B^+ \to K^{*+} \phi$         &$35.2$&$18.8$&$1.27$  &$ 62$&$ 65$&$ 114$  &$42.9$&$17.9$&$1.20$  &$ 54$&$ 58$&$ 101$  \\
$ B^0 \to K^{*0} \phi$         &$33.2$&$17.7$&$1.20$  &$ 62$&$ 65$&$ 113$  &$40.4$&$16.9$&$1.13$  &$ 54$&$ 58$&$ 101$  \\
$ B^+ \to \rho^+ \phi$         &$0.10$&$0.004$&$0.67$  &$ 59$&$ 1.9$&$ 42$  &$0.10$&$0.004$&$0.64$  &$ 50$&$ 1.9$&$ 36$  \\
$ B^0 \to \rho^0 \phi$         &$0.05$&$0.002$&$0.32$  &$ 59$&$ 1.9$&$ 42$  &$0.05$&$0.002$&$0.30$  &$ 50$&$ 1.9$&$ 36$  \\
$ B^0 \to \omega \phi$         &$0.05$&$0.002$&$0.32$  &$ 59$&$ 1.9$&$ 42$  &$0.05$&$0.002$&$0.30$  &$ 50$&$ 1.9$&$ 36$  \\
\hline
\end{tabular}
\end{center} \end{table}

\begin{table}[t]
\begin{center}
\caption{$B\to V V$ branching ratios (in units of $10^{-6}$) using the BSW
form factors in models I and II, assuming $\mhp=200$ GeV, $N^{eff}=2,3,\infty$ and
$\tan{\beta}=2$. }
\label{bvvm2}
\vspace{0.2cm}
\begin{tabular} {l| c|c|c| c|c|c| c|c|c |c|c|c} \hline
 & \multicolumn{3}{c|}{Model I }& \multicolumn{3}{c|}{$\delta {\cal  B}\; [\%]$}
 & \multicolumn{3}{c|}{Model II}   & \multicolumn{3}{c}{$\delta {\cal  B}\; [\%]$}  \\ \cline{2-13}
Channel & $2$ & $3$ & $\infty$ & $2$& $3$ & $\infty$ & $2$ & $3$ & $\infty$ & $2$ & $3$ & $\infty$    \\ \hline
$ B^0 \to \rho^+ \rho^-$       &$17.7$&$19.8$&$24.4$&$ 0.0$&$ 0.0$&$ 0.0$  &$17.7$&$20.2$&$24.3$&$-0.1$&$-0.1$&$-0.1$   \\
$ B^0 \to \rho^0 \rho^0 $      &$0.53$&$0.10$&$1.21$&$ 0.1$&$ 0.6$&$ 0.1$  &$0.52$&$0.08$&$1.20$&$-1.8$&$-15$&$-1.1$   \\
$ B^0 \to \omega \omega$       &$0.90$&$0.16$&$0.93$&$ 0.6$&$ 1.1$&$ 0.0$  &$0.86$&$0.14$&$0.93$&$-4.2$&$-9.9$&$-0.2$   \\
$ B^+ \to \rho^+ \rho^0 $      &$13.5$&$10.7$&$6.04$&$ 0.0$&$ 0.0$&$ 0.0$  &$13.6$&$10.1$&$6.04$&$ 2.1$&$ 0.0$&$ 0.0  $   \\
$ B^+ \to \rho^+ \omega$       &$14.6$&$11.6$&$6.50$&$-10.9$&$-9.6$&$-7.4$  &$14.6$&$10.9$&$6.50$&$-11$&$-11$&$-7.4$   \\
$ B^0 \to K^{*+} \rho^- $      &$6.42$&$7.13$&$8.66$&$ 2.1$&$ 1.9$&$ 1.6$  &$5.70  $&$6.25$&$7.71$&$-9.3$&$-10$&$-9.6$   \\
$ B^0 \to K^{*0} \rho^0 $      &$1.70$&$1.97$&$2.72$&$ 0.3$&$ 0.5$&$ 0.7$  &$1.47$&$1.70$&$2.36$&$-13$&$-16$&$-13$   \\
$ B^+ \to K^{*+} \rho^0 $      &$5.44$&$6.08$&$7.63$&$ 2.8$&$ 2.8$&$ 2.7$  &$4.90$&$5.55$&$6.83$&$-7.5$&$-7.0$&$-8.0$   \\
$ B^+ \to K^{*0} \rho^+ $      &$6.55$&$8.13$&$11.8$&$ 1.8$&$ 1.9$&$ 2.1$  &$5.78$&$7.51$&$10.5$&$-10$&$-10$&$-9.3$   \\
$ B^+ \to K^{*+} \bar{K}^{*0}$ &$0.37$&$0.46$&$0.67$&$ 1.8$&$ 1.9$&$ 2.1$  &$0.33$&$0.43$&$0.60$&$-10$&$-10$&$-9.1$   \\
$ B^0 \to K^{*0} \bar{K}^{*0}$ &$0.35$&$0.45$&$0.65$&$ 1.8$&$ 1.9$&$ 2.1$  &$0.32$&$0.41$&$0.57$&$-10$&$-10$&$-9.1$   \\
$ B^0 \to \rho^0 \omega$       &$0.41$&$0.23$&$0.02$&$ 1.1$&$ 1.0$&$ 0.4$  &$0.37$&$0.21$&$0.02$&$-10$&$-12$&$-12$    \\
$ B^0 \to K^{*0} \omega$       &$12.3$&$4.15$&$0.92$&$ 1.3$&$ 1.3$&$ 1.3$  &$11.1$&$4.07$&$0.83$&$-9.3$&$-9.9$&$-8.2$    \\
$ B^+ \to K^{*+} \omega$       &$12.6$&$3.85$&$2.39$&$ 1.6$&$ 1.4$&$ 1.6$  &$11.3$&$3.52$&$2.24$&$-9.4$&$-11$&$-4.9$    \\
$ B^+ \to K^{*+} \phi$         &$20.4$&$10.8$&$0.65$&$ 1.4$&$ 1.4$&$ 2.2$  &$18.0$&$10.0$&$0.54$&$-11$&$-12$&$-16$    \\
$ B^0 \to K^{*0} \phi$         &$19.2$&$10.1$&$0.61$&$ 1.1$&$ 1.0$&$ 0.5$  &$17.0$&$9.41$&$0.51$&$-11$&$-12$&$-16$    \\
$ B^+ \to \rho^+ \phi$         &$0.1 $&$0.01$&$0.44$&$-0.7$&$ 9.7$&$ 2.4$  &$0.05$&$0.005$&$0.40$&$-11$&$ 11$&$-7.5$   \\
$ B^0 \to \rho^0 \phi$         &$0.02$&$0.002$&$0.21$&$-0.7$&$ 9.7$&$ 2.4$  &$0.02$&$0.002$&$0.19$&$-11$&$ 11$&$-7.5$   \\
$ B^0 \to \omega \phi$         &$0.02$&$0.002$&$0.21$&$-0.7$&$ 9.7$&$ 2.4$  &$0.02$&$0.002$&$0.19$&$-11$&$ 11$&$-7.5$   \\
\hline
\end{tabular}\end{center}
\end{table}

\newpage

\newpage

\newpage

\newpage

\newpage

\newpage

\newpage

\newpage

\newpage

\newpage

\newpage

\newpage

\newpage

\newpage

\newpage

\newpage

\newpage

\end{document}